\documentclass[fleqn,usenatbib]{mnras}  

\usepackage{newtxtext,newtxmath}

\usepackage[T1]{fontenc}
\usepackage{ae,aecompl}

\usepackage{graphicx}	
\usepackage{amsmath}	
\usepackage{booktabs}
\usepackage{threeparttable}
\usepackage{xcolor}

\newcommand{\rep}[1]{#1}
\newcommand{\repp}[1]{#1}

\title[CrA-9 b, B or BKG?]{A faint companion around CrA-9: protoplanet or obscured binary?}

\author[V.~Christiaens et al.]{V.~Christiaens$^{1}$\thanks{E-mail: Valentin.Christiaens@monash.edu},  M.-G.~Ubeira-Gabellini$^{2}$, H.~C\'anovas$^{3}$, P.~Delorme$^{4}$, B.~Pairet$^{5}$,
\newauthor O.~Absil$^{6}$, S.~Casassus$^{7}$, J.~H.~Girard$^{8}$, A.~Zurlo$^{9,10}$, Y.~Aoyama$^{11,12}$, G-D.~Marleau$^{13,14,15}$, \newauthor L.~Spina$^{16}$, N.~van~der~Marel$^{17}$, L.~Cieza$^{9,10}$, G.~Lodato$^{2}$, S. P\'erez$^{18}$, C.~Pinte$^{1,4}$, \newauthor D.~J.~Price$^{1}$, M.~Reggiani$^{19}$
\\
$^{1}$School of Physics and Astronomy, Monash University, Clayton Vic 3800, Australia\\
$^{2}$Dipartimento di Fisica, Università degli Studi di Milano, Via Celoria 16, 20133 Milano MI, Italy\\
$^{3}$Aurora Technology for ESA/ESAC, Camino bajo del Castillo s/n, Urbanizaci\'on Villafranca del Castillo, Villanueva de la Ca\~nada, 28692 Madrid, Spain\\
$^{4}$Institut de Plan\'etologie et d'Astrophysique de Grenoble, Universit\'e Grenoble Alpes, 38058 Grenoble, France\\
$^{5}$ISPGroup, ELEN/ICTEAM, UCLouvain, Belgium\\
$^{6}$Departamento de Astronom\'ia, Universidad de Chile, Casilla 36-D, Santiago, Chile\\
$^{7}$Space sciences, Technologies \& Astrophysics Research (STAR) Institute, Universit\'e de Li\`ege, All\'ee du Six Ao\^ut 19c, B-4000 Sart Tilman, Belgium\\
$^{8}$Space Telescope Science Institute, 3700 San Martin Dr. Baltimore, MD 21218, USA\\
$^{9}$N\'ucleo de Astronom\'ia, Facultad de Ingenier\'ia, Universidad Diego Portales, Av. Ejercito 441, Santiago, Chile\\
$^{10}$Escuela de Ingenier\'ia Industrial, Facultad de Ingenier\'ia y Ciencias, Universidad Diego Portales, Av. Ejercito 441, Santiago, Chile\\
$^{11}$Institute for Advanced Study, Tsinghua University, Beijing 100084, People's Republic of China\\
$^{12}$Department of Astronomy, Tsinghua University, Beijing 100084, People's Republic of China\\
$^{13}$Institut f\"ur Astronomie und Astrophysik, Universit\"at T\"ubingen, Auf der Morgenstelle 10,
D-72076 T\"ubingen, Germany\\
$^{14}$Physikalisches Institut, Universit\"{a}t Bern, Gesellschaftsstr.~6, CH-3012 Bern, Switzerland\\
$^{15}$Max-Planck-Institut f\"ur Astronomie, K\"onigstuhl 17, D-69117 Heidelberg, Germany\\
$^{16}$INAF-Osservatorio Astronomico di Padova, Vicolo dell'Osservatorio 5, Padova, Italy, 35122-I\\
$^{17}$Physics \& Astronomy Department, University of Victoria, 3800 Finnerty Road, Victoria, BC, V8P 5C2, Canada\\
$^{18}$Departamento de F\'isica, Universidad de Santiago de Chile. Avenida Ecuador 3493, Estaci\'on Central, Santiago, Chile\\
$^{19}$Institute of astrophysics, KU Leuven, Celestijnlaan 200D, 3001 Leuven, Belgium
}

\date{Accepted 2021 February 16. Received 2021 February 15; in original form 2020 December 01}

\pubyear{2021}

\begin{document}
\label{firstpage}
\pagerange{\pageref{firstpage}--\pageref{lastpage}}
\maketitle

\begin{abstract}
Understanding 
how giant planets form
requires observational input from directly imaged
protoplanets. 
We used VLT/NACO and VLT/SPHERE to search for companions in the transition disc of 2MASS J19005804-3645048 (hereafter CrA-9), 
an accreting M0.75 dwarf 
with an estimated age of 1--2 Myr. We found a faint point source at $\sim$0.7\arcsec separation from CrA-9 ($\sim$108 au projected separation). Our 3-epoch astrometry rejects a fixed background star with a $5\sigma$ significance. 
The near-IR absolute magnitudes of the object
point towards a planetary-mass companion. \rep{However}, our analysis of the 1.0--3.8$\mu$m spectrum extracted for the companion suggests \rep{it is a young M5.5 dwarf}, based on \rep{both} the 1.13-$\mu$m Na index \rep{and comparison with templates of the Montreal Spectral Library}. 
The observed 
spectrum is best reproduced with high effective temperature ($3057^{+119}_{-36}$K) BT-DUSTY and BT-SETTL models, \rep{but} the corresponding photometric radius required to match the measured flux is only $0.60^{+0.01}_{-0.04}$ Jovian radius.
We discuss possible explanations to reconcile our measurements, including an M-dwarf companion obscured by an edge-on circum-secondary disc or the shock-heated part of the photosphere of an accreting protoplanet. Follow-up observations  
covering a larger wavelength range and/or at finer spectral resolution 
are required to 
discriminate these two scenarios. 
\end{abstract}

\begin{keywords}
protoplanetary discs -- planet-disc interactions -- techniques: image processing -- planets and satellites: formation
\end{keywords}



\section{Introduction}
The classical debate on the formation of giant planets confronts \emph{core accretion} 
\citep[][]{Mizuno1980,Pollack1996} to \emph{gravitational instability} 
\citep[][]{Boss1998,Kratter2016}.
While the majority of the detected population of short-orbit mature exoplanets appears consistent with predictions from core accretion models \citep[e.g.][]{Winn2015,Mordasini2018}, \rep{it is unclear whether} the properties of young giant planets that have been directly imaged at large orbital separations \rep{are also consistent with formation through core accretion} 
\citep[e.g.~ the HR 8799 planets, $\beta$ Pic~b and c, HIP 65426~b;][]{Marois2008,Lagrange2009,Bonnefoy2013,Marleau2014, Chauvin2017,Marleau2019a, Nowak2020}.
How did these adolescent 5--12 Jupiter mass ($M_J$) planets found at up to $\sim$100 au separation form in the first place?
\rep{If similar planetary-mass companions are also found at large separations at very young ages ($\sim$1~Myr), this could be a challenge for core accretion, even assisted with pebble accretion \citep[e.g.][]{Paardekooper2018}}.
Detections of nascent giant planets at the youngest ages and at multiple wavelengths are required to break the degeneracy between predictions from different models \citep[e.g.][]{Spiegel2012,Mordasini2012a,Zhu2015b,Mordasini2017}. 
In this context, protoplanetary discs with large cavities, also known as \emph{transition discs}, constitute prime targets to search for nascent giant planets, \rep{since they} may be carving the cavity \citep[e.g.][]{Espaillat2014,Casassus2016,Owen2016,van-der-Marel2021}.


High-contrast imaging in IR is one of the most powerful technique to detect those young companions \citep[e.g.][]{Absil2010,Bowler2016}. A particularly suited observing strategy to reach high contrast is angular differential imaging \citep[ADI;][]{Marois2006}.
When coupled with an appropriate post-processing algorithm such as principal component analysis \citep[PCA;][]{Amara2012, Soummer2012}, this technique can efficiently model and suppress the bright stellar halo of the star, while preserving that of the planet.
Nevertheless, in the presence of a bright circumstellar disc, aggressive ADI filtering can create point-like artefacts which can be confused with substellar companions \citep[e.g.][]{Milli2012, Rich2019, Currie2019}.
This has led to a number of protoplanet detection claims whose authenticity has subsequently been debated in the recent years \citep[e.g][]{Quanz2013,Sallum2015,Guidi2018}.
To add to the confusion, even faint companions imaged at a location external to the circumstellar disc can also be misclassified. 
The IR magnitudes of companions FW~Tau~C and CS~Cha~B suggested a planetary mass \citep[][]{Kraus2014,Ginski2018}, however recent studies have shown that they were more likely to be obscured M-dwarf companions \citep[][]{Wu2017a,Haffert2020}.

So far, the only confirmed detection of protoplanets was made in the transition disc of PDS 70, with multiple independent detections in the IR \citep{Keppler2018,Muller2018,Christiaens2019a,Christiaens2019b}, in the H$\alpha$ line \citep{Haffert2019}, and at sub-mm wavelengths \citep{Isella2019}. 
This further motivates the search for young companions in other discs harbouring large cavities. 
In particular, a statistically significant number of detections at very young age could constrain the planet formation mechanisms that are indeed at work, and the connection with their natal disc.

In this work we focus on 2MASS J19005804-3645048 \citep[hereafter {\bf CrA-9}, as in][]{Peterson2011,Romero2012,Cazzoletti2019}, a young accreting T-Tauri star in the Corona Australis (CrA) molecular cloud, surrounded by a transition disc.
We report the discovery and characterisation of a faint point source at 0\farcs7 separation from CrA-9. In Section \ref{sec:CrA9}, we summarise the known properties of the system. In Section~\ref{sec:Obs+DataRed} we describe the observations and data reduction methods used in this work. In Section~\ref{sec:Results}, we present our final images and spectral characterisation of the point source. In Section~\ref{sec:Discussion}, we discuss its possible nature. Finally, we summarise our conclusions in Section~\ref{sec:Conclusion}.

\section{The CrA-9 system} \label{sec:CrA9}

\begin{table} 
\begin{center}
\caption{Characteristics of CrA-9 and its protoplanetary disc.}
\label{tab:CrA9_props}
\begin{tabular}{lcc}
\hline
\hline
Parameter & Value & Reference \\
\hline
2MASS name & J19005804-3645048 & \\
Right ascension & $19^{\rm h}00^{\rm m}58^{\rm s}.044$ & 1\\
Declination & $-36\degr45\arcmin04\farcs883$ & 1\\
Distance [pc] & $153.1 \pm 1.2$ & 1 \\
Spectral type & M0.75$\pm$0.5 & 2 \\
$T_{\mathrm{eff}}$ [K] & $3720\pm150^{\rm (a)}$ & 3 \\
Log($g$) & $3.5\pm0.2^{\rm (b)}$ & \\
Luminosity $L_{\star}$ [$L_{\odot}$] & 0.46 & 2\\
Age [Myr] & 1--2 & Sec.~\ref{sec:age} \\
Mass [$M_{\sun}$] & $0.45$ & 3\\
Accretion rate [Log($M_{\sun}$ yr$^{-1}$)]] & -8.6 & 2 \\
Li EW [\AA] & 0.48 & 2 \\
$A_{\rm V}$ [mag] & 1.8--2.1 & 4, 5 \\
Disc luminosity [Log($L_{\rm d}/L_{\star}$)] & -2.4 & 2\\
Dust mass [$M_{\earth}$] & $3.70 \pm 0.12$& 3 \\
\hline
\end{tabular}
\end{center}
{\bf Notes}: $^{\rm (a)}$Based on the empirical relation to convert from spectral type to effective temperature in \citet{Herczeg2014}.
$^{\rm (b)}$Based on effective temperature and age estimates, and the isochrones of either \citet{Tognelli2011} or \citet{Baraffe2015}.
\\
{\bf References}: (1) \citet{Gaia-Collaboration2018} (2) \citet{Romero2012}; (3) \citet{Cazzoletti2019}; 
(4) \citet{Dunham2015}; (5) \citet{vanderMarel2016c}. 
\end{table}

\subsection{Stellar properties}

Table~\ref{tab:CrA9_props} summarises the known physical properties of CrA-9. 
CrA-9 is located at the edge of the R CrA dark cloud \citep[also referred to as the \emph{Coronet};][]{Taylor1984}, a highly obscured and very young region of the CrA molecular cloud \citep[Figure~\ref{fig:CrA9}a and b;][]{Gutermuth2009,Peterson2011,Bresnahan2018}.
The Gaia DR2 parallax for CrA-9 corresponds to a distance of $153.1\pm1.2$ pc \citep{Gaia-Collaboration2018}, 
which is in agreement with the median value of 154 $\pm$ 4 pc obtained from the parallax of all members of the CrA molecular cloud in the Gaia Data Release 2 catalogue \citep{Dzib2018}.

\citet{Romero2012} inferred a spectral type of M0.75$\pm$0.5 for CrA-9 based on the strength of the TiO$_5$ molecular band \citep{Cruz2002}. 
Using the empirical relation in \citet{Herczeg2014}, this converts to an effective temperature of 3720$\pm$150K, where the quoted uncertainty reflects both the uncertainty on the subclass and systematic uncertainties affecting the empirical relation.
By fitting the SED of CrA-9, \citet{Romero2012} inferred a total stellar luminosity of 0.46 $L_{\odot}$. 



An extinction of $A_V \sim 2.1$ mag was estimated from dust extinction maps \citep{Dunham2015}. This value may be slightly overestimated due to the lack of an absolute calibration \citep[][]{Peterson2011}. \citet{vanderMarel2016c} found $A_V \sim 1.8$ mag from SED fitting, 
which we use as prior for our SED fitting 
(Section~\ref{sec:specCrA9}), 
adopting an uncertainty of 0.3 mag.

The star is actively accreting. \citet{Romero2012} measured a velocity width of 440 km s$^{-1}$ for the H$\alpha$ line, which converts into an accretion rate $\dot{M} \approx 2.5 \times 10^{-9} M_{\odot} {\rm yr}^{-1}$ \citep{Natta2004}. This corresponds to an accretion luminosity $L_{\rm acc} \approx 0.035 L_{\odot}$, hence about 8\% of the stellar luminosity.


\subsection{Age} \label{sec:age}

\citet{Romero2012} measured an equivalent width (EW) of 0.48 \AA~for the Li I line (6707 \AA).
Considering the effective temperature of the star, the non-depletion of lithium points towards an age younger than $\sim$ 4 Myr \citep[e.g.][]{Baraffe2015}. 
A more quantitative estimate can be obtained from comparison to the distribution of Li I EWs in the $\rho$ Ophiucus dark cloud. 
Both the measured 0.48 \AA~EW and the Li~I EWs of other members of the CrA cloud \citep[e.g.][]{Sicilia-Aguilar2008} are consistent to what is seen in other stellar forming regions with ages $\sim$ 1--3 Myr \citep{Spina2017}. 

In addition to the presence of significant accretion and the measured Li I EW, 
another clue for the youth of CrA-9 is its location at the edge of the R CrA core ($\lesssim$ 1 pc projected distance; Figure~\ref{fig:CrA9}b).
The presence of several Class 0 candidates in the Coronet suggests that it is actively forming new stars, hence that it is very young \citep{Wang2004,Nutter2005,Sicilia-Aguilar2013}.
Isochrone fitting to extinction-corrected NIR photometry of R CrA members yielded very young age estimates: 0.5--2 Myr \citep{Nisini2005}; 0.3--3 Myr \citep{Meyer2009}; and 0.5--1 Myr \citep{Sicilia-Aguilar2011}.
All these studies considered however a distance of 130~pc, i.e.~smaller than inferred by Gaia DR2. Their luminosity estimates are thus slightly underestimated, hence the quoted ages are to be considered upper limits for the R CrA dark cloud. 

\citet{Neuhauser2000} and \citet{Peterson2011} suggested the presence of two populations in the CrA molecular cloud: a younger (the R CrA dark cloud) and an older one.
\citet{Cazzoletti2019} found that the measured dust masses are relatively low for a 1--3 Myr-old region compared to other young star-forming regions, possibly in agreement with 
the presence of an old population.
However, \citet{Cazzoletti2019} 
also found evidence for a single coeval population with an age of 1--3 Myr based on comparison of their HR diagram to isochrones of \citet{Baraffe2015}.
We used the same evolutionary models to estimate the age to be 1--2 Myr, based on the combination of effective temperature (3720$\pm$150 K) and luminosity ($\sim 0.46 L_{\odot}$) inferred for CrA-9 assuming a distance of 150 pc \citep{Romero2012}, hence consistent with all other estimates discussed above.



\subsection{Protoplanetary disc}

The SED of CrA-9 is 
characteristic of a protoplanetary disc with a large dust cavity \citep{Romero2012, vanderMarel2016c}, i.e. a \emph{transition disc}.
Based on both the positive slope of the IR excess and the high stellar accretion rate, \citet{Romero2012} recognised it as one of the transition discs of their sample whose properties are
the most compatible with giant planet(s) dynamically carving the dust depletion.
\citet{vanderMarel2016c} inferred a cavity radius of $\sim$14~au based on SED modeling.

\citet{Sicilia-Aguilar2008} had already noted the larger fraction of transition discs in the R CrA cloud core ($\sim 50$\%) compared to other dark clouds of similarly young age, suggesting they were not tracing a short-lived transition phase but rather that they formed with these ``transition-like'' feature, e.g. due to binarity.
The fractional luminosity of the disc with respect to the star \citep[$L_{\rm d}/L_{\star} \approx 10^{-2.4}$;][]{Romero2012} also suggests an evolutionary stage earlier than debris discs \citep[typically $L_{\rm d}/L_{\star} < 10^{-3}$; e.g.][]{Cieza2010}. This is also consistent with the fact that the star is still accreting.

The disc has been detected by ALMA 1.3mm continuum observations, 
as part of a survey of the whole CrA cloud, with an estimated dust disc radius of $\sim 0\farcs39$ \citep{Cazzoletti2019}. 
Considering standard assumptions on dust opacity at mm wavelengths \citep{Beckwith1990} and a constant dust temperature of 20 K, the measured 1.3-mm flux of $5.07\pm0.16$ mJy translates to $3.7\pm0.1 M_{\earth}$ of dust for an optically thin disc \citep{Cazzoletti2019}. Under the standard (but highly uncertain) assumption of a gas-to-dust mass ratio of 100:1 \citep{Williams2011}, this corresponds to a disc mass of $\sim 1.2 M_J$.   

\section{Methods}

\begin{table*} 
\caption{Summary of the observations on CrA-9 used in this work.}
\label{tab:CrA9_obs}
\begin{threeparttable}
\centering
\begin{tabular}{lllcccccccccl}
\hline
\hline
Date & Program & Instrument & Filter & Mode & Plate scale & DIT$^{\rm a}$ & NDIT$^{\rm b}$ & NEXP$^{\rm c}$ & T$_{\rm int}^{\rm d}$ & $<\beta>^{\rm e}$ & $\Delta$PA$^{\rm f}$ & Notes\\
& &  & & & [mas px$^{-1}$] & [s] & & & [min] & [\arcsec] & [\degr] \\ 
\hline
2017-08-26 & 099.C-0883 & NACO & $L'$ & ADI & 27.2 & 0.1 & 453 & 62 & 44.7 & 0.52 & 25.3 & non-sat.\\
2018-07-02 & 0101.C-0924 & NACO & $L'$ & ADI & 27.2 & 0.1 & 453 & 124 & 88.0 & 0.95 & 42.3 & non-sat.\\
2019-07-02 & 0103.C-0865 & NACO & $H$ & PDI & 27.15 & 0.8 & 80 & 56 & 56 & 0.57 & -- & non-sat.\\
2019-09-\rep{28} & 103.2036.001 & IFS & $YJH$ & ADI & 7.46 & 16 & 11 & 32 & 80.3 & 0.75 & 30.5 & non-sat.\\
2019-09-\rep{28} & 103.2036.001 & IRDIS & $K12$ & ADI & 12.256 & 8 & 22 & 32 & 71.7 & 0.75 & 30.5 & sat.~core\\
2019-09-\rep{28} & 103.2036.001 & IRDIS & $K12$ & ADI & 12.256 & 2 & 24 & 2 & 1.1 & 0.75 & 31.8 & non-sat.\\
\hline
\end{tabular}
\begin{tablenotes}
\item[a] Detector integration time.
\item[b] Number of co-add images in each exposure. 
\item[c] Number of exposures.
\item[d] Total integration times (excluding overheads), calculated after bad frame removal as explained in Section~\ref{sec:SPHEREcalib}.
\item[e] Average seeing at $\lambda=500$nm achieved during the sequence, as returned by ESO Paranal DIMM station.
\item[f] Parallactic angle variation achieved during the observed sequence.
\end{tablenotes}
\end{threeparttable}
\end{table*}

\subsection{Observations and data reduction}\label{sec:Obs+DataRed}
We observed CrA-9 with VLT instruments NACO \citep{Rousset2003,Lenzen2003} and SPHERE \citep{Beuzit2008,Claudi2008,Dohlen2008} at four different epochs (ESO programs 099.C-0883, 0101.C-0924, 0103.C-0865, and 103.2036.001, respectively). Owing to the faintness of the source, all observations were obtained without coronagraph. Table~\ref{tab:CrA9_obs} summarises the observations. 

\subsubsection{VLT/NACO $H$-band polarimetric dataset}\label{app:NACO-PDI}

We observed CrA-9 with the NACO instrument \citep{Rousset2003,Lenzen2003} at
the VLT on 8 June 2019 in service mode (ESO programme 0103.C-0865). The observations were
performed in polarimetric mode using the broad-band NACO $H$-band filter ($\lambda_c = 1.66\, \mu$m). In this
observing mode a half-wave plate (HWP) rotates the polarisation plane of the incoming light before a Wollaston
prism splits the light into two orthogonally polarised beams that are projected on different detector regions.
The CCD pixel size was set to $0\farcs027 \, \mathrm{px}^{-1}$, the readout mode to {\tt Double RdRstRd},
and the detector mode to {\tt HighDynamic}, while we used the {\tt K} dichroic that splits the incoming light
between the CONICA system and the wavefront-sensor.

The observations consisted in multiple polarimetric cycles where each cycle contains four datacubes, one
per HWP position angle (at $0\degr, 22.5\degr, 45\degr$, and $67.5\degr$, measured on sky east from north).
We used detector integration times (DIT's) of $0.8$ seconds, with a total exposure time of $11898$ seconds
($3.3$ hours). During that time the airmass ranged from $1.0$ to $1.1$ and the seeing was mostly good and
stable with an average value of  $0\farcs56 \pm 0\farcs13$. Standard calibrations including darks and flat fields
were provided by the ESO observatory. 

The two simultaneous, orthogonally polarised images recorded on the detector when the HWP is at $0\degr
(45\degr)$ were subtracted to produce the Stokes parameter $Q^+(Q^-)$. This process was repeated for the
$22.5\degr (67.5\degr)$ angles to produce the Stokes $U^+ (U^-)$ images. The total intensity (Stokes I) was
computed by combining all the images. We used the imaging polarimetry pipeline described by
\citet{Canovas2011} and \citet{Canovas2015} to process the raw data. Each science frame
was dark-current subtracted and flat-field corrected. Hot and dead pixels were identified with a $\sigma$-clipping
algorithm and masked out using the average of their surrounding good pixels. The two images recorded in each
science frame were aligned with an accuracy of 0.05 pixels. This process was applied to every science frame
resulting in a datacube for each Stokes $Q^\pm, U^\pm$ parameter. We applied the double-difference method as
described in \citet{Canovas2011} to correct for instrumental polarisation the final, median-combined images.
Finally, we derived the polarised intensity ($P_{I} = \sqrt{Q^2 + U^2}$) and the $Q_{\phi}$ and $U_{\phi}$ images
\citep[see][]{Schmid2006}. 

\subsubsection{VLT/NACO $L'$-band datasets}\label{sec:NACOobs}

Both the 2017 and 2018 NACO datasets were acquired using the $L'$ filter ($\lambda \sim 3.8 \mu$m). The first dataset was obtained in service mode in excellent conditions on 26 August 2017 (stable DIMM seeing $\sim$0\farcs5). 
The second sequence was obtained on \rep{2} July 2018 in visitor mode. Conditions were mediocre with a variable seeing oscillating between $\sim$0\farcs7 and $\sim$2\arcsec. To compensate, we acquired a twice longer integration than at the first epoch (88 min vs. 45 min). Both observations were obtained in cube mode, which allows one to save each individual co-add image instead of median-combining them. We opted for a pupil-tracking observing strategy to enable angular differential imaging \citep[ADI;][]{Marois2006}, achieving 25\degr~and 42\degr~field rotation in the 2017 and 2018 datasets, respectively. We used a 2-point dithering pattern in the two good quadrants of NACO's detector, excluding the bottom-left and top-right quadrants to avoid bad columns and higher dark current noise, respectively. The stellar PSF did not saturate the detector in either observation.

We reduced both datasets in the same way, using a custom-made pipeline built on {\sc Python} routines from the Vortex Image Processing package ({\sc vip}\footnote{Available from \url{https://github.com/vortex-exoplanet/vip}.}; \citealt{GomezGonzalez2017}). 
In brief, we (i) found the approximate stellar position in each science cube and record the quadrant; (ii) subtracted an estimate of the sky background for each image using images where the star is in a different quadrant; (iii) flat-fielded each image; (iv) corrected for NaN values and bad pixels; (v) found the stellar centroid with a 2D Moffat fit and shift all images to place the star on the central pixel; (vi) combined all images in a master cube and compute the associated parallactic angles; (vii) identified and remove bad frames; (viii) median-combined together sets of 10 (resp.~16) consecutive images in the 2017 (resp.~2018) dataset; (ix) measured the FWHM in the median image and the stellar flux in each image of the cube; and (x) finally post-processed the calibrated cube using median-ADI \citep{Marois2006} and principal component analysis \citep[PCA-ADI;][]{Amara2012,Soummer2012}.
Appendix~\ref{app:NACOpip} gives the details of each step.

\rep{
We searched for NACO $L'$ observations of standard stars in the ESO archive in order to provide an absolute photometric calibration of CrA-9 in the $L'$ band.
No standard star was observed the same night as our observations of CrA-9, therefore we extended our search to standard stars observed in the same airmass and seeing conditions. We considered standard stars HD~205772 and HD~75223 observed on 27 August 2017 and 4 December 2018, for the NACO 2017 and 2018 CrA-9 datasets, respectively. We applied steps (i) to (v) of our NACO reduction pipeline for sky+dark subtraction, flat-fielding, bad pixel correction and recentering of the standard stars data (Appendix~\ref{app:NACOpip}). Visual inspection of the PSF of the STD stars suggests similar Strehl ratios as achieved for CrA-9, for each respective pair of observations. We measured the average flux (in ADUs) in a 1-FWHM aperture, and used the physical $L'$ fluxes tabulated in \citet{van-der-Bliek1996} to compute zero points. 
The absolute $L'$ fluxes calculated for CrA-9 in the 2017 and 2018 datasets are $1.63 \pm 0.16 \times 10^{-14}$ W m$^{-2}$ $\mu$m$^{-1}$ and $1.08 \pm 0.11 \times 10^{-14}$ W m$^{-2}$ $\mu$m$^{-1}$ respectively, where we considered a 10\% relative uncertainty based on our procedure (e.g.~possible small differences in achieved Strehl ratios for the standard stars and CrA-9 observations). Despite the uncertainties involved with our procedure, we note that these different absolute fluxes for the star appear to compensate exactly the discrepant contrasts measured at the two NACO epochs for the faint point source (i.e.~they lead to the same $L'$ flux Section~\ref{sec:contrast_spectrum}). 
}

\subsubsection{VLT/SPHERE dataset} \label{sec:SPHEREcalib}

We observed CrA-9 on \rep{28} September 2019 with VLT/SPHERE \citep{Beuzit2008} in IRDIFS-EXT mode, i.e.~with both the IFS and IRDIS sub-instruments acquiring images simultaneously \citep{Claudi2008,Dohlen2008}. 
IFS has a spectral resolution of 54, covering the $YJH$ bands. It acquired 32 datacubes with 11 co-adds, each containing 39 spectral channels ranging from 0.95 to 1.68 $\mu$m. The stellar PSF did not saturate in any of the spectral channels.
The same number of datacubes was obtained with IRDIS in the $K1$ and $K2$ filters ($\lambda \approx 2.11 \mu$m and $2.25 \mu$m, respectively), with an integration time of 8s. The core of the stellar PSF saturated in that sequence. We also acquired two datacubes at the beginning and end of the observation with the integration time set to 2s in order to measure the unsaturated stellar flux. Conditions were average and variable throughout the sequence (average seeing of $\sim$0\farcs76), which combined with the faintness of the source ($R \approx 13.6$mag) led to fluctuating levels of adaptive optics correction and variations of up to a factor $\sim$2 in the measured stellar flux in the IFS channels. Three sets of sky background images were obtained for both IFS and IRDIS. 
A total of $\sim$31\degr~field rotation was achieved throughout the pupil-stabilised sequence.

We implemented two new pipelines to reduce our non-coronagraphic IFS and IRDIS data, respectively. Both of them are divided in three parts: basic calibration, pre-processing and post-processing. 
For both pipelines, the basic calibration relied mostly on ESO's Common Pipeline Library {\sc esorex} recipes (version 3.13.2), while both the pre- and post-processing parts made use of routines from the {\sc vip} package. 

Our reduction pipeline for IRDIS data consists of: (i) sky background subtraction; (ii) flat-fielding; (iii) bad pixel correction; (iv) centering based on 2D Moffat fits of the stellar halo; (v) trimming of bad frames; (vi) correction for the anamorphism present in the IRDIS images \citep{Maire2016}; and (vii) median-ADI, PCA-ADI and sPCA post-processing \citep{Absil2013}.
Appendix~\ref{app:IRDISpip} details the different steps of the pipeline. 

Our IFS reduction pipeline involves more steps than for IRDIS, owing to both the complexity of the IFS instrument and the identification of some sub-optimal features in the {\sc esorex} calibration recipes. In short, the pipeline first computes all master calibration files (darks, coloured and white detector flat fields, spectra positions, total instrument flat, wavelength solution, IFU flat) and uses them to reduce the science images and convert them into spectral cubes of 39 channels. Next, {\sc vip} routines deal with the bad pixel correction, centering on the star in each frame, bad frame identification and removal, and anamorphism correction. Finally the pipeline also relies on {\sc vip} for post-processing of the cubes leveraging on either spectral differential imaging \citep[SDI;][]{SparksFord2002} and/or angular differential imaging (ADI). More specifically, we used PCA-SDI on individual spectral cubes, PCA-ADI on 3D cubes for each spectral channel sampled in the temporal dimension, and \rep{PCA-ASDI} on the 4D cubes leveraging on both SDI and ADI with a single PCA library.
Details on each step of the pipeline and on the post-processing are provided in Appendix~\ref{app:IFSpip}. In particular, we detail the calibration steps that enabled us to mitigate bright stripes in the final images.

We double-checked the performance of our IRDIS and IFS reduction pipelines by comparing our calibrated cubes to the outputs of an independent calibration made by the SPHERE data center \citep{Delorme2017a,Galicher2018}. We found consistent astrometric and contrast estimates for the point source in the IRDIS data (Section~\ref{sec:characterisation_point_source}).
However the SPHERE data center calibration of the IFS data led to bright stripes in the final post-processed images. We also identified sub-optimal steps regarding the dark subtraction and distortion correction, which are yet to be implemented. We therefore favoured the results obtained by our pipeline for the rest of this work.

\rep{
Contrary to our NACO data, we do not expect to achieve a good absolute flux calibration of our SPHERE stellar flux measurements based on STD stars. This is because CrA-9 is significantly fainter than the nominal R mag limit for the visible wavefront sensor of SPHERE to provide a good AO correction (R$\sim$13.6 $>$ 12). This is confirmed by visual inspection of the PSF in the IFS channels, which suggests a poor Strehl ratio was achieved despite good observing conditions. The STD stars observed by SPHERE that we found in the ESO archive are all significantly brighter than CrA-9 in $R$ band, and are hence expected to achieve a better Strehl ratio. 
We therefore considered literature flux measurements of CrA-9 itself for absolute flux calibration (Section~\ref{sec:specCrA9}).}

\subsection{Improvements to the {\sc negfc} module of {\sc vip}}\label{sec:MCMC-NEGFC}

The negative fake companion technique \citep[NEGFC; e.g.][]{Lagrange2010,Marois2010b}
enables to extract reliable astrometry and photometry for faint 
point sources found in images obtained using ADI-related post-processing algorithms. 
The principle of {\sc negfc} is to inject a negative companion in the calibrated cube (i.e.~before ADI post-processing), and find the position and flux that minimise the residuals in the final ADI image in an aperture centered on the location of the companion candidate. This forward modeling approach allows us to alleviate the biases that would affect astrometric and photometric estimates made directly in the final ADI image, i.e. geometric biases and flux losses.
In this work, we have updated the {\sc negfc} module implemented in {\sc vip}
\citep{Wertz2017,GomezGonzalez2017} in order to improve the astrometric and contrast estimates of the faint point source presented in Section~\ref{sec:detection}.

The {\sc negfc} module in {\sc vip} relies on
PCA-ADI in a single 3-FWHM wide annulus including the companion candidate. 
By default, the default PCA algorithm used in NEGFC does not consider a threshold in PA to build the PCA library (for computation efficiency).
We have now added the option to use a threshold in PA (which is used in this work for the non-saturated IRDIS dataset).
Three consecutive steps are involved for refined estimates of the companion's radial separation, PA and flux: (1) a grid search on the negative flux alone, at companion coordinates provided by the user; (2) a Nelder-Mead downhill simplex on the three free parameters, using the estimates in step 1 \citep{NelderMead1965}; and (3) a Markov Chain Monte Carlo (MCMC) algorithm sampling the probability distribution of the companion's 3D parameter space, using the simplex result as initial guess. 
The MCMC algorithm relies on {\sc emcee} \citep{Foreman-Mackey2013}, which is a {\sc Python} implementation of the affine-invariant ensemble sampler proposed in \citet{Goodman2010}, and allows to infer uncertainties on each of the three parameters.

We applied three changes to the MCMC algorithm compared to the version presented in \citet{Wertz2017}: (i) we now use a new expression for the log-probability provided to the MCMC sampler; (ii) we now check the convergence of the MCMC algorithm based on the integrated auto-correlation time instead of a Gelman-Rubin test \citep{Gelman1992}; and (iii) we have now added the option to inject different negative companion fluxes in the different frames of the ADI cube, according to weights reflecting varying observing conditions throughout the ADI sequence. These changes are detailed in Appendix~\ref{app:MCMC-NEGFC_improvements}.

\begin{figure*}
	\centering
	\includegraphics[width=\textwidth]{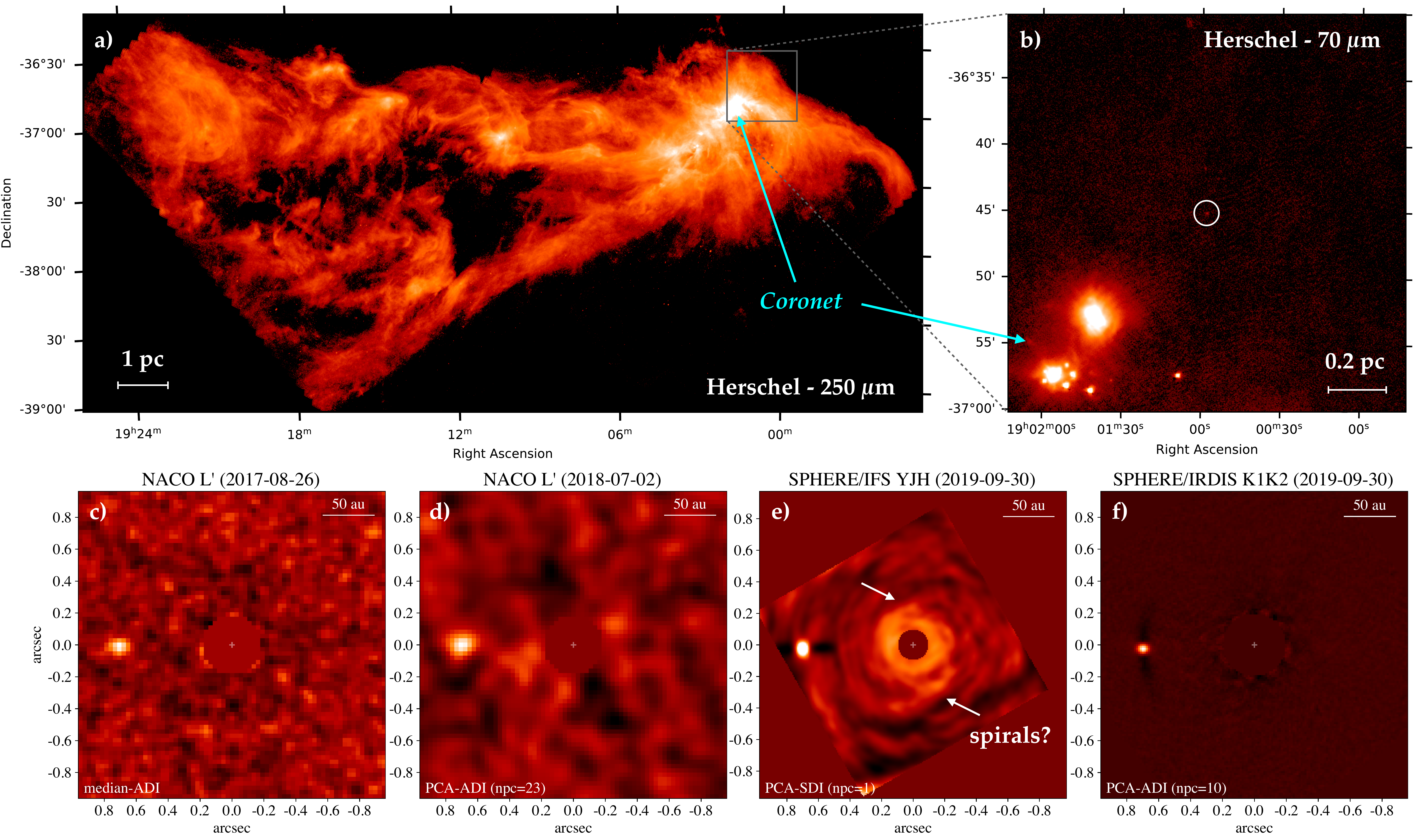}
    \caption{{\bf a--b)} Herschel images showing the CrA-9 system (circled) within the Corona Australis star forming region \citep{Bresnahan2018}. CrA-9 is located within 1 pc projected separation from the dark cloud (\emph{Coronet}), hence suggesting a very young age. {\bf c--f)} Images of the CrA-9 system obtained with VLT/NACO (c and d) and VLT/SPHERE (e and f) after subtraction of stellar emission using either angular differential imaging (c, d and f) or spectral differential imaging (e). The color scale of panels c-f) is linear and the cuts used are min/max except for panel e), where we used a scale spanning 0.1--99.9 percentiles in order to highlight a tentative spiral pattern possibly connected to the companion.
    }
    \label{fig:CrA9}
\end{figure*}

\subsection{{\sc specfit}: A new module for the spectral characterisation of point sources} 

We implemented {\sc specfit}, a new {\sc vip} \citep{GomezGonzalez2017} module which provides the tools to perform the spectral characterisation of directly imaged companions in a Bayesian framework.
The core routine of the module, \verb+mcmc_spec_sampling+, is a wrapper of the {\sc emcee} package \citep{Foreman-Mackey2013}, adapted to sample the probability distribution of the free parameters associated to the models that are fitted to the observed spectrum. 
Any grid of model can be used for the fit as long as a snippet function to read input grid files is provided as argument to \verb+mcmc_spec_sampling+.

Apart from the parameters associated to the model grid such as the effective temperature ($T_e$) and surface gravity ($\log(g)$), additional free parameters include (i) the photometric radius ($R$) used to scale the model along with the provided distance to the system; (ii) optionally the optical extinction $A_V$, treated as in \citet{Cardelli1989}; (iii) optionally the ratio of total to selective extinction $R_V$, set by default to the diffuse interstellar medium (ISM) value $R_V = 3.1$ if not a free parameter; (iv) optionally the flux of specific emission lines, provided as an optional input dictionary; (v) optionally additional black body components, each characterised by an effective temperature and radius. For MCMC samples falling between grid points, linear interpolation is performed using only the closest two points in each dimension of the grid. The fit can also be performed just with black body component(s) if no model grid is provided.
Uniform or Gaussian priors can be provided for each free parameter. Furthermore, a prior on the mass of the object can also be provided, which will be taken into account through the radius and surface gravity values (if the latter is a grid parameter).

For the MCMC sampler, we used a log-likelihood 
which (1) uses the spectral covariance \citep[between the IFS channels only; e.g.][]{Greco2016,Delorme2017} and (2) assigns weights to all spectrometric or photometric points that are proportional to the relative channel width or filter FWHM, respectively 
\citep[e.g.][]{Olofsson2013,Christiaens2019b}:
\begin{align}
\label{Eq:logL}
\log \mathcal{L}(D|M) &= - \frac{1}{2} \chi^2 \\
\chi^2 &= \big[\mathbf{W}(\mathbf{F_{\rm obs}}-\mathbf{F_{\rm mod}})^T\big] \mathbf{C^{-1}} \big[\mathbf{W}^T(\mathbf{F_{\rm obs}}-\mathbf{F_{\rm mod}})\big]
\label{Eq:GoodnessOfFit}
\end{align}
where $\mathbf{W}$ is the vector of weights $w_i \propto \Delta\lambda_i/\lambda_i$, with $\Delta\lambda_i$ the FWHM of the filter (for IRDIS and NACO points) or spectral channel width (for IFS points); $\mathbf{F_{\rm obs}}$ and $\mathbf{F_{\rm mod}}$ are the fluxes of the observed and model spectra; and $\mathbf{C}$ is the spectral covariance matrix. 

$\mathbf{W}$ is normalised so that $\sum_i w_i^2 = N$, where N is the number of points in the spectrum. 
The inclusion of $\mathbf{W}$ in the expression of the log-likelihood makes it different to that used 
in recent MCMC-based SED modeling implementations \citep[e.g.][]{Wang2020,Stolker2020a,Wang2021}.
The motivation behind the use of $\mathbf{W}$ is to assign a weight proportional to the amount of ``spectral information'' contained by each point. Without $\mathbf{W}$, all points would contribute in the same way to the likelihood. For our spectral sampling of the point source, this would bias the algorithm in trying to better reproduce the $YJH$ points, where a higher density of measurements is available, at the expense of the $K1$, $K2$ and $L'$ photometric points, although the latter cover a larger bandpass. 

The model flux points $F_{\rm mod, i}$ were obtained after convolution with the filter of the respective instrument they are compared to. 
In the case of the IFS channels, we considered a 17.33-nm 
FWHM based on the specifications of the IFS prism provided in the ESO manual, 
while for the $K1$, $K2$ and $L'$ points we used the filter transmission curves provided by the observatory.
The values of $C_{ij}$ for $i$ and $j < 39$ are calculated as in \citet{Delorme2017} on the PCA-ADI images obtained with the different IFS spectral channels, and $C_{ij}$ = $\delta_{ij}$ for $i$ or $j > 39$ (i.e. the IRDIS and NACO points), where $\delta_{ij}$ is the Kroenecker symbol. 

\rep{The} \verb+mcmc_spec_sampling+ \rep{routine allows to infer the most likely parameter values for a given parametric model grid. However, for fits to non-parametric libraries, we have implemented} \verb+best_fit_tmp+, \rep{a routine to search for the most similar template to an input spectrum, which is agnostic of the chosen spectral library. The only requirement is to provide a snippet function to} \verb+best_fit_tmp+ \rep{in order to read the template files. Either one or two free parameters can be considered to find the best match: a flux scaling factor and, optionally, optical extinction. Two options are available for the search of these optimal values: either a grid search (with a user-provided range) or a Nelder-Mead simplex algorithm, which is faster and can also be constrained to an allowed range of values. The routine then returns a user-defined number of best-fit templates minimizing the goodness-of-fit. As for} \verb+mcmc_spec_sampling+, \rep{the goodness of fit takes into account spectral covariance and weights (i.e.~it is given by Equation~\ref{Eq:GoodnessOfFit}). Depending on the spectral resolution of the template, it is either interpolated or convolved with the filter(s) used for the observed spectrum.}

\section{Results}
\label{sec:Results}
\subsection{Detection of a point source and tentative spirals}\label{sec:detection}

\begin{figure}
	\centering
	\includegraphics[width=\columnwidth]{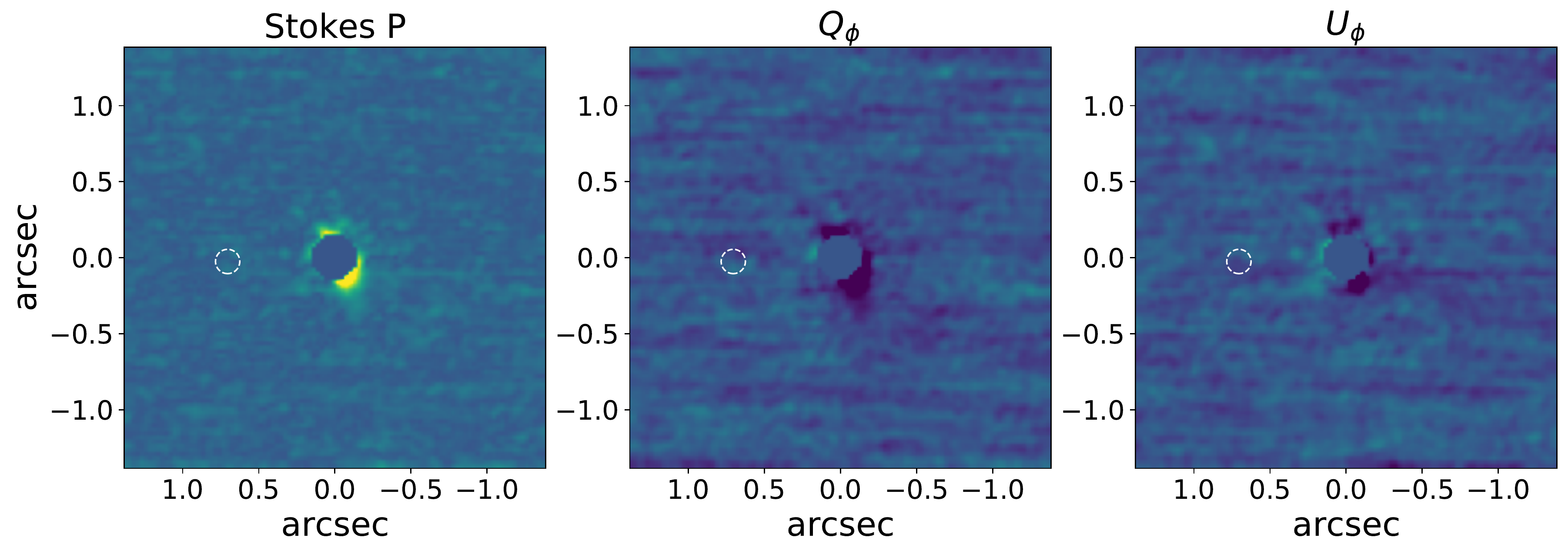}
    \caption{Stokes $P$ (\emph{left}), $Q_{\phi}$ (\emph{middle}) and $U_{\phi}$ (\emph{right}) $H$ band images obtained with VLT/NACO using polarimetric differential imaging. No significant polarised signal is detected. The expected location of the companion is shown with a white circle.
    }
    \label{fig:NACOPDI}
\end{figure}

Figure~\ref{fig:CrA9}c shows the final NACO image obtained for the 2017 dataset. Given the excellent and stable observing conditions, median-ADI was sufficient to reveal a point source at a significant level at a separation of 0\farcs7 to the east of the star. We measured a signal-to-noise ratio (SNR) of $\sim$7.8 in the median-ADI image, considering the penalty for small number statistics \citep{Mawet2014}. This corresponds to a 5.8$\sigma$ detection. 

The 2017 detection motivated us to follow up the source in 2018. The conditions during the 2018 visitor observing run were mediocre and the seeing (at $\lambda = 500$nm) varied between 0\farcs6 and 1\farcs8 throughout the sequence. Consequently, we only managed to redetect the point source using PCA-ADI. Figure~\ref{fig:CrA9}d shows the PCA-ADI image obtained with the number of principal components that maximises the SNR of the companion candidate (SNR$\sim$7.5 with $n_{\rm pc} = 23$).
The re-detection was obtained at approximately the same location. However, ruling out a possible background star required a more accurate astrometry at a third epoch, and color information.

Figure~\ref{fig:CrA9}e and f shows the final PCA-SDI and PCA-ADI images obtained with IFS and IRDIS respectively, upon follow-up of the source with SHERE in 2019. We also redetected the companion candidate after applying PCA-ADI in individual IFS spectral channels. The point source was recovered at an SNR ranging between 10 and 30 in the 39 IFS channels, with the minimum SNR in the first two spectral channels, and maximum SNR in the middle of the $H$ band. We measured an SNR of 91 and 58 in the IRDIS $K1$ and $K2$ PCA-ADI images obtained with the optimal number of principal components, respectively. Figure~\ref{fig:CrA9}f shows the average of the final images obtained for $K1$ and $K2$. The companion was also recovered by applying sPCA on the short non-saturated set of images, which consists of two individual cubes acquired at the beginning and end of the observation respectively (Table~\ref{tab:CrA9_obs}). 

\begin{figure}
	\centering
	\includegraphics[width=\columnwidth]{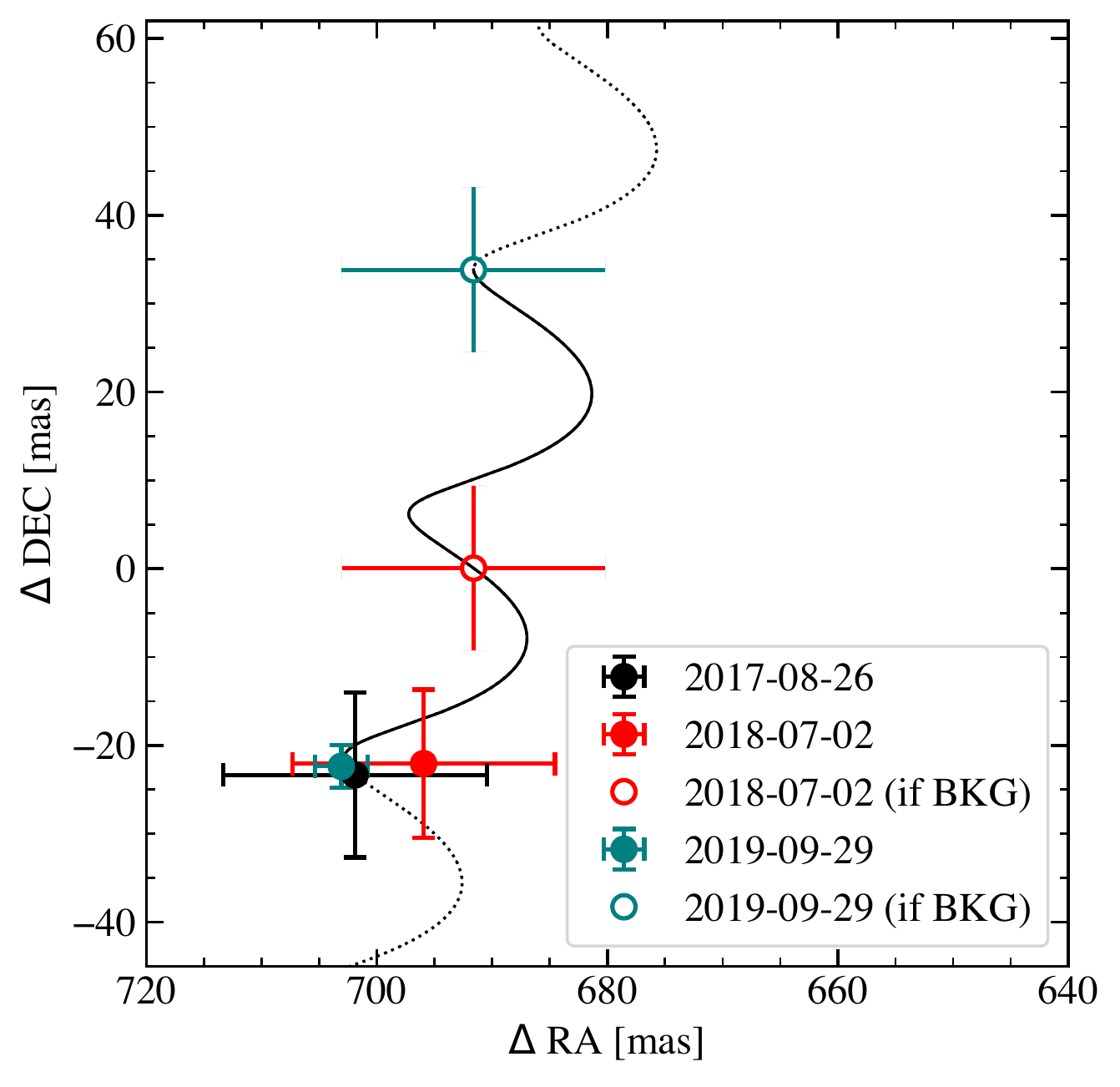}
    \caption{Multi-epoch astrometry of the companion extracted using MCMC-NEGFC on the NACO 2017, NACO 2018 and IRDIS 2019 datasets, compared to predictions for a fixed background object based on the proper motion of CrA-9 measured by Gaia \citep{Gaia-Collaboration2018}. A fixed background object can be rejected at a 5$\sigma$ confidence level. 
    }
    \label{fig:NEGFC_results_astrom}
\end{figure}

In addition to the point source, the 
PCA-SDI image reveals a tentative spiral pattern (Figure~\ref{fig:CrA9}e). A possible primary arm extending to the south of the disc appears to point towards the companion candidate, as expected from hydro-dynamical simulations \citep[e.g.][]{Dong2015}. We measure a signal-to-noise ratio (SNR) of $\sim$3 in the spiral feature in the PCA-SDI ($n_{\rm pc} = 1$ image) using the SNR definition in \citet{Mawet2014}. The feature appears increasingly self-subtracted for larger values of $n_{\rm pc}$ used. It is also tentatively seen using PCA-\rep{ASDI}, although self-subtraction of azimuthally extended structures due to angular differential imaging may account for the differences between the two images. 
PCA-SDI and PCA-\rep{ASDI} images obtained with different $n_{\rm pc}$ subtracted are provided in Appendix~\ref{app:PCA-SDI_PCA-SADI}.
We also computed a standardized trajectory intensity mean map \citep[STIM map;][]{Pairet2019a}, using the residual cube after subtraction of the PCA-SDI model \citep[as in][]{Christiaens2019a}. The STIM map does not reveal the spiral feature conspicuously, as the inverse STIM map (i.e.~obtained with opposite derotation angles) reveals pixels of similar intensity as the spiral feature seen in the (regular) STIM map. Finally, we also applied the {\sc mayonnaise} algorithm on the core-saturated IRDIS dataset \citep{Pairet2020}. While the {\sc mayonnaise} image may suggest some extended disc signal to be present, the spiral pattern seen with IFS is not recovered (Figure~\ref{fig:MAYO}). New observations are thus required to confirm the spiral pattern.

\begin{table*} 
\begin{center}
\caption{Parameters inferred for the companion in our different datasets.}
\label{tab:CrA9b_props}
\begin{threeparttable}
\begin{tabular}{llcccccc}
\hline
\hline
Date & Instrument & Filter & r & PA & contrast & App.~Mag\tnote{a}
& Abs.~Mag\tnote{b} \\
& &  & [mas] & [\degr] & ($\times 10^{-4}$) & [mag] & [mag] \\ 
\hline
2017-08-26 & NACO & $L'$ & $702.3\pm11.1$ & $91.9\pm0.5$ & $9.6\pm1.5$ & $16.39\pm0.18$ & $10.38\pm0.19$\\
2018-07-02 & NACO & $L'$ & $696.3\pm11.1$ & $91.8\pm0.4$ & $14.2\pm1.1$ & $16.39\pm0.10$ & $10.38\pm0.11$\\
2019-09-30 & IFS\tnote{c} & --- & $699.1\pm5.4$ & $92.0\pm0.2$ & $7.6\pm0.6$ & --- & ---\\
2019-09-30 & IFS\tnote{d} & $J$ & --- & --- & $8.1\pm0.1$ & $18.37\pm0.02$ & $11.90\pm0.10$\\
2019-09-30 & IRDIS & $K1$ & $703.7\pm2.0$ & $92.0\pm0.2$ & $8.8\pm0.2$ & $16.75\pm0.03$ & $10.59\pm0.06$\\
2019-09-30 & IRDIS & $K2$ & $704.3\pm2.1$ & $92.0\pm0.2$ & $9.0\pm0.4$ & $16.71\pm0.05$ & $10.58\pm0.07$\\
\hline
\end{tabular}
\begin{tablenotes}
\item[a] Apparent magnitude. Uncertainties include the uncertainties on the contrast, the stellar flux photon noise and \rep{either the uncertainties on the stellar spectrum model (for the SPHERE measurements) or the uncertainties on the calibrated $L'$ flux (for the NACO measurements)}. 
\item[b] Absolute magnitude after dereddening assuming $A_V = 2.07\pm0.1$mag (Table~\ref{tab:CrA9b_spec_props}). 
\item[c] Reported uncertainties include both systematic and statistical (i.e.~dispersion over all spectral channels) uncertainties. 
\item[d] Reported magnitude is integrated over the 2MASS $J$ band filter transmission curve. Only the $J$-band filter is used since the IFS channels only partially cover the $Y$- and $H$-band filter transmission curves.
\end{tablenotes}
\end{threeparttable}
\end{center}
\end{table*}

Figure~\ref{fig:NACOPDI} shows the final $Q_{\phi}$ and $U_{\phi}$ images. The central $r<0\farcs2$ is dominated by noise. No polarised signature was detected neither around the primary nor at the location of the companion (indicated by a white circle). However, it is unclear how the achieved sensitivity to circumstellar disc signal compares to that obtained with SPHERE/IFS, considering the image obtained by the latter is in total intensity, suffers from less systematic biases than NACO, and combines images from a larger spectral bandwidth ($Y$ to $H$ bands instead of only $H$).

\subsection{Characterisation of the point source} \label{sec:characterisation_point_source}

We applied MCMC-NEGFC individually to each spectral channel of the IFS, the $K1$ and $K2$ bands of IRDIS and both NACO $L'$ datasets. We show in Figure~\ref{fig:MCMC-NEGFC_results} 
three example corner plots obtained by MCMC-NEGFC among our 43 ADI sequences, 
for the NACO 2017, IFS 2019 (first spectral channel) and IRDIS 2019 ($K1$ filter) observations, respectively. For all datasets, we used the posterior distributions to infer the most likely value and uncertainties on the radial separation, PA and contrast of the companion candidate. We fitted a gaussian to the marginalised posteriors in order to infer the 1-sigma uncertainties, as in \citet{Wertz2017}. 

\subsubsection{Astrometry}


Figure~\ref{fig:NEGFC_results_astrom} shows the astrometric points retrieved by MCMC-NEGFC for the companion candidate in the NACO 2017, NACO 2018 and SPHERE 2019 datasets. 
For the SPHERE 2019 epoch, we only considered the IRDIS measurement given both the higher astrometric accuracy and higher SNR of the point source than in the IFS images. To make sure our astrometric extraction was accurate, we also got our dataset reduced by the SPHERE data center, and inferred consistent astrometric estimates within 10\% of our reported uncertainties. The IFS data were plagued by significant stripes hence not further considered in this work).

We considered four sources of uncertainty: (i) the residual speckle or background noise uncertainty captured by the variance of the MCMC-NEGFC posterior distribution on $r$ and PA; (ii) a stellar centering uncertainty conservatively assumed to be 0.1 pixel in the NACO and IRDIS datasets and 0.5 pixel in the IFS dataset (where the Strehl ratio was significantly lower); (iii) the systematic uncertainty on the plate scale, affecting the radial separation estimate; and (iv) the uncertainty on the PA of true north, affecting the PA estimate. We combined the different sources of uncertainty in quadrature for each parameter. For NACO, \rep{\citet{Launhardt2020} reported plate scale and true north measurements of $27.2\pm0.1$ mas/px and $0.57\pm0.12$\degr~based on all their astrometric measurements between December 2015 and March 2018, respectively. We adopted these values, but conservatively adopted an uncertainty of 0.5\degr~for the PA of TN, to account for any difference between for the different epochs of observations. We expect this uncertainty to be conservative considering the consistent independent PA of TN estimate presented in \citet{Milli2017} based on 2016/09 data ($0.58\degr\pm0.10\degr$), and the maximum variation of 0.3\degr~for the PA of TN of NACO for all astrometric measurements within 2 years time reported in \citet{Chauvin2012}}. 
For IRDIS, we adopted the systematic uncertainties quoted in \citet{Maire2016}: a true north of $-1.75\pm0.08$, and plate scales of 12.267 and 12.263 mas/px with the $K1$ and $K2$ filters, respectively.

In 
Figure~\ref{fig:NEGFC_results_astrom}, we compare our astrometric measurements to the expected trajectory of a fixed background star based on the proper motion of CrA-9 measured by Gaia \citep{Gaia-Collaboration2018}. Considering the 2017 and 2019 epochs, we can rule out at a 5$\sigma$ confidence level that the object is a background star with null proper motion. 
Instead, the measurements are consistent with negligible orbital motion over the course of $\sim$2.1 years, as would be expected for a physically bound companion given the projected radial separation of $\sim$108 au.
Our astrometric measurements are provided in Table~\ref{tab:CrA9b_props}.

\subsubsection{Contrast spectrum}\label{sec:contrast_spectrum}

\begin{figure*}
	\centering
	\includegraphics[width=0.93\textwidth]{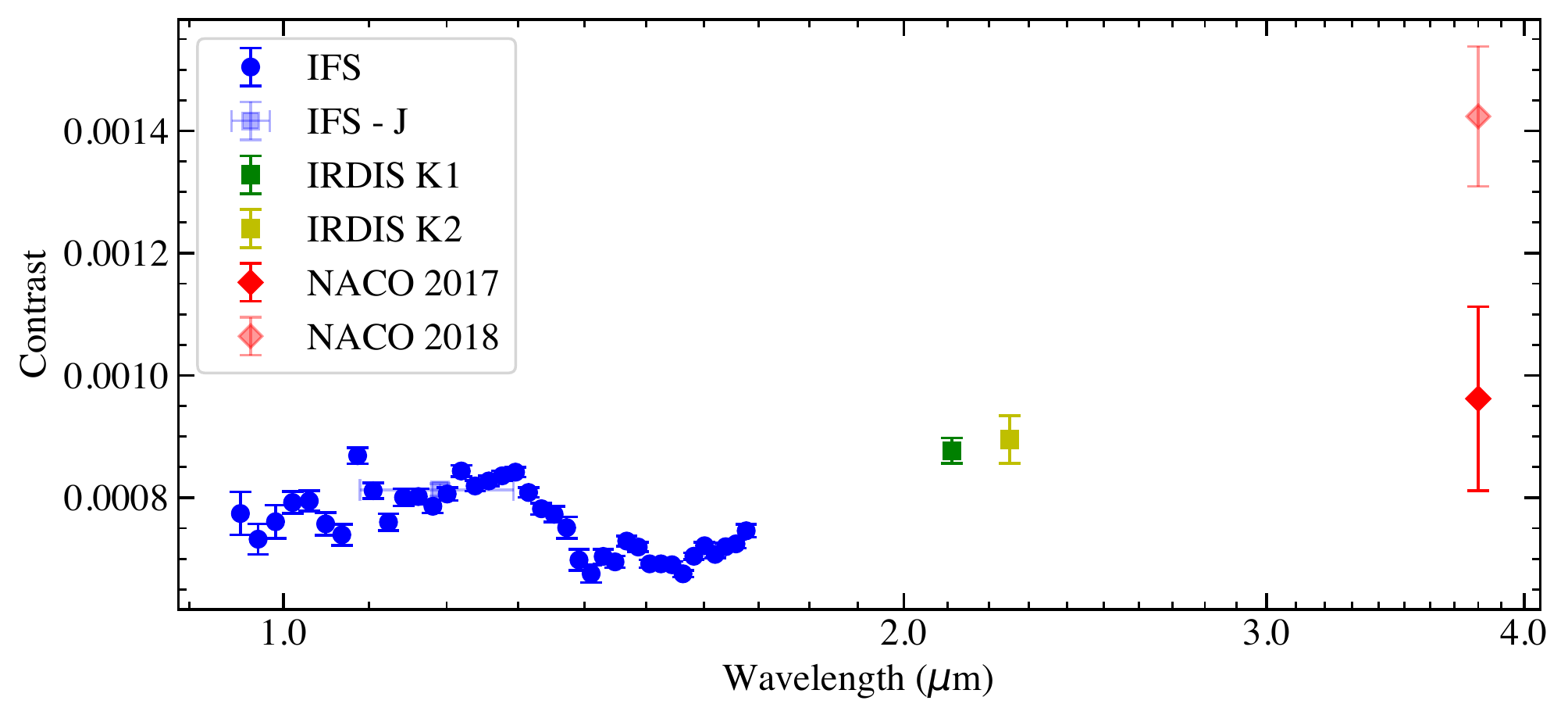}
    \caption{Contrast spectrum of the companion extracted using MCMC-NEGFC on all unsaturated datasets. The overall positive slope with wavelength suggests a redder companion than the central star. \rep{The two discrepant $L'$ contrast estimates appear consistent with the variation in absolute stellar flux measured for CrA-9 in 2017 and 2018 (Section~\ref{sec:NACOobs}).}
    }
    \label{fig:NEGFC_results_contrast}
\end{figure*}

Figure~\ref{fig:NEGFC_results_contrast} shows the contrast spectrum of the point source, i.e. the flux ratio with respect to the star at each wavelength.
Our third modification to MCMC-NEGFC allows temporal variations throughout the different observed sequences to be accounted for, hence enabling us to reach high precision on the estimated contrast of the point source. The accuracy of the contrast spectrum only depends on the residual speckle and background noise level at the separation of the point source, which is captured by the variance of the posterior distribution for the contrast (Figure~\ref{fig:MCMC-NEGFC_results}).
Although the SNR of the point source is higher in the IRDIS images obtained with the stellar-core saturated dataset, it was not used to infer the contrast of the point source given the ignorance on the temporal variation of the stellar flux. Instead, the contrast in the $K1$ and $K2$ filters was inferred from the two unsaturated cubes acquired at the beginning and end of the sequence. Considering both the stellar flux variations within the two unsaturated IRDIS cubes and throughout the IFS sequence, we expect the 
stellar flux variations to lead to larger uncertainties on the $K1$ and $K2$ contrast had we used the core-saturated sequence.

The contrast spectrum is the relevant quantity to infer stellar-model independent conclusions for the point source. Furthermore, it may be used in future studies to re-estimate the companion's (flux) spectrum if a higher-resolution calibrated stellar spectrum in the IR becomes available. 
Two features can be noted from the contrast spectrum:
\begin{enumerate}
    \item The two NACO $L'$ points are discrepant. To test a possible bias related to the poorer quality of the NACO 2018 images, we ran MCMC-NEGFC again with a variation of our third modification: instead of injecting the median unsaturated stellar PSF with varying fluxes in the individual images of the cube, the injection was directly made with the corresponding stellar PSF, scaled to the tested contrast. This led to a consistent contrast estimate for the 2018 point, i.e.~to the same level of discrepancy with the 2017 point. 
    This suggests that the discrepancy may rather be due to variability of the primary star and/or the companion.
    \rep{The $\sim$40\% relative difference between the 2017 and 2018 contrasts is consistent with the difference in absolute flux obtained for CrA-9 using STD stars at these 2 epochs: 23\% larger and 19\% smaller than the expected flux at $L'$ band based on the WISE W1 measurement of CrA-9 \citep{Wright2010}. This suggests that the contrast discrepancy may be assigned entirely to the variability of the primary star, as it leads to a consistent $L'$ flux for the point source in 2017 and 2018.}
    \item We used the 2MASS $J$-band filter transmission curve to infer the contrast that would have been measured in that broad band filter (light blue point in 
    Figure~\ref{fig:NEGFC_results_contrast}). Comparison of the $J$-band contrast to the $K1$ and $K2$ measurements, acquired at the same epoch, indicates a redder spectral slope for the point source than the primary star, suggesting either a later spectral type than M0.75 and/or a larger extinction towards the companion candidate.
\end{enumerate}



\rep{\subsection{Spectrum of CrA-9} \label{sec:specCrA9}}

\begin{figure*}
	\centering
	\includegraphics[width=\textwidth]{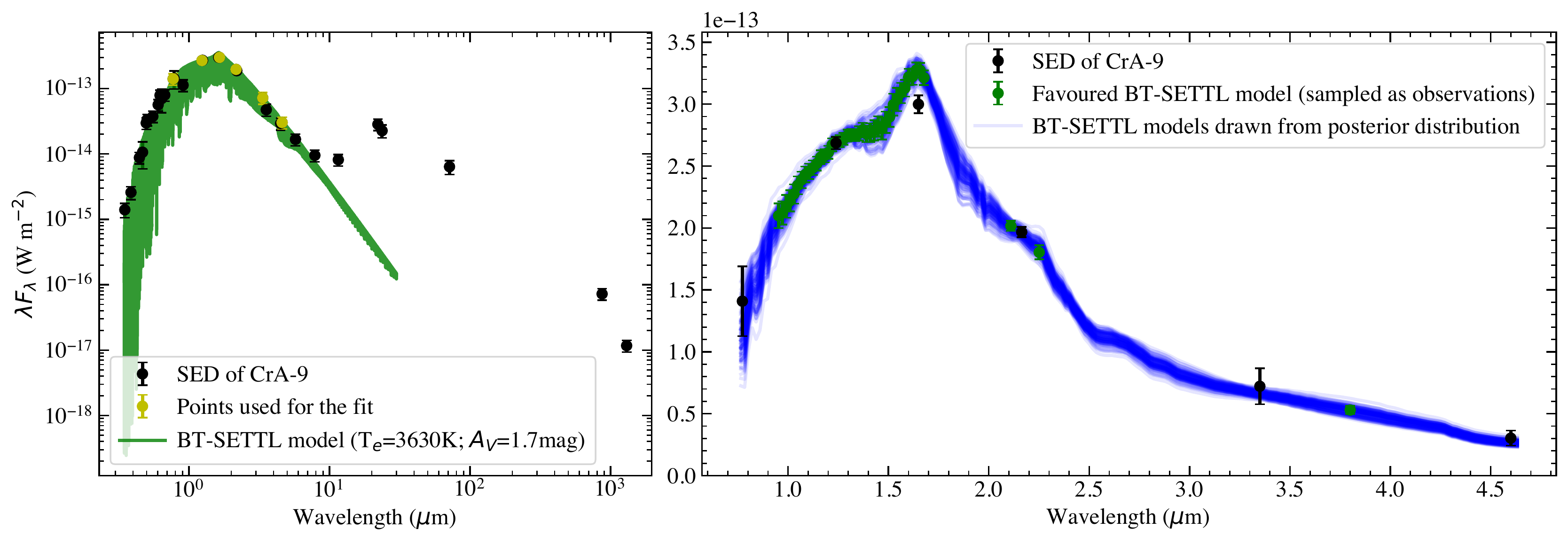}
    \caption{(\emph{left}) Spectral energy distribution of CrA-9 (\emph{black} points) and favoured BT-SETTL model 
    as inferred by \texttt{specfit} (\emph{dark green} curve). The \emph{light green} points are the ones used for the fit. (\emph{right}) Zoom on the wavelength range relevant for our observations. The \emph{dark green} points indicate the flux of the favoured BT-SETTL model at the sampling of our IFS, IRDIS and NACO observations. The \emph{blue} curves are randomly sampled models from the inferred posterior distribution (Figure~\ref{fig:SED_cornerplot}), which are used to estimate the uncertainties on the model spectrum of the star at each wavelength.
    }
    \label{fig:SED}
\end{figure*}

In order to infer the spectrum of the point source in contrast of that of the star, \rep{an absolute calibration of the stellar flux measurements is required first. This can be obtained through flux calibrators observed in similar conditions as the star of interest. Alternatively a reliable model spectrum for the star can be used. Given the impossibility to obtain a good absolute calibration for the SPHERE stellar flux measurements (Section~\ref{sec:SPHEREcalib}), we opted for the second option}. Although past studies have estimated the spectral type and effective temperature of CrA-9, the spectrum of the star is currently poorly sampled at IR wavelengths. Therefore, we used {\sc specfit} in combination with the BT-SETTL grid of models to infer the most likely SED for CrA-9 in the 0.9--4.0 $\mu$m range. BT-SETTL models consider a parameter-free cloud prescription to account for dust formation, coagulation and settling \citep{Allard2012,Allard2014}.
We only considered measurements made by (1) Gaia (DR2) in its $G_{RP}$ filter \citep{Gaia-Collaboration2016,Gaia-Collaboration2018}, (2) 2MASS in the \emph{JHK} bands \citep{Cutri2003}, and (3) WISE at 3.3 and 4.6 $\mu$m \citep[W1 and W2 bands;][]{Wright2010}. 
These instruments have the smallest reported photometric uncertainties in their respective wavelength range.
We did not extend the wavelength range to avoid the need to add extra components to the model to account for either accretion luminosity or disc emission, and to be less affected by the poorer knowledge of the extinction law at short wavelength. 
Including more photometric points at either shorter or longer wavelengths 
would involve more biases and likely lead to a poorer model in the range of interest. 

T-Tauri stars are known to show significant variability over time due to chromospheric activity and/or accretion \citep[e.g.][and references therein]{Bouvier2004,Hartmann2016}. \rep{Our $L'$ stellar flux measurements calibrated using standard stars suggest that CrA-9 is no exception (Section~\ref{sec:NACOobs}). This can also} be seen from the vertical scatter of points at different optical to near-IR wavelengths in the left panel of Figure~\ref{fig:SED}. 
In order to take into account the effect of variability, we considered the reported photometric uncertainties of the $JHK$ 2MASS measurements, but assigned a 20\% relative uncertainty on both the GAIA and WISE photometric measurements, taken at different epochs.

We set Gaussian priors on the free parameters according to literature estimates (Table~\ref{tab:CrA9_props}):
$T_{\rm e} \approx 3720\pm150$ K, $\log(g) \approx 3.5 \pm 0.2$, and $A_V \approx 1.8 \pm 0.3$mag. 
We allowed for variable $R_V$ in our model, to account for possibly different grain sizes in the line of sight than in the diffuse ISM \citep[e.g.][]{Weingartner2001,Calvet2004}. 
Figure~\ref{fig:SED} shows the model favoured by {\sc specfit} in green. The right panel of Figure~\ref{fig:SED} provides a zoom on the wavelength range of interest and shows 60 sample models from the posterior distribution (in light blue). The associated corner plot is shown in Figure~\ref{fig:SED_cornerplot}. For each parameter, the quoted uncertainties correspond to the 34th and 66th percentile of the posterior distribution. We find an effective temperature $T_{\rm e} \approx 3598^{+189}_{-137}$ K, 
a surface gravity $\log(g) \approx 3.8\pm0.2$, a photometric radius $R \approx 1.53^{+0.09}_{-0.14} R_{\odot}$, an optical extinction $A_V \approx 1.5^{+0.3}_{-0.2}$ mag and a broad constraint on the ratio of total-to-selective optical extinction that is consistent with the diffuse ISM value. 

We drew 1000 sample spectra from our posterior distribution in order to estimate uncertainties on the stellar spectrum at the wavelengths of our observations. We fitted a Gaussian profile to the distribution of sample values at each relevant wavelengths, and considered the standard deviation of each Gaussian fit as the uncertainty in stellar flux, for error propagation to the companion candidate spectrum. 

\rep{We note that the $L'$ flux of the posterior BT-SETTL models ($1.39 \pm 0.09 \times 10^{-14}$ W m$^{-2} \mu$m$^{-1}$) is consistent with the average of the two absolute $L'$ fluxes estimated in the 2017 and 2018 datasets through standard stars ($1.35 \pm 0.14 \times 10^{-14}$ W m$^{-2} \mu$m$^{-1}$; Section~\ref{sec:NACOobs}).}

\begin{figure*}
	\centering
	\includegraphics[width=\textwidth]{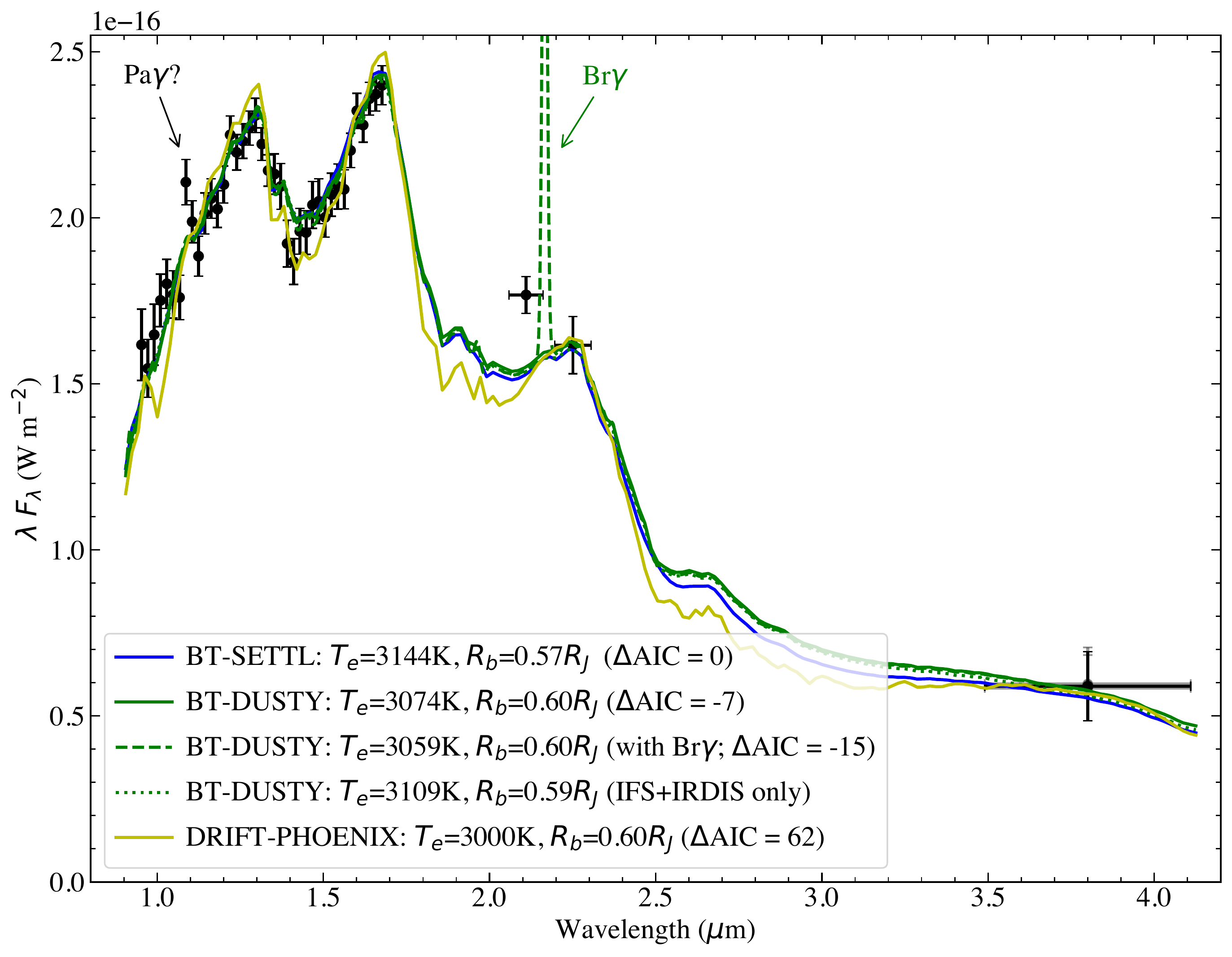}
    \caption{Measured spectrum of CrA-9B/b (\emph{black points}) compared to the favoured BT-SETTL (\emph{blue}), BT-DUSTY (\emph{green}) \rep{and DRIFT-PHOENIX (\emph{yellow}) models returned by \texttt{specfit}. The 
    results are obtained with $R_V$ set to 3.1, although the curves obtained when $R_V$ is set as a free parameter are visually identical.}
    The \emph{dashed green} curve considers the luminosity of a Br$\gamma$ emission line as free parameter. The \emph{dotted green} curve is obtained considering only the IFS+IRDIS points for the fit. The measurements considered for the fit are shown in \emph{black}. \rep{At 3.8~$\mu$m, we consider the weighted average of the} NACO 2017 and 2018 individual measurements shown in \emph{gray}. \rep{Including a Br$\gamma$ emission line reproduces better} all measurements, including the $K1$ photometric point. 
    }
    \label{fig:SED_CrA9B}
\end{figure*}

\rep{\subsection{Spectrum of the faint point source} \label{sec:specCrA9b}}

The spectrum of the point source is obtained by multiplying our contrast spectrum (Figure~\ref{fig:NEGFC_results_contrast}) to \rep{(i)} the favoured BT-SETTL model of the star (Figure~\ref{fig:SED}) \rep{for the SPHERE points}, after the model is convolved with the relevant filters used in the SPHERE observation; \rep{and (ii) 
the absolute $L'$ flux measurements for the star for the 2017 and 2018 NACO points. For the rest of our analysis, we adopt the weighted average of the two $L'$ flux values, which are consistent with each other.} 
The spectrum of the faint point source is shown with \emph{black points} in Figure~\ref{fig:SED_CrA9B}, \rep{with the two individual $L'$ measurements for 2017 and 2018 shown with \emph{gray points}}.
Our final uncertainties on the spectrum include three contributions combined in quadrature: (1) the uncertainties on the contrast inferred for the point source by NEGFC-MCMC (Section~\ref{sec:contrast_spectrum}); (2) the photon noise on the measured flux of the star in each dataset; and (3) the uncertainty on the best-fit BT-SETTL model of the star (Section~\ref{sec:specCrA9}). The latter dominate the uncertainty budget, being an order of magnitude larger than the uncertainties on the contrast and the stellar photon noise at all wavelengths.
We report in Table~\ref{tab:CrA9b_props} the apparent magnitudes inferred for the point source in the $J$, $K1$, $K2$ and $L'$ filters considering the above uncertainties.

\rep{\subsubsection{Empirical comparison} \label{sec:specCrA9b}}

\begin{figure*}
	\centering
	\includegraphics[width=0.9\textwidth]{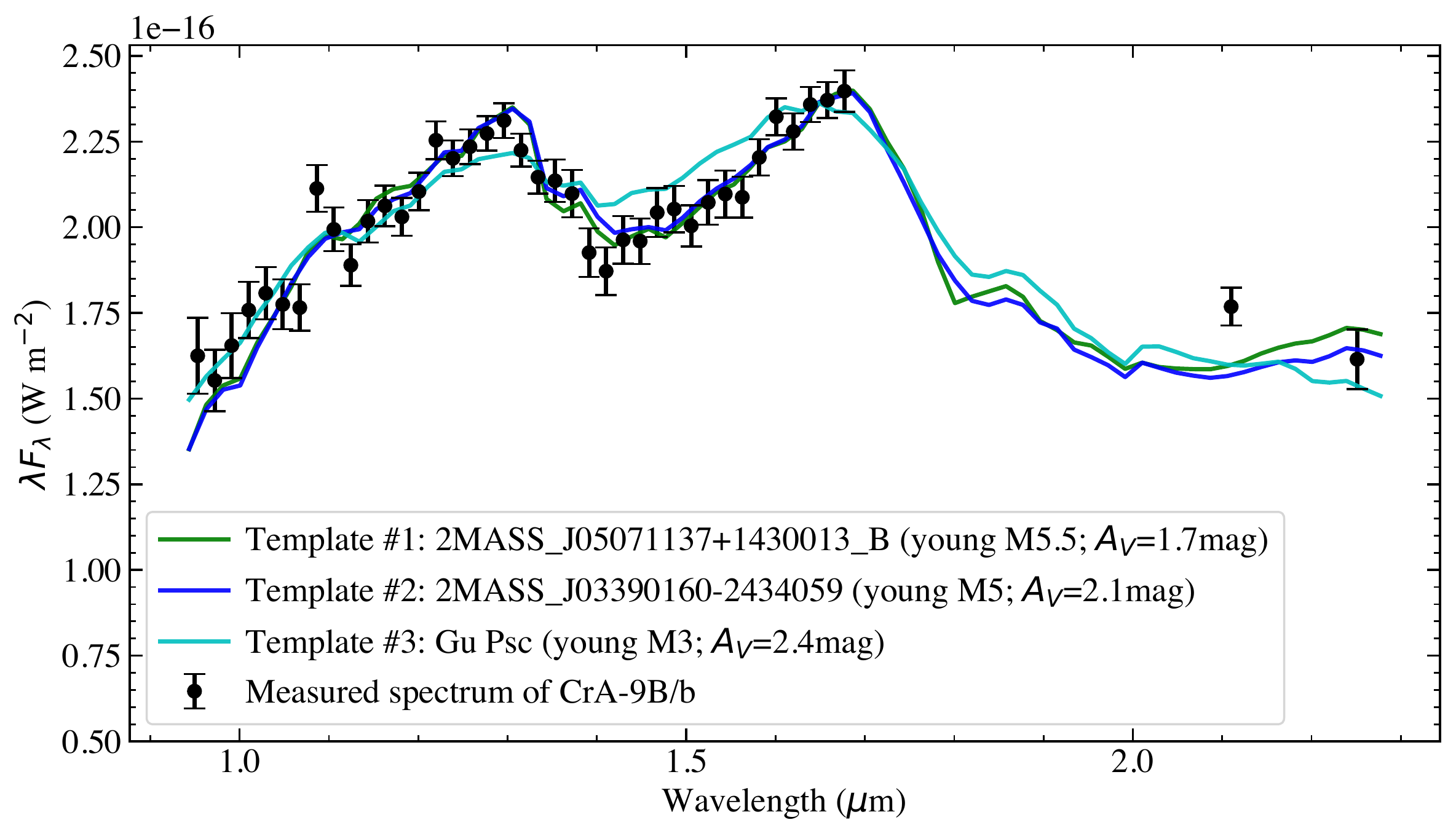}
    \caption{\rep{SPHERE (IFS+IRDIS) spectrum of CrA-9 B/b (\emph{black} points) compared to the three best-fit templates of the Montreal Spectral Library: 2MASS J05071137+1430013 B ($\chi^2_r \sim 1.7$; \citealt{Gagne2015}), 2MASS J03390160-2434059 ($\chi^2_r \sim 1.7$; \citealt{Gagne2015}) and GU Psc A ($\chi^2_r \sim 1.9$; \citealt{Naud2014}). The fit considered two free parameters: a scaling factor and optical extinction (whose best-fit value is provided in parenthesis).}
    }
    \label{fig:template_fit}
\end{figure*}

The most significant feature in the spectrum of the point source is the broad H$_2$O absorption band spanning $\sim$1.33 to 1.51$\mu$m \citep[][]{Auman1967}, which has been observed in a number of young low-mass companions \citep[e.g.][]{Bonnefoy2014}. We compute the H$_2$O and Na indices introduced in \citet{Allers2007} to estimate the spectral type and gravity of M5 to L5 \repp{dwarfs}, respectively. The H$_2$O index is insensitive to gravity and the Na index is relatively insensitive to spectral type, contrary to most other spectral indices which are sensitive to both \citep{Allers2007, Bonnefoy2014}. We measured an H$_2$O index of $\langle F_{\lambda=1.55-1.56}\rangle/\langle F_{\lambda=1.492-1.502}\rangle = 0.993\pm0.007$ and a Na index of $\langle F_{\lambda=1.15-1.16}\rangle/\langle F_{\lambda=1.134-1.144}\rangle = 1.027\pm0.007$. The former suggests a spectral type of M5.5$\pm$0.9 and the latter points to a surface gravity consistent with that of either young standards in the Cha I star-forming region or red giants \citep{Allers2007}. The median age of the Cha I dark cloud is $\sim$2 Myr \citep[e.g.][]{Luhman2008}, hence similar to the estimate for CrA-9. 
The spectral indices were calculated after dereddening of the spectrum considering an extinction value of 2.0 mag,
although it is worth noting that their value is only minimally affected by the assumed amount of extinction
(e.g.~using $A_V = 0$ mag leads to consistent spectral index and spectral type estimates).

\rep{To further constrain the nature of the point source, we compared its spectrum to templates of the Montreal Spectral Library \citep[MSL;][]{Gagne2014}, a library of 424 observed spectra of M, L and T} \repp{\repp{dwarfs}}, \rep{including $\sim$1/3 of young objects. Since no template spectrum has $L'$ measurements, we only considered the IFS+IRDIS spectrum of CrA-9 and compared it to all MSL templates (both field and young) with a wavelength range covering 0.9--2.3$\mu$m (i.e.~326 templates in total). We let both scaling and extinction as free parameters in our fit, and used the simplex search mode of the} \verb+best_fit_tmp+ \rep{routine. Figure~\ref{fig:template_fit} shows the best three fits ($\chi_r^2 \sim 1.7$--$1.9$), which all correspond to young mid-M dwarf templates reddened with $A_V \sim 2$ mag. In particular, we notice that the best-fit template, 2MASS J05071137+1430013 B, is a young M5.5 dwarf companion from the $\beta$ Pic moving group \citep{Gagne2015}.
These results are consistent with the spectral type and low-gravity inferred using the H$_2$O index and Na index.}


\rep{\subsubsection{Atmospheric models} \label{sec:specCrA9b}}

\rep{To retrieve physical parameters for the companion candidate, we subsequently ran {\sc specfit} with the BT-SETTL, BT-DUSTY and DRIFT-PHOENIX grids of models. We considered four free parameters for the BT-SETTL and BT-DUSTY models ($T_{\rm e}$, $\log(g)$, $R$ and $A_V$), and five free parameters, adding the metallicity $\log(Z/Z_{\odot})$, for the DRIFT-PHOENIX models. 
All three grids are based on the PHOENIX atmosphere models \citep{Hauschildt1992}.}
Compared to the BT-SETTL grid, BT-DUSTY models consider the maximum amount of dust allowed in equilibrium with the gas phase, without a cloud prescription \citep[][]{Allard2011,Allard2012}. 
\rep{The DRIFT-PHOENIX models use a kinetic approach to model dust formation, growth, settling and advection \citep{Woitke2003,Woitke2004,Helling2006,Helling2008}.}

For all grids, we considered uniform priors on all parameters.
In the case of the BT-SETTL grid, we considered all available models with $T_e \in [1200,5000]$K in steps of 100~K, and $\log(g) \in [2.5,5.5]$ in steps of 0.5dex. For the BT-DUSTY grid, we considered all models with $T_e \in [3000,5000]$K and $\log(g) \in [2.0,5.5]$, owing to incomplete wavelength coverage for models at low temperatures. Similar results were obtained with either grids.
\rep{We considered all available DRIFT-PHOENIX models with $T_e \in [1000,3000]$K in steps of 100~K, $\log(g) \in [3.0,5.0]$ in steps of 0.5dex, and metallicity $\log(Z/Z_{\odot}) \in [-0.3,0,0.3]$.}

The favoured models for the \rep{different} grids are shown in Figure~\ref{fig:SED_CrA9B}, and the favoured parameters are provided in Table~\ref{tab:CrA9b_spec_props}. 
For \rep{all} grids of models, the inferred effective temperature (3000--3200K) \rep{is consistent with a mid-M dwarf}. However, the photometric radius required to account for the faint measured flux is only $\sim$0.6 Jovian radius, which is highly inconsistent with a young M dwarf. 
We also note that all measurements are well reproduced by the BT-SETTL and BT-DUSTY favoured models (within $2 \sigma$), except for the $K1$ photometric point which stands out as a $\sim3 \sigma$ outlier and the 1.09$\mu$m IFS channel (2--2.5$\sigma$).
\rep{The best-fit DRIFT-PHOENIX models hit the upper bound of the grid (3000~K), which accounts for the poorer visual fit (Figure~\ref{fig:SED_CrA9B}) and underestimated uncertainties on the different parameters (Table~\ref{tab:CrA9b_spec_props}).}
The favoured value of extinction ($A_V \approx 2.0\pm0.1$ mag) is similar to that inferred for the star ($A_V \approx 1.5^{+0.3}_{-0.2}$ mag; Section~\ref{sec:specCrA9}). 
\rep{The surface gravity values appear slightly larger for the BT-SETTL and BT-DUSTY models than expected from a 1--2 Myr-old mid-M dwarf ($\sim$3.7; \citealt{Tognelli2011,Baraffe2015}). We show however in Appendix~\ref{app:specfit_tests} that the support for models with lower surface gravity, compatible with a young age, is only marginally lower}. 

In order to improve the interpretation of our results, we explored the effects of (i) \rep{fixing the photometric radius to the expected physical radius of a planet with the measured luminosity of the point source; (ii) using a different extinction law; (iii) removing the NACO points from the fit (i.e.~testing possible variability); and (iv) including} a possible Br$\gamma$ emission line affecting the $K1$ photometric point. We summarise the results of these tests below, in light of the Akaike Information Criterion. 

We used the Akaike Information Criterion \citep[AIC;][]{Akaike1974} in order to determine which of the BT-SETTL or BT-DUSTY models reproduced better the observed spectrum, and whether the addition of extra parameters to the models was useful. The AIC considers both the maximum likelihood achieved with a specific type of model and the number of free parameters involved in that model. This trade off prevents overfitting and informs on whether the addition of an additional free parameter is necessary. For each type of model, we calculated the difference $\Delta$AIC between its AIC and the AIC obtained with the grid of BT-SETTL models using 4 free parameters ($T_e$, $\log(g)$, $R$ and $A_V$), and reported that difference in Table~\ref{tab:CrA9b_spec_props}. The lower the value of AIC (hence $\Delta$AIC), the higher the likelihood.

\rep{Our first test consisted in fixing the photometric radius to 1.8 $R_{\rm Jup}$, letting the other parameters free. The best-fit BT-SETTL and DRIFT-PHOENIX models obtained with such constraint are shown in Figure~\ref{fig:specfit_test_R}. These models are poor fits to the data, with $\Delta$AIC values larger than 2000. This suggests that the solution found without constraint ($T_{\rm e} \sim$ 3000--3200~K and $R_p \sim 0.6 R_{\rm Jup}$) results indeed from the lack of good solutions with larger photometric radii.}

\begin{table*} 
\caption{Physical properties of CrA-9B/b inferred from \rep{models with high support retrieved by \texttt{specfit} for the different atmospheric grids}.}
\label{tab:CrA9b_spec_props}
\begin{threeparttable}
\centering
\begin{tabular}{lccccccccc}
\hline
\hline
Parameter & From spectral & BT-SETTL & BT-DUSTY & \rep{DRIFT-PHOENIX} & BT-DUSTY & BT-DUSTY & BT-DUSTY \\
 & indices & (fixed $R_V$) & (fixed $R_V$) & \rep{(fixed $R_V$)} & (free $R_V$) & (free $L_{\rm Br\gamma}$) & (SPHERE only) \\
\hline
Spectral type & M5.5$\pm$0.9$^{\rm (a)}$ &  --- &  --- &  --- &  --- &  --- &  ---\\
$T_{\mathrm{e}}$ [K]  & $2910\pm180^{\rm (b)}$ & $3148^{+47}_{-52}$  & $3074\pm43$ & $3000^{+0}_{-5}$ $^{\rm (e)}$  & $3064^{+56}_{-36}$ & $3057^{+49}_{-36}$ & $3114^{+37}_{-59}$ \\
Log($g$) & $3.7\pm0.3^{\rm (c)}$ & $4.6^{+0.3}_{-0.2}$ & $4.5^{+0.2}_{-0.4}$ & $4.0\pm0.1$ & $4.5^{+0.2}_{-0.4}$ & $4.5^{+0.2}_{-0.3}$ & $4.5^{+0.2}_{-0.5}$ \\
Radius [$R_J$] & $7.6\pm2.2^{\rm (d)}$ & $0.57\pm0.01$ & $0.60^{+0.01}_{-0.02}$ & $0.60\pm0.01$ & $0.60^{+0.01}_{-0.02}$ & $0.60\pm0.01$ & $0.58\pm0.02$\\ 
$A_{\rm V}$ [mag] & --- & $2.0\pm0.1$ & $2.0\pm0.1$ & $1.7^{+0.4}_{-0.1}$ & $1.8^{+0.4}_{-0.2}$ & $2.0\pm0.1$ & $1.9\pm0.1$\\
$R_{\rm V}$ &  --- & (3.1) & (3.1) & (3.1) & $1.8^{+3.0}_{-0.2}$ & (3.1) & $1.7^{+3.2}_{-0.2}$\\
$\log(Z/Z_{\odot})$ & --- & --- & --- & $-0.30^{+0.02}_{-0.00}$ $^{\rm (e)}$ & --- & --- & ---\\
$\log(\frac{L_{\rm Br\gamma}}{L_{\odot}})$ & --- & --- & --- & --- & --- & $-5.89^{+0.06}_{-0.10}$ & --- \\
\hline
$\Delta$AIC & --- & 0 & $-7$ & $62$ & $-5$ & $-15$ & ---\\
\hline
\end{tabular}
\begin{tablenotes}
\item[a] Based on the H$_2$O index and empirical relation derived in \citet{Allers2007}.
\item[b] Based on the empirical relation to convert from spectral type to effective temperature in \citet{Herczeg2014}.
\item[c] Considering the Na index value, an age of 2 Myr \citep[since the Na index is consistent with that measured for Cha I members;][]{Allers2007}, and the isochrones of either \citet{Tognelli2011} or \citet{Baraffe2015} to estimate a $\log(g)$ value.
\item[d] Considering $T_{\mathrm{e}} = 2910\pm170$K, an age of 2 Myr, and the evolutionary models of either \citet{Tognelli2011} or \citet{Baraffe2015}.
\item[e] \rep{Parameter hits bound of the grid; reported uncertainties for all parameters of this model are to be considered lower limits.} 
%
\end{tablenotes}
\end{threeparttable}
\end{table*}

\begin{figure}
	\centering
	\includegraphics[width=0.92\columnwidth]{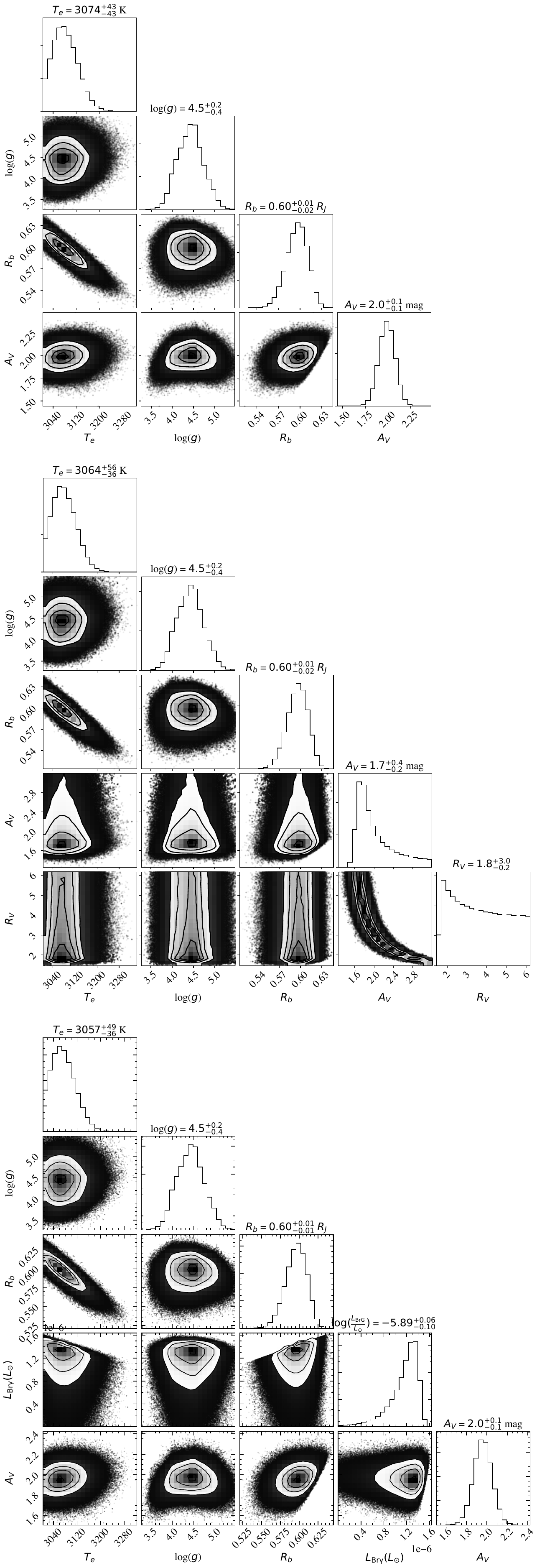}
    \caption{Corner plots retrieved by \texttt{specfit} using the BT-DUSTY models to reproduce the spectrum of CrA-9~B/b, when (\emph{top}) fixing $R_V$ to the diffuse ISM value, (\emph{middle}) setting $R_V$ as a free parameter, and (\emph{bottom}) considering the Br$\gamma$ luminosity as a free parameter. 
    }
    \label{fig:CrA9B_corner_plots}
\end{figure}

\rep{Next we tested the addition of another free parameter to model dust extinction: $R_V$.}
Since the system is young, a bound companion may be surrounded by its own disc of gas and dust. Dust growth in discs may affect the extinction law, possibly leading to a different value of $R_V$ than in the diffuse ISM \citep[e.g.][]{Weingartner2001,Calvet2004}. 
We compare in Figure~\ref{fig:CrA9B_corner_plots} the posterior distributions of the BT-DUSTY model parameters yielded by {\sc specfit} when $R_V$ is set as a free parameter or not (similar results are obtained with BT-SETTL). The favoured values of $T_e$, $\log(g)$ and $R$ are consistent with the results obtained when $R_V$ is fixed (to 3.1).
The only difference is a significant degeneracy between the values of $A_V$ and $R_V$, with a range of values for ($A_V$, $R_V$) pairs all leading to similar quality fits.
Considering or not $R_V$ as a free parameter led to similar AIC values for either BT-SETTL or BT-DUSTY models (Table~\ref{tab:CrA9b_spec_props}),
and none of the favoured models is able to reproduce the $K1$ photometric point (Figure~\ref{fig:SED_CrA9B}). 
Given that $\Delta$AIC$<10$ between all models which fixed $R_V$ or not, there is no significant support for one of these models over the others \citep[e.g.][]{Burnham2002}.

\rep{Given the possibility for the companion to be variable like the host star,}
it is unclear \rep{whether its} $L'$ flux is \rep{the same} at the epoch of the SPHERE dataset \rep{and at the epochs of the NACO datasets.
Furthermore, our absolute flux calibration of the SPHERE measurements is based on CrA-9 which is known to be variable, and could hence lead to a shift with respect to the $L'$ measurement}.
Therefore we also ran {\sc specfit} on the IFS+IRDIS points only.
In that case, we found that the favoured BT-SETTL and BT-DUSTY models are consistent with the ones obtained when including the $L'$ point. 
This suggests that the puzzling parameter values found above and the impossibility for the models to reproduce the $K1$ point are due to \rep{neither variability of the point source nor an inadequate absolute flux calibration of the SPHERE measurements}. 


Our last test consisted in considering 
the luminosity of a putative Br$\gamma$ emission line ($\lambda = 2.166 \mu$m) as a free parameter, in an attempt to account for the observed discrepancy between the $K1$ photometric measurement and the favoured BT-SETTL/BT-DUSTY models from all previous tests 
(Figure~\ref{fig:SED_CrA9B}).
This test is physically motivated by CrA-9 being a known accretor \citep{Romero2012}, suggesting that a putative companion would likely be accreting material as well. 
We therefore ran {\sc specfit} again, this time adding an extra free parameter for the luminosity of the Br$\gamma$ emission, but fixing $R_V$ to 3.1 and including the weighted-average NACO point. 
{\sc specfit} assumes a Gaussian profile for injected lines. 
To limit the line injection process to a single free parameter -- its flux, we set the full width at 10\% height to 100 km s$^{-1}$, based on the observed H recombination line width of the PDS 70 and Delorme 1 (AB) b planets \citep[][]{Haffert2019,Eriksson2020}. Given the low resolution of our spectrum and the $K1$ filter transmission curve, we do not expect any significant change for a different line width.
The favoured BT-SETTL and BT-DUSTY models when considering the Br$\gamma$ emission line enable us to reproduce the observed $K1$ flux with a Br$\gamma$ luminosity $\log(\frac{L_{\rm Br\gamma}}{L_{\odot}})\approx-5.9\pm0.1$, and lead to similar physical parameters as estimated without including that free parameter (Figure~\ref{fig:SED_CrA9B}; Table~\ref{tab:CrA9b_spec_props}). The values of $\Delta$AIC for these types of model ($-8$ and $-13$ for BT-SETTL and BT-DUSTY, respectively) suggests marginal support in favour of including the Br$\gamma$ luminosity as a free parameter. 
The favoured BT-DUSTY model including Br$\gamma$ line emission is shown with a \emph{green dashed line} in Figure~\ref{fig:SED_CrA9B}, \rep{and the corresponding corner plot is shown in Figure~\ref{fig:CrA9B_corner_plots}}.

\section{Discussion}\label{sec:Discussion}

In this section, we discuss several possible interpretations for our results, their respective likelihood, and the kind of observations required to discrimate our two leading hypotheses.

\subsection{What is the faint point source?}

\subsubsection{Is it a background star?}

A background star scenario would explain why the BT-SETTL and BT-DUSTY models favoured by {\sc specfit} have an anomalously small photometric radius ($\sim0.6 R_J$ at the distance of CrA-9), and a larger surface gravity ($\log(g) \sim 4.5$) than expected for an object of the age of CrA-9 ($\log(g) \sim 3.7$).
Considering the effective temperature of 3000--3200~K inferred by {\sc specfit} based on the BT-SETTL and BT-DUSTY model grid, if the point source were a background field M-dwarf one would expect a radius of $\sim$ 2.5--3.0 $R_J$ \citep{Baraffe2015}, hence lying at a distance several times that of CrA-9.
However, the three-epoch astrometric positions we measured are consistent with a companion co-moving with CrA-9 (Figure~\ref{fig:NEGFC_results_astrom}).
We can consequently reject a background object with negligible proper motion with a 5$\sigma$ confidence.
Furthermore, 
it is unlikely that a background star several times further than CrA-9 had a high proper motion perfectly matching the significant foreground motion of CrA-9.
A high-proper motion background star scenario is also inconsistent with the measured 1.13$\mu$m-Na spectral index (Section~\ref{sec:specCrA9b}). The value of $1.027\pm0.007$ is consistent with a young very low-gravity object and rejects at a 5$\sigma$ confidence the same gravity as field dwarfs.

The Galactic latitude of CrA-9 ($b \approx -17\degr$) further makes a background star several times the distance to CrA-9 unlikely. Our estimated probability of the point source being a background star is $\lesssim$0.2\%, considering both its separation of $\sim$0\farcs7 and its apparent magnitude in $L'$ band. We estimated this by considering a spatial homogeneous Poisson point process with the rate set by the number density of objects brighter than $L' = 16.7$mag and the area set to a disc with radius equal to the separation of the point source \citep[e.g.][]{Ubeira-Gabellini2020}. The number density was evaluated in that region of the sky using the TRILEGAL model of the galaxy \citep{Girardi2012}.

Considering the above, the evidence 
is in favour of a bound companion. We thereafter refer to the point source as \emph{companion}.

\subsubsection{Is it a protoplanet?}  \label{sec:discu_protoplanet}

Considering a distance of $d=153.1\pm1.2$~pc \citep{Gaia-Collaboration2018}, an optical extinction of $1.9\pm0.2$ mag and $R_V=3.1$ (Table~\ref{tab:CrA9b_spec_props}), we estimate the absolute magnitudes of the companion to be $11.90\pm0.10$ mag in the $J$ band, $10.59 \pm 0.07$ mag with the $K1$/$K2$ filter, and either $10.63 \pm 0.19$ mag or $10.21\pm0.11$ mag in the $L'$ band, based on the 2017 and 2018 epochs data respectively (Table~\ref{tab:CrA9b_props}). Considering an age of 1--2 Myr, these absolute magnitudes would correspond to masses of $\sim$2--5 $M_J$ according to either the COND evolutionary models \citep{Baraffe2003} or the hot-start models presented in \citet{Spiegel2012}. However the $J-L'$ color of 1.3--1.7 mag appears bluer than predicted by those models ($>2.1$ mag). The inferred radius of 0.57--0.60 $R_J$ is also significantly smaller than expected for 
\rep{a gas-dominated planet}, and the estimated effective temperature of the companion (3000--3200 K) appears to defy even the most optimistic hot-start models for a 10 $M_J$ planet \citep[e.g.][]{Baraffe2003,Spiegel2012,Mordasini2012a}. 
The companion also shows a significantly bluer spectrum than PDS~70~b \citep{Muller2018,Christiaens2019b}, 
although this might be due to PDS~70~b being more significantly enshrouded by dust   \citep{Christiaens2019b,Wang2020,Stolker2020b}. 

The arguments above 
\emph{a priori} suggest that CrA-9B/b is unlikely to be a protoplanet.
Nonetheless, an alternative explanation for the observed spectrum is that it traces the fraction of a protoplanet photosphere that is heated by an accretion shock \citep[e.g.][]{Zhu2015b,Aoyama2020}. This scenario would be compatible with a significant Br$\gamma$ emission line (2.166 $\mu$m), possibly required to account for the observed $K1$ photometric measurement (Table~\ref{tab:CrA9b_spec_props}).
Our spectral fit suggests a Br$\gamma$ luminosity 
$\log(L_{\rm Br\gamma}/L_{\odot}) \approx -5.9\pm0.1$ 
in order to make up for the $\gtrsim 3\sigma$ difference between the favoured BT-DUSTY and BT-SETTL models (without Br$\gamma$) and the measured $K1$ flux (Figure~\ref{fig:SED_CrA9B}). %
The Br$\gamma$ line is a known tracer of gas accretion for classical T-Tauri stars \citep{Muzerolle1998,Calvet2004}, and also possibly for protoplanets \citep{Aoyama2020}.
According to the protoplanet accretion shock models presented in \citet{Aoyama2020}, a $\log(L_{\rm Br\gamma}/L_{\odot}) \approx -5.9$ would suggest a mass accretion rate of $\sim 10^{-5} M_J$ yr$^{-1}$.
This estimate is significantly larger than inferred for the PDS~70 planets \citep[][]{Haffert2019,Hashimoto2020}, although the latter may be at a more advanced stage considering they have already cleared a wide and deep gap \citep[e.g.][]{Hashimoto2011, Dong2012,Keppler2020}.
The inferred mass accretion rate is similar to what has been predicted from magneto-hydrodynamical simulations of accreting sub-Jovian planets embedded in the protoplanetary disc \citep[e.g.][]{Gressel2013}. This would require the companion to still lie within the gaseous component of the protoplanetary disc. Although the estimated dust disc radius \citep[$\sim 0\farcs39$;][]{Cazzoletti2019} is smaller than the separation of the companion, there is no current constraint on the extent of the gas disc which is likely to extend to larger separations than the mm-size dust, hence possibly up to the separation of the companion.

If we are indeed witnessing an accreting protoplanet, a significant fraction of the observed luminosity would be expected to come from accretion luminosity \citep[potentially larger than the photospheric contribution; e.g.][]{Mordasini2017,Marleau2017,Marleau2019}. 
Whether the accretion stream shocks on the circumplanetary disc or directly onto the photosphere of the protoplanet, the surface is expected to reach $\gtrsim 3000$K  \citep[e.g.][]{Lovelace2011,Zhu2015b,Szulagyi2017a}.
In this scenario, the BT-SETTL/BT-DUSTY models may be inappropriate to represent the continuum emission.
\citet{Zhu2015b} presented three-component model SEDs for accreting protoplanets, including the contribution from the photosphere, the fraction of the photosphere that is heated by the accretion shock(s), and the accreting circumplanetary disc. 
In presence of accretion shocks, \citet{Zhu2015b} showed that the fraction of the photosphere that is heated by the shock can become orders of magnitude brighter than the rest of the photosphere. 
The former contribution dominates at short wavelength, from optical to near-IR wavelengths, while the circumplanetary disc emission may dominate at mid- to far-IR wavelengths. 
The effective temperature to which the photosphere is heated at the accretion shock can be $\gtrsim 3000$K for a moderately accreting protoplanet. 
Furthermore, considering a mass accretion rate of $\sim 10^{-5} M_J$ yr$^{-1}$ (based on $L_{\rm Br \gamma}$ and the models in \citealt{Aoyama2020}), the observed blue slope in the 1.0--3.8$\mu$m range may be compatible with a subset of the models presented in \citet{Zhu2015b} with different combinations of inner truncation radius for the circumplanetary disc and filling factors between 1--10\% (i.e. the fraction of the surface covered by accretion shocks). 
Finally, the above scenario would account for the small photometric radius inferred in this work; a 0.6 $R_J$ photometric radius makes for 2--10\% of the area of a 1.6--5 $R_J$ protoplanet predicted from cold- to warm-start models in \citet[][]{Aoyama2020}.

A potential caveat of the above hypothesis is that one may also expect to observe Pa$\beta$ (1.282 $\mu$m) and Pa$\gamma$ (1.094 $\mu$m) recombination lines in emission \citep[e.g.][]{Aoyama2020}. 
A 2.5$\sigma$ excess can be seen at $\sim 1.09 \mu$m but no excess is seen at $\sim$1.28 $\mu$m.
However, our conclusions on the H recombination lines of the companion are limited by our ignorance of the emission lines affecting the star. Indeed, the spectrum of the companion is obtained in contrast of a model SED for the star which does not include its emission lines (i.e. only the physics included in the BT-SETTL/BT-DUSTY models). The star is known to have a significant H$\alpha$ emission line though \citep{Romero2012}, which suggests it also harbours other H recombination lines such as Pa$\beta$, Pa$\gamma$ or Br$\gamma$. 
Not including those in the model SED of the star leads to lower limits on the inferred flux of those lines in the companion spectrum. 
It is nevertheless possible to discuss the relative line ratios between the star and the companion. If the excesses at $\sim$1.09~$\mu$m and in the $K1$ band are to be attributed to Pa$\gamma$ and Br$\gamma$ emission lines, respectively, the lack of excess at $\sim$1.28~$\mu$m suggests that both the Pa$\gamma$/Pa$\beta$ and Br$\gamma$/Pa$\beta$ line ratios are larger for the companion than for the star.
This translates to Pa$\gamma_{b}$/Pa$\beta_{b} \gtrsim 0.8$ and Br$\gamma_{b}$/Pa$\beta_{b} \gtrsim 0.3$, where we considered the reported luminosity and mass accretion rate of CrA-9 to estimate its expected line ratios using the models in \citet{Edwards2013}. 
The former requirement may be met by either an accreting T-Tauri star or protoplanet for reasonable assumptions on the shock velocity and H number density \citep[][]{Edwards2013,Aoyama2020}, but the latter condition appears more difficult to reconcile with predictions for an accreting protoplanet (Br$\gamma_{b}$/Pa$\beta_{b}\approx$ 0.1--0.2; \citealt{Aoyama2020}).
The above reasoning is nonetheless subject to a number of assumptions, e.g.~that the star had a similar accretion rate $\sim$10 years earlier \citep[H$\alpha$ measurement presented in][]{Romero2012} than our 2019 SPHERE observations; and that the filling factors assumed for line ratio predictions were $\sim$10\% \citep{Edwards2013} and $\sim$100\% \citep{Aoyama2020} for the accretion on the star and protoplanet cases, respectively. A similar analysis should therefore be performed using a higher resolution (and contemporary) IR spectrum for both the star and the companion to reach a more definitive conclusion. 



\subsubsection{Is it an obscured stellar binary?}\label{sec:discu_binary}

The recent examples of FW~Tau~C \citep{Kraus2014,Wu2017a} and CS~Cha~B \citep{Ginski2018,Haffert2020} show that highly extinct stellar binary companions may mimic the signal of a planetary-mass companion, in particular when the only available information is a near-IR flux and/or colour.
In the case of the faint companion around CrA-9, the optical extinction derived by the spectral analysis appears moderate ($A_V \sim 1.9$ mag), and appears well-constrained by the slope of the blue end of the spectrum. This would 
argue against the possibility of a highly extinct binary.
An alternative, however, is that the stellar flux is completely blocked by an edge-on disc\rep{, and re-emitted by the disc only at mid-IR wavelengths.} \rep{In that case the near-IR} signal from the point source would only correspond to scattered light by the surface of the (circum-secondary) disc.
Only a few percents of the total flux would come through and the optical extinction would become irrelevant.
This scenario would account for the small photometric radius inferred for the companion.
This would make the companion similar to HK~Tau~B, an M-dwarf binary companion with a circum-secondary disc seen almost edge-on \rep{and a near-IR spectrum consistent with pure (albeit underluminous) photospheric emission 
\citep{Stapelfeldt1998,McCabe2011}. Another similar case in the literature is the wide-separation (3400~AU) young M-dwarf binary TWA 30 B, whose faint photometry and high-resolution spectrum suggests the presence of an edge-on disc obscuring stellar light \citep{Looper2010}.} 



An effective temperature of 3000--3200 K would suggest a 0.1--0.2 $M_\odot$ M-dwarf for an age of 1--2 Myr \citep[e.g.][]{Baraffe2015}. Considering the $L_{\rm Br\gamma}$--$L_{\rm acc}$ empirical relations calibrated on low-mass T-Tauri stars \citep{Muzerolle1998}, the Br$\gamma$ luminosity of $\log(L_{\rm Br\gamma}/L_{\odot}) \approx -5.9$
inferred for the companion would suggest a total accretion luminosity $\log(L_{\rm acc}/L_{\odot}) \approx -3.0$.
This converts to a mass accretion rate of $\sim 2\times 10^{-9} M_{\odot}$ yr$^{-1} $for a young 0.15-$M_{\odot}$ M dwarf. This is 
similar to the average accretion rate of low-mass T-Tauri stars \citep[$6 \times 10^{-9} M_{\odot}$ yr$^{-1}$;][]{Gullbring1998,Calvet2004}, and to the mass accretion rate expected for CrA-9 based on its measured H$\alpha$ luminosity ($2.5\times 10^{-9} M_{\odot}$ yr$^{-1}$).
A significant Br$\gamma$ emission line has also been observed for HK Tau B \citep[][]{McCabe2011}, which is another property shared with the companion seen around CrA-9.

The obscured binary scenario may appear slightly at odds with the lack of disc signal from that location. Our NACO PDI observations did not detect any polarised signal at the location of the companion, despite good and stable observing conditions and a $\sim$1 h integration (Appendix~\ref{app:NACO-PDI}). However, it is unclear whether the observation was sensitive enough to detect the polarised fraction of the faint flux received from the companion. 
No sub-mm continuum emission was detected at the location of the companion either \citep{Cazzoletti2019}. Considering their RMS noise and a dust temperature of 20~K, this corresponds to a 3$\sigma$ upper limit on the dust mass of $\sim 0.72 M_{\earth}$ \rep{(or $\sim 0.2 M_J$ total disc mass assuming a standard 100:1 gas-to-dust ratio). The lack of resolved near-IR emission around the point source translates into an upper limit of $\sim$12 au (considering a 1.3FWHM resolution power achieved with IRDIS) for the extent of the circum-secondary disc, which corresponds to a fraction of the Hill radius of a 0.1 $M_{\odot}$ companion orbiting at $\gtrsim$108 au separation ($\gtrsim 44$ au Hill radius). This limit may be consistent with the expected size of a small edge-on scattering disc producing all the flux observed from the companion. To obtain an order of magnitude estimate we consider the example of the well-characterised HK~Tau~B edge-on circum-secondary disc. The disc of HK Tau B has a radius of 104~au which leads to a $\sim$10 times underluminous M2 dwarf spectrum; we would thus expect a $\sim$ 10 times smaller disc radius for the $\sim$1000 times underluminous spectrum of CrA-9 B/b; i.e. a radius of $\sim$ 10 au.}

A similar caveat affects the obscured accreting binary and accreting protoplanet scenarios: the lack of significant Pa$\beta$ in the IFS spectrum. The latter is observed to be brighter than the Br$\gamma$ line for most CTTS \citep[e.g.][]{Calvet2004,Edwards2013}. In the case of an obscured binary, it is also surprising that such a significant Br$\gamma$ emission line would not be as affected by obscuration as the continuum flux. This may require an unlikely viewing geometry.
Alternatively, this leaves the door open to the possibility of the mismatch between the $K1$ photometric measurement and the BT-DUSTY/BT-SETTL models having an origin other than Br$\gamma$ line emission. Either the assumptions made in these models not applying to the case of CrA-9~B/b (e.g.~assumptions on atmosphere microphysics, opacity sources, and/or metallicity), or another emission line may be affecting the $K1$ measurement. 

Another result of our spectral characterisation of CrA-9B/b using BT-DUSTY and BT-SETTL models, is that the favoured surface gravity values appear larger than expected based on the age of that system \citep[$\log(g) \gtrsim 4.5$ instead of $\log(g) \approx 3.7$;][]{Tognelli2011,Baraffe2015}.
A similar conclusion was reached upon characterisation of the $H$+$K$ spectrum of the young accreting M-dwarf companion HD~142527~B 
\citep[$\log(g) \gtrsim 4.5$;][]{Christiaens2018}.
Could the large surface gravity values inferred for both CrA-9 B/b and HD 142527 B be due to the BT-SETTL and BT-DUSTY models not including the effect of magnetic fields? 
When inferring physical parameters of low-mass T-Tauri stars from their optical/near-IR spectrum, not considering the effect of their strong magnetic field \citep[e.g.][]{Johns-Krull1999,Johns-Krull2007} is known to lead to significant discrepancies in effective temperature and surface gravity estimates \citep[see e.g.][for the case of TW Hya]{Sokal2018}.
However, to our knowledge this effect has not been investigated on near-IR continuum shape, it is thus unclear whether this could account for the large inferred surface gravity. 




Finally, it is worth noting that if the point source is an obscured M-dwarf, it could either trace a bound companion or a stellar fly-by. Either scenarios would be compatible with the tentative spirals seen in the IFS image and the presence of a possibly large amount of obscuring material  \citep[e.g.][]{Zhu2015a,Cuello2020}. Moreover, both scenarios are common outcomes in hydrodynamical simulations of star formation in relatively dense environments \citep[e.g.][]{Bate2018}. 

\subsection{How to disentangle a protoplanet from an obscured binary?}

The following types of follow-up observations would be the most useful to constrain the nature of the companion:
\begin{enumerate}
    \item VLT/MUSE observations would set independent constraints on the accretion luminosity and mass accretion rate of the companion and potentially provide a spectrum at visible wavelengths for the companion \citep[e.g.][]{Haffert2019,Hashimoto2020,Haffert2020}. 
    \item Either VLT/GRAVITY or Keck/KPIC observations could provide a medium resolution spectrum of the companion in the $K$ band, hence confirm whether a Br$\gamma$ emission line is present \citep{Gravity-Collaboration2019,Mawet2016}. \rep{Furthermore,} GRAVITY could also constrain the size of the emitting region, as it was recently done for PDS~70~b \citep{Wang2021}. \rep{In particular, it would be able to resolve an edge-on circum-secondary disc as small as $\sim$0.4 au at the distance of CrA-9, hence possibly confirm the obscured M-dwarf scenario}. It would also provide a high-accuracy astrometric point which would constrain the proper motion \rep{of the companion}.
    \item High-sensitivity IR observations at longer thermal-IR wavelengths (e.g.~with VLT/VISIR, the upcoming VLT/ERIS instrument or the James Webb Space Telescope) would allow us to set constraints on the presence of a potential circum-planetary/secondary disc \citep[e.g.][for PDS 70 b]{Stolker2020b}.
    \item Sub-mm wavelength observations with ALMA to search for a circum-companion disc, either in the continuum to probe mm-size dust grains \citep[e.g.][]{Isella2019}, or in $^{12}$CO to probe the gaseous component. The latter could provide an estimate of the mass of the companion based on the disc kinematics \citep[e.g.][for the case of FW Tau C]{Wu2017a}.
    These observations may also confirm substructures in the circumprimary disc (such as the tentative IR spirals), which could then be used to derive independent mass estimates on the companion through hydro-dynamical modeling \citep[e.g.][]{Price2018,Calcino2020}. 
    \item \rep{A combination of high-precision radial velocity and Gaia astrometric measurements for CrA-9, together with our direct imaging constraints, would enable to set independent constraints on the mass of the companion \citep[e.g.][]{Brandt2019,Brandt2020}.}
\end{enumerate}
Each of these observations could then be compared to the predictions from the scenarios presented in Sections~\ref{sec:discu_protoplanet} \& \ref{sec:discu_binary}, in order to constrain the nature of the companion.

\section{Conclusions}\label{sec:Conclusion}

In this work, we developed the following methods:
\begin{enumerate}
    \item We implemented new reduction pipelines for non-coronagraphic data obtained with VLT instruments NACO, IRDIS and IFS. Each of these pipelines makes use of {\sc vip} routines, while the latter two also make use of {\sc esorex} recipes.
    \item We adapted the {\sc negfc} module of the open-source package {\sc vip} \citep{GomezGonzalez2017} in order to refine the extraction of the astrometry and contrast of directly imaged companions.
    \item We implemented {\sc specfit}, a new module for the spectral characterisation of both stellar and substellar objects in a Bayesian framework, added it to {\sc vip}, and used it to infer the most likely physical parameters for both the star and the faint companion discovered in this work.
\end{enumerate}

We have applied the methods listed above to analyse our data on CrA-9. Our scientific results are summarised as follow:
\begin{enumerate}
    \item We observed the T-Tauri star CrA-9, a known transition disc with a dust cavity, and detected a faint point source at $0\farcs7$ separation from the star in VLT/SPHERE and VLT/NACO observations obtained at three different epochs. 
    \item We also report the tentative detection of a spiral pattern, possibly connected to the point source.
    \item Our NACO polarised intensity observations did not detect any scattered-light signal neither from the location of the point source, nor from the protoplanetary disc.
    \item The multi-epoch astrometry we inferred for the point source rejects a fixed background star at a $5\sigma$ confidence level, and is consistent with a bound companion.
    \item We determined that the companion was 7.1--7.9 mag fainter than the star in the 1.0--3.8 $\mu$m wavelength range, leading to absolute magnitude estimates consistent with a planetary-mass companion (2--5 $M_J$ considering the COND models and an age of 1--2 Myr; \citealt{Baraffe2003}). 
    \item \rep{We fitted our spectrum with all available templates from the Montreal Spectral Library and found the best match with templates of young mid-M} \repp{dwarfs}. \rep{This result is consistent with the measured} $J$-band spectral indices suggesting a spectral type of M5.5$\pm$0.9 and a low surface gravity consistent with that of Cha I dark cloud members. 
    \item The models favoured by {\sc specfit} point towards an effective temperature of 3000--3200 K and a photometric radius of only 0.56--0.61 $R_J$ (considering all models within AIC-AIC$_{\rm{min}} < 10$). 
    The discrepancy between the inferred $K1$ photometric point and both BT-DUSTY and BT-SETTL models may suggest the presence of a significant Br$\gamma$ emission line, which would indicate on-going mass accretion.
    \item Our two leading hypotheses regarding the nature of the companion are: (1) an accreting protoplanet, from which we are probing the fraction of the photosphere that is heated by accretion shocks; (2) an obscured stellar binary harbouring an edge-on disc such that only a small fraction of its light, scattered from the disc surface, is reaching us. 
\end{enumerate}

New observations at either shorter (including the H$\alpha$ line) or longer wavelengths (mid-infrared and sub-mm) could confirm whether the point source is an accreting substellar companion or an obscured stellar binary.



\section*{Acknowledgments}

We thank Matthias Schreiber and Alejandro Melo for sharing the data presented in \citet{Romero2012}, Rebecca Jensen-Clem for suggesting the use of a criterion based on the autocorrelation time for MCMC convergence, and Julien Milli for useful discussions regarding the degree of polarisation of companions.
We acknowledge funding from the Australian Research Council via DP180104235, FT130100034 and FT170100040.
Part of this work has received funding from the European Research Council (ERC) under the European Union's Horizon 2020 research and innovation programme (grant agreement No 819155), and by the Wallonia-Brussels Federation (grant for Concerted Research Actions.
This project has received funding from the European Union's Horizon 2020 research and innovation programme under the Marie Skłodowska-Curie grant agreement No 823823 (DUSTBUSTERS).
G-DM acknowledges the support of the German Science Foundation (DFG) priority program SPP~1992 ``Exploring the Diversity of Extrasolar Planets'' (KU 2849/7-1) and from the Swiss National Science Foundation under grant BSSGI0\_155816 ``PlanetsInTime''. Parts of this work have been carried out within the framework of the NCCR PlanetS supported by the Swiss National Science Foundation. P.D. acknowledges the support of the French National Research Agency in the framework of the Investissements d'Avenir program (ANR-15-IDEX-02), through the funding of the "Origin of Life" project of the Univ. Grenoble-Alpes.
This work has made use of the SPHERE Data Centre, jointly operated by OSUG/IPAG (Grenoble), PYTHEAS/LAM/CeSAM (Marseille), OCA/Lagrange (Nice), Observatoire de Paris/LESIA (Paris), and Observatoire de Lyon/CRAL, and supported by a grant from Labex OSUG@2020 (Investissements d'avenir – ANR10 LABX56).
This work has made use of data from the European Space Agency (ESA) mission
{\it Gaia} (\url{https://www.cosmos.esa.int/gaia}), processed by the {\it Gaia}
Data Processing and Analysis Consortium (DPAC,
\url{https://www.cosmos.esa.int/web/gaia/dpac/consortium}). Funding for the DPAC
has been provided by national institutions, in particular the institutions
participating in the {\it Gaia} Multilateral Agreement. This work has made use of the Multi-modal Australian ScienceS Imaging and Visualisation Environment (MASSIVE) (\url{www.massive.org.au}). \rep{This research has benefitted from the Montreal Brown Dwarf and Exoplanet Spectral Library, maintained by Jonathan Gagné}.

\section*{Data Availability}
The data underlying this article will be made available on CDS.

\bibliographystyle{mnras}
\bibliography{CrA9} 

\begin{thebibliography}{}
\makeatletter
\relax
\def\mn@urlcharsother{\let\do\@makeother \do\$\do\&\do\#\do\^\do\_\do\%\do\~}
\def\mn@doi{\begingroup\mn@urlcharsother \@ifnextchar [ {\mn@doi@}
  {\mn@doi@[]}}
\def\mn@doi@[#1]#2{\def\@tempa{#1}\ifx\@tempa\@empty \href
  {http://dx.doi.org/#2} {doi:#2}\else \href {http://dx.doi.org/#2} {#1}\fi
  \endgroup}
\def\mn@eprint#1#2{\mn@eprint@#1:#2::\@nil}
\def\mn@eprint@arXiv#1{\href {http://arxiv.org/abs/#1} {{\tt arXiv:#1}}}
\def\mn@eprint@dblp#1{\href {http://dblp.uni-trier.de/rec/bibtex/#1.xml}
  {dblp:#1}}
\def\mn@eprint@#1:#2:#3:#4\@nil{\def\@tempa {#1}\def\@tempb {#2}\def\@tempc
  {#3}\ifx \@tempc \@empty \let \@tempc \@tempb \let \@tempb \@tempa \fi \ifx
  \@tempb \@empty \def\@tempb {arXiv}\fi \@ifundefined
  {mn@eprint@\@tempb}{\@tempb:\@tempc}{\expandafter \expandafter \csname
  mn@eprint@\@tempb\endcsname \expandafter{\@tempc}}}

\bibitem[\protect\citeauthoryear{{Absil} \& {Mawet}}{{Absil} \&
  {Mawet}}{2010}]{Absil2010}
{Absil} O.,  {Mawet} D.,  2010, \mn@doi [\aapr] {10.1007/s00159-009-0028-y},
  \href {https://ui.adsabs.harvard.edu/abs/2010A&ARv..18..317A} {18, 317}

\bibitem[\protect\citeauthoryear{{Absil} et~al.,}{{Absil}
  et~al.}{2013}]{Absil2013}
{Absil} O.,  et~al., 2013, \mn@doi [\aap] {10.1051/0004-6361/201322748}, \href
  {https://ui.adsabs.harvard.edu/abs/2013A&A...559L..12A} {559, L12}

\bibitem[\protect\citeauthoryear{{Akaike}}{{Akaike}}{1974}]{Akaike1974}
{Akaike} H.,  1974, IEEE Transactions on Automatic Control, \href
  {https://ui.adsabs.harvard.edu/abs/1974ITAC...19..716A} {19, 716}

\bibitem[\protect\citeauthoryear{{Allard}}{{Allard}}{2014}]{Allard2014}
{Allard} F.,  2014, in {Booth} M.,  {Matthews} B.~C.,   {Graham} J.~R.,  eds,
  IAU Symposium Vol. 299, Exploring the Formation and Evolution of Planetary
  Systems. pp 271--272, \mn@doi{10.1017/S1743921313008545}

\bibitem[\protect\citeauthoryear{{Allard}, {Homeier}  \& {Freytag}}{{Allard}
  et~al.}{2011}]{Allard2011}
{Allard} F.,  {Homeier} D.,   {Freytag} B.,  2011, in {Johns-Krull} C.,
  {Browning} M.~K.,   {West} A.~A.,  eds,  Astronomical Society of the Pacific
  Conference Series Vol. 448, 16th Cambridge Workshop on Cool Stars, Stellar
  Systems, and the Sun. p.~91 (\mn@eprint {arXiv} {1011.5405})

\bibitem[\protect\citeauthoryear{{Allard}, {Homeier}  \& {Freytag}}{{Allard}
  et~al.}{2012}]{Allard2012}
{Allard} F.,  {Homeier} D.,   {Freytag} B.,  2012, \mn@doi [Philosophical
  Transactions of the Royal Society of London Series A]
  {10.1098/rsta.2011.0269}, \href
  {https://ui.adsabs.harvard.edu/abs/2012RSPTA.370.2765A} {370, 2765}

\bibitem[\protect\citeauthoryear{{Allers} et~al.,}{{Allers}
  et~al.}{2007}]{Allers2007}
{Allers} K.~N.,  et~al., 2007, \mn@doi [\apj] {10.1086/510845}, \href
  {https://ui.adsabs.harvard.edu/abs/2007ApJ...657..511A} {657, 511}

\bibitem[\protect\citeauthoryear{{Amara} \& {Quanz}}{{Amara} \&
  {Quanz}}{2012}]{Amara2012}
{Amara} A.,  {Quanz} S.~P.,  2012, \mn@doi [\mnras]
  {10.1111/j.1365-2966.2012.21918.x}, \href
  {http://cdsads.u-strasbg.fr/abs/2012MNRAS.427..948A} {427, 948}

\bibitem[\protect\citeauthoryear{{Aoyama}, {Marleau}, {Mordasini}  \&
  {Ikoma}}{{Aoyama} et~al.}{2020}]{Aoyama2020}
{Aoyama} Y.,  {Marleau} G.-D.,  {Mordasini} C.,   {Ikoma} M.,  2020, arXiv
  e-prints, \href {https://ui.adsabs.harvard.edu/abs/2020arXiv201106608A} {p.
  arXiv:2011.06608}

\bibitem[\protect\citeauthoryear{{Auman}}{{Auman}}{1967}]{Auman1967}
{Auman} Jason J.,  1967, \mn@doi [\apjs] {10.1086/190153}, \href
  {https://ui.adsabs.harvard.edu/abs/1967ApJS...14..171A} {14, 171}

\bibitem[\protect\citeauthoryear{{Baraffe}, {Chabrier}, {Barman}, {Allard}  \&
  {Hauschildt}}{{Baraffe} et~al.}{2003}]{Baraffe2003}
{Baraffe} I.,  {Chabrier} G.,  {Barman} T.~S.,  {Allard} F.,   {Hauschildt}
  P.~H.,  2003, \mn@doi [\aap] {10.1051/0004-6361:20030252}, \href
  {http://adsabs.harvard.edu/abs/2003A%26A...402..701B} {402, 701}

\bibitem[\protect\citeauthoryear{{Baraffe}, {Homeier}, {Allard}  \&
  {Chabrier}}{{Baraffe} et~al.}{2015}]{Baraffe2015}
{Baraffe} I.,  {Homeier} D.,  {Allard} F.,   {Chabrier} G.,  2015, \mn@doi
  [\aap] {10.1051/0004-6361/201425481}, \href
  {http://cdsads.u-strasbg.fr/abs/2015A%26A...577A..42B} {577, A42}

\bibitem[\protect\citeauthoryear{{Bate}}{{Bate}}{2018}]{Bate2018}
{Bate} M.~R.,  2018, \mn@doi [\mnras] {10.1093/mnras/sty169}, \href
  {https://ui.adsabs.harvard.edu/abs/2018MNRAS.475.5618B} {475, 5618}

\bibitem[\protect\citeauthoryear{{Beckwith}, {Sargent}, {Chini}  \&
  {Guesten}}{{Beckwith} et~al.}{1990}]{Beckwith1990}
{Beckwith} S.~V.~W.,  {Sargent} A.~I.,  {Chini} R.~S.,   {Guesten} R.,  1990,
  \mn@doi [\aj] {10.1086/115385}, \href
  {http://adsabs.harvard.edu/abs/1990AJ.....99..924B} {99, 924}

\bibitem[\protect\citeauthoryear{{Beuzit} et~al.,}{{Beuzit}
  et~al.}{2008}]{Beuzit2008}
{Beuzit} J.-L.,  et~al., 2008, in Society of Photo-Optical Instrumentation
  Engineers (SPIE) Conference Series. , \mn@doi{10.1117/12.790120}

\bibitem[\protect\citeauthoryear{{Bonnefoy} et~al.,}{{Bonnefoy}
  et~al.}{2013}]{Bonnefoy2013}
{Bonnefoy} M.,  et~al., 2013, \mn@doi [\aap] {10.1051/0004-6361/201220838},
  \href {http://cdsads.u-strasbg.fr/abs/2013A%26A...555A.107B} {555, A107}

\bibitem[\protect\citeauthoryear{{Bonnefoy}, {Chauvin}, {Lagrange}, {Rojo},
  {Allard}, {Pinte}, {Dumas}  \& {Homeier}}{{Bonnefoy}
  et~al.}{2014}]{Bonnefoy2014}
{Bonnefoy} M.,  {Chauvin} G.,  {Lagrange} A.-M.,  {Rojo} P.,  {Allard} F.,
  {Pinte} C.,  {Dumas} C.,   {Homeier} D.,  2014, \mn@doi [\aap]
  {10.1051/0004-6361/201118270}, \href
  {http://cdsads.u-strasbg.fr/abs/2014A%26A...562A.127B} {562, A127}

\bibitem[\protect\citeauthoryear{{Boss}}{{Boss}}{1998}]{Boss1998}
{Boss} A.~P.,  1998, \mn@doi [\apj] {10.1086/306036}, \href
  {http://adsabs.harvard.edu/abs/1998ApJ...503..923B} {503, 923}

\bibitem[\protect\citeauthoryear{{Bouvier}, {Dougados}  \& {Alencar}}{{Bouvier}
  et~al.}{2004}]{Bouvier2004}
{Bouvier} J.,  {Dougados} C.,   {Alencar} S.~H.~P.,  2004, \mn@doi [\apss]
  {10.1023/B:ASTR.0000045072.68528.7b}, \href
  {http://adsabs.harvard.edu/abs/2004Ap%26SS.292..659B} {292, 659}

\bibitem[\protect\citeauthoryear{{Bowler}}{{Bowler}}{2016}]{Bowler2016}
{Bowler} B.~P.,  2016, \mn@doi [\pasp] {10.1088/1538-3873/128/968/102001},
  \href {http://cdsads.u-strasbg.fr/abs/2016PASP..128j2001B} {128, 102001}

\bibitem[\protect\citeauthoryear{{Brandt}, {Dupuy}  \& {Bowler}}{{Brandt}
  et~al.}{2019}]{Brandt2019}
{Brandt} T.~D.,  {Dupuy} T.~J.,   {Bowler} B.~P.,  2019, \mn@doi [\aj]
  {10.3847/1538-3881/ab04a8}, \href
  {https://ui.adsabs.harvard.edu/abs/2019AJ....158..140B} {158, 140}

\bibitem[\protect\citeauthoryear{{Brandt}, {Dupuy}, {Bowler}, {Bardalez
  Gagliuffi}, {Faherty}, {Brandt}  \& {Michalik}}{{Brandt}
  et~al.}{2020}]{Brandt2020}
{Brandt} T.~D.,  {Dupuy} T.~J.,  {Bowler} B.~P.,  {Bardalez Gagliuffi} D.~C.,
  {Faherty} J.,  {Brandt} G.~M.,   {Michalik} D.,  2020, \mn@doi [\aj]
  {10.3847/1538-3881/abb45e}, \href
  {https://ui.adsabs.harvard.edu/abs/2020AJ....160..196B} {160, 196}

\bibitem[\protect\citeauthoryear{{Bresnahan} et~al.,}{{Bresnahan}
  et~al.}{2018}]{Bresnahan2018}
{Bresnahan} D.,  et~al., 2018, \mn@doi [\aap] {10.1051/0004-6361/201730515},
  \href {https://ui.adsabs.harvard.edu/abs/2018A&A...615A.125B} {615, A125}

\bibitem[\protect\citeauthoryear{Burnham \& Anderson}{Burnham \&
  Anderson}{2002}]{Burnham2002}
Burnham K.~P.,  Anderson D.~R.,  eds, 2002, Information and Likelihood Theory:
  A Basis for Model Selection and Inference.
Springer New York, New York, NY, pp 49--97,
  \mn@doi{10.1007/978-0-387-22456-5_2}, \url
  {https://doi.org/10.1007/978-0-387-22456-5_2}

\bibitem[\protect\citeauthoryear{{Calcino}, {Christiaens}, {Price}, {Pinte},
  {Davis}, {van der Marel}  \& {Cuello}}{{Calcino} et~al.}{2020}]{Calcino2020}
{Calcino} J.,  {Christiaens} V.,  {Price} D.~J.,  {Pinte} C.,  {Davis} T.~M.,
  {van der Marel} N.,   {Cuello} N.,  2020, \mn@doi [\mnras]
  {10.1093/mnras/staa2468}, \href
  {https://ui.adsabs.harvard.edu/abs/2020MNRAS.498..639C} {498, 639}

\bibitem[\protect\citeauthoryear{{Calvet}, {Muzerolle}, {Brice{\~n}o},
  {Hern{\'a}ndez}, {Hartmann}, {Saucedo}  \& {Gordon}}{{Calvet}
  et~al.}{2004}]{Calvet2004}
{Calvet} N.,  {Muzerolle} J.,  {Brice{\~n}o} C.,  {Hern{\'a}ndez} J.,
  {Hartmann} L.,  {Saucedo} J.~L.,   {Gordon} K.~D.,  2004, \mn@doi [\aj]
  {10.1086/422733}, \href {http://cdsads.u-strasbg.fr/abs/2004AJ....128.1294C}
  {128, 1294}

\bibitem[\protect\citeauthoryear{{Canovas}, {Rodenhuis}, {Jeffers}, {Min}  \&
  {Keller}}{{Canovas} et~al.}{2011}]{Canovas2011}
{Canovas} H.,  {Rodenhuis} M.,  {Jeffers} S.~V.,  {Min} M.,   {Keller} C.~U.,
  2011, \mn@doi [\aap] {10.1051/0004-6361/201116918}, \href
  {https://ui.adsabs.harvard.edu/abs/2011A&A...531A.102C} {531, A102}

\bibitem[\protect\citeauthoryear{{Canovas} et~al.,}{{Canovas}
  et~al.}{2015}]{Canovas2015}
{Canovas} H.,  et~al., 2015, \mn@doi [\apj] {10.1088/0004-637X/805/1/21}, \href
  {https://ui.adsabs.harvard.edu/abs/2015ApJ...805...21C} {805, 21}

\bibitem[\protect\citeauthoryear{{Cardelli}, {Clayton}  \& {Mathis}}{{Cardelli}
  et~al.}{1989}]{Cardelli1989}
{Cardelli} J.~A.,  {Clayton} G.~C.,   {Mathis} J.~S.,  1989, \mn@doi [\apj]
  {10.1086/167900}, \href
  {https://ui.adsabs.harvard.edu/abs/1989ApJ...345..245C} {345, 245}

\bibitem[\protect\citeauthoryear{{Casassus}}{{Casassus}}{2016}]{Casassus2016}
{Casassus} S.,  2016, \mn@doi [PASA] {10.1017/pasa.2016.7}, \href
  {http://cdsads.u-strasbg.fr/abs/2016PASA...33...13C} {33, e013}

\bibitem[\protect\citeauthoryear{{Cazzoletti} et~al.,}{{Cazzoletti}
  et~al.}{2019}]{Cazzoletti2019}
{Cazzoletti} P.,  et~al., 2019, \mn@doi [\aap] {10.1051/0004-6361/201935273},
  \href {https://ui.adsabs.harvard.edu/abs/2019A&A...626A..11C} {626, A11}

\bibitem[\protect\citeauthoryear{{Chauvin} et~al.,}{{Chauvin}
  et~al.}{2012}]{Chauvin2012}
{Chauvin} G.,  et~al., 2012, \mn@doi [\aap] {10.1051/0004-6361/201118346},
  \href {https://ui.adsabs.harvard.edu/abs/2012A&A...542A..41C} {542, A41}

\bibitem[\protect\citeauthoryear{{Chauvin} et~al.,}{{Chauvin}
  et~al.}{2017}]{Chauvin2017}
{Chauvin} G.,  et~al., 2017, \mn@doi [\aap] {10.1051/0004-6361/201731152},
  \href {https://ui.adsabs.harvard.edu/abs/2017A&A...605L...9C} {605, L9}

\bibitem[\protect\citeauthoryear{{Christiaens} et~al.,}{{Christiaens}
  et~al.}{2018}]{Christiaens2018}
{Christiaens} V.,  et~al., 2018, \mn@doi [\aap] {10.1051/0004-6361/201629454},
  \href {http://adsabs.harvard.edu/abs/2018A%26A...617A..37C} {617, A37}

\bibitem[\protect\citeauthoryear{{Christiaens} et~al.,}{{Christiaens}
  et~al.}{2019a}]{Christiaens2019a}
{Christiaens} V.,  et~al., 2019a, \mn@doi [\mnras] {10.1093/mnras/stz1232},
  \href {https://ui.adsabs.harvard.edu/abs/2019MNRAS.486.5819C} {486, 5819}

\bibitem[\protect\citeauthoryear{{Christiaens}, {Cantalloube}, {Casassus},
  {Price}, {Absil}, {Pinte}, {Girard}  \& {Montesinos}}{{Christiaens}
  et~al.}{2019b}]{Christiaens2019b}
{Christiaens} V.,  {Cantalloube} F.,  {Casassus} S.,  {Price} D.~J.,  {Absil}
  O.,  {Pinte} C.,  {Girard} J.,   {Montesinos} M.,  2019b, \mn@doi [\apjl]
  {10.3847/2041-8213/ab212b}, \href
  {https://ui.adsabs.harvard.edu/abs/2019ApJ...877L..33C} {877, L33}

\bibitem[\protect\citeauthoryear{{Cieza} et~al.,}{{Cieza}
  et~al.}{2010}]{Cieza2010}
{Cieza} L.~A.,  et~al., 2010, \mn@doi [\apj] {10.1088/0004-637X/712/2/925},
  \href {https://ui.adsabs.harvard.edu/abs/2010ApJ...712..925C} {712, 925}

\bibitem[\protect\citeauthoryear{{Claudi} et~al.,}{{Claudi}
  et~al.}{2008}]{Claudi2008}
{Claudi} R.~U.,  et~al., 2008, in Ground-based and Airborne Instrumentation for
  Astronomy II. p. 70143E, \mn@doi{10.1117/12.788366}

\bibitem[\protect\citeauthoryear{{Cruz} \& {Reid}}{{Cruz} \&
  {Reid}}{2002}]{Cruz2002}
{Cruz} K.~L.,  {Reid} I.~N.,  2002, \mn@doi [\aj] {10.1086/339973}, \href
  {https://ui.adsabs.harvard.edu/abs/2002AJ....123.2828C} {123, 2828}

\bibitem[\protect\citeauthoryear{{Cuello} et~al.,}{{Cuello}
  et~al.}{2020}]{Cuello2020}
{Cuello} N.,  et~al., 2020, \mn@doi [\mnras] {10.1093/mnras/stz2938}, \href
  {https://ui.adsabs.harvard.edu/abs/2020MNRAS.491..504C} {491, 504}

\bibitem[\protect\citeauthoryear{{Currie} et~al.,}{{Currie}
  et~al.}{2019}]{Currie2019}
{Currie} T.,  et~al., 2019, \mn@doi [\apjl] {10.3847/2041-8213/ab1b42}, \href
  {https://ui.adsabs.harvard.edu/abs/2019ApJ...877L...3C} {877, L3}

\bibitem[\protect\citeauthoryear{{Cutri} et~al.,}{{Cutri}
  et~al.}{2003}]{Cutri2003}
{Cutri} R.~M.,  et~al., 2003, VizieR Online Data Catalog, \href
  {https://ui.adsabs.harvard.edu/abs/2003yCat.2246....0C} {p. II/246}

\bibitem[\protect\citeauthoryear{{Delorme} et~al.,}{{Delorme}
  et~al.}{2017a}]{Delorme2017a}
{Delorme} P.,  et~al., 2017a, in {Reyl{\'e}} C.,  {Di Matteo} P.,  {Herpin} F.,
   {Lagadec} E.,  {Lan{\c{c}}on} A.,  {Meliani} Z.,   {Royer} F.,  eds,
  SF2A-2017: Proceedings of the Annual meeting of the French Society of
  Astronomy and Astrophysics. p.~Di (\mn@eprint {arXiv} {1712.06948})

\bibitem[\protect\citeauthoryear{{Delorme} et~al.,}{{Delorme}
  et~al.}{2017b}]{Delorme2017}
{Delorme} P.,  et~al., 2017b, \mn@doi [\aap] {10.1051/0004-6361/201731145},
  \href {http://cdsads.u-strasbg.fr/abs/2017A%26A...608A..79D} {608, A79}

\bibitem[\protect\citeauthoryear{{Dohlen} et~al.,}{{Dohlen}
  et~al.}{2008}]{Dohlen2008}
{Dohlen} K.,  et~al., 2008, in {McLean} I.~S.,  {Casali} M.~M.,  eds,  Society
  of Photo-Optical Instrumentation Engineers (SPIE) Conference Series Vol.
  7014, Ground-based and Airborne Instrumentation for Astronomy II. p. 70143L,
  \mn@doi{10.1117/12.789786}

\bibitem[\protect\citeauthoryear{{Dong} et~al.,}{{Dong}
  et~al.}{2012}]{Dong2012}
{Dong} R.,  et~al., 2012, \mn@doi [\apj] {10.1088/0004-637X/760/2/111}, \href
  {http://adsabs.harvard.edu/abs/2012ApJ...760..111D} {760, 111}

\bibitem[\protect\citeauthoryear{{Dong}, {Zhu}, {Rafikov}  \& {Stone}}{{Dong}
  et~al.}{2015}]{Dong2015}
{Dong} R.,  {Zhu} Z.,  {Rafikov} R.~R.,   {Stone} J.~M.,  2015, \mn@doi [\apjl]
  {10.1088/2041-8205/809/1/L5}, \href
  {https://ui.adsabs.harvard.edu/abs/2015ApJ...809L...5D} {809, L5}

\bibitem[\protect\citeauthoryear{{Dunham} et~al.,}{{Dunham}
  et~al.}{2015}]{Dunham2015}
{Dunham} M.~M.,  et~al., 2015, \mn@doi [\apjs] {10.1088/0067-0049/220/1/11},
  \href {http://cdsads.u-strasbg.fr/abs/2015ApJS..220...11D} {220, 11}

\bibitem[\protect\citeauthoryear{{Dzib}, {Loinard}, {Ortiz-Le{\'o}n},
  {Rodr{\'\i}guez}  \& {Galli}}{{Dzib} et~al.}{2018}]{Dzib2018}
{Dzib} S.~A.,  {Loinard} L.,  {Ortiz-Le{\'o}n} G.~N.,  {Rodr{\'\i}guez} L.~F.,
   {Galli} P. A.~B.,  2018, \mn@doi [\apj] {10.3847/1538-4357/aae687}, \href
  {https://ui.adsabs.harvard.edu/abs/2018ApJ...867..151D} {867, 151}

\bibitem[\protect\citeauthoryear{{Edwards}, {Kwan}, {Fischer}, {Hillenbrand},
  {Finn}, {Fedorenko}  \& {Feng}}{{Edwards} et~al.}{2013}]{Edwards2013}
{Edwards} S.,  {Kwan} J.,  {Fischer} W.,  {Hillenbrand} L.,  {Finn} K.,
  {Fedorenko} K.,   {Feng} W.,  2013, \mn@doi [\apj]
  {10.1088/0004-637X/778/2/148}, \href
  {https://ui.adsabs.harvard.edu/abs/2013ApJ...778..148E} {778, 148}

\bibitem[\protect\citeauthoryear{{Eriksson}, {Asensio Torres}, {Janson},
  {Aoyama}, {Marleau}, {Bonnefoy}  \& {Petrus}}{{Eriksson}
  et~al.}{2020}]{Eriksson2020}
{Eriksson} S.~C.,  {Asensio Torres} R.,  {Janson} M.,  {Aoyama} Y.,  {Marleau}
  G.-D.,  {Bonnefoy} M.,   {Petrus} S.,  2020, \mn@doi [\aap]
  {10.1051/0004-6361/202038131}, \href
  {https://ui.adsabs.harvard.edu/abs/2020A&A...638L...6E} {638, L6}

\bibitem[\protect\citeauthoryear{{Espaillat} et~al.,}{{Espaillat}
  et~al.}{2014}]{Espaillat2014}
{Espaillat} C.,  et~al., 2014, \mn@doi [Protostars and Planets VI]
  {10.2458/azu_uapress_9780816531240-ch022}, \href
  {http://cdsads.u-strasbg.fr/abs/2014prpl.conf..497E} {pp 497--520}

\bibitem[\protect\citeauthoryear{{Foreman-Mackey}, {Hogg}, {Lang}  \&
  {Goodman}}{{Foreman-Mackey} et~al.}{2013}]{Foreman-Mackey2013}
{Foreman-Mackey} D.,  {Hogg} D.~W.,  {Lang} D.,   {Goodman} J.,  2013, \mn@doi
  [\pasp] {10.1086/670067}, \href
  {https://ui.adsabs.harvard.edu/abs/2013PASP..125..306F} {125, 306}

\bibitem[\protect\citeauthoryear{{Gagn{\'e}}, {Lafreni{\`e}re}, {Doyon}, {Malo}
   \& {Artigau}}{{Gagn{\'e}} et~al.}{2014}]{Gagne2014}
{Gagn{\'e}} J.,  {Lafreni{\`e}re} D.,  {Doyon} R.,  {Malo} L.,   {Artigau}
  {\'E}.,  2014, \mn@doi [\apj] {10.1088/0004-637X/783/2/121}, \href
  {https://ui.adsabs.harvard.edu/abs/2014ApJ...783..121G} {783, 121}

\bibitem[\protect\citeauthoryear{{Gagn{\'e}} et~al.,}{{Gagn{\'e}}
  et~al.}{2015}]{Gagne2015}
{Gagn{\'e}} J.,  et~al., 2015, \mn@doi [\apjs] {10.1088/0067-0049/219/2/33},
  \href {https://ui.adsabs.harvard.edu/abs/2015ApJS..219...33G} {219, 33}

\bibitem[\protect\citeauthoryear{{Gaia Collaboration} et~al.,}{{Gaia
  Collaboration} et~al.}{2016}]{Gaia-Collaboration2016}
{Gaia Collaboration} et~al., 2016, \mn@doi [\aap]
  {10.1051/0004-6361/201629272}, \href
  {https://ui.adsabs.harvard.edu/abs/2016A&A...595A...1G} {595, A1}

\bibitem[\protect\citeauthoryear{{Gaia Collaboration} et~al.,}{{Gaia
  Collaboration} et~al.}{2018}]{Gaia-Collaboration2018}
{Gaia Collaboration} et~al., 2018, \mn@doi [\aap]
  {10.1051/0004-6361/201833051}, \href
  {https://ui.adsabs.harvard.edu/abs/2018A%26A...616A...1G} {616, A1}

\bibitem[\protect\citeauthoryear{{Galicher} et~al.,}{{Galicher}
  et~al.}{2018}]{Galicher2018}
{Galicher} R.,  et~al., 2018, \mn@doi [\aap] {10.1051/0004-6361/201832973},
  \href {https://ui.adsabs.harvard.edu/abs/2018A&A...615A..92G} {615, A92}

\bibitem[\protect\citeauthoryear{{Gelman} \& {Rubin}}{{Gelman} \&
  {Rubin}}{1992}]{Gelman1992}
{Gelman} A.,  {Rubin} D.~B.,  1992, \mn@doi [Statistical Science]
  {10.1214/ss/1177011136}, \href
  {https://ui.adsabs.harvard.edu/abs/1992StaSc...7..457G} {7, 457}

\bibitem[\protect\citeauthoryear{{Ginski} et~al.,}{{Ginski}
  et~al.}{2018}]{Ginski2018}
{Ginski} C.,  et~al., 2018, \mn@doi [\aap] {10.1051/0004-6361/201732417}, \href
  {https://ui.adsabs.harvard.edu/abs/2018A&A...616A..79G} {616, A79}

\bibitem[\protect\citeauthoryear{{Girardi} et~al.,}{{Girardi}
  et~al.}{2012}]{Girardi2012}
{Girardi} L.,  et~al., 2012, \mn@doi [Astrophysics and Space Science
  Proceedings] {10.1007/978-3-642-18418-5_17}, \href
  {http://adsabs.harvard.edu/abs/2012ASSP...26..165G} {26, 165}

\bibitem[\protect\citeauthoryear{{Gomez Gonzalez} et~al.,}{{Gomez Gonzalez}
  et~al.}{2017}]{GomezGonzalez2017}
{Gomez Gonzalez} C.~A.,  et~al., 2017, \mn@doi [\aj]
  {10.3847/1538-3881/aa73d7}, \href
  {http://cdsads.u-strasbg.fr/abs/2017AJ....154....7G} {154, 7}

\bibitem[\protect\citeauthoryear{{Goodman} \& {Weare}}{{Goodman} \&
  {Weare}}{2010}]{Goodman2010}
{Goodman} J.,  {Weare} J.,  2010, \mn@doi [Communications in Applied
  Mathematics and Computational Science] {10.2140/camcos.2010.5.65}, \href
  {https://ui.adsabs.harvard.edu/abs/2010CAMCS...5...65G} {5, 65}

\bibitem[\protect\citeauthoryear{{Gravity Collaboration} et~al.,}{{Gravity
  Collaboration} et~al.}{2019}]{Gravity-Collaboration2019}
{Gravity Collaboration} et~al., 2019, \mn@doi [\aap]
  {10.1051/0004-6361/201935253}, \href
  {https://ui.adsabs.harvard.edu/abs/2019A&A...623L..11G} {623, L11}

\bibitem[\protect\citeauthoryear{{Greco} \& {Brandt}}{{Greco} \&
  {Brandt}}{2016}]{Greco2016}
{Greco} J.~P.,  {Brandt} T.~D.,  2016, \mn@doi [\apj]
  {10.3847/1538-4357/833/2/134}, \href
  {http://cdsads.u-strasbg.fr/abs/2016ApJ...833..134G} {833, 134}

\bibitem[\protect\citeauthoryear{{Gressel}, {Nelson}, {Turner}  \&
  {Ziegler}}{{Gressel} et~al.}{2013}]{Gressel2013}
{Gressel} O.,  {Nelson} R.~P.,  {Turner} N.~J.,   {Ziegler} U.,  2013, \mn@doi
  [\apj] {10.1088/0004-637X/779/1/59}, \href
  {http://adsabs.harvard.edu/abs/2013ApJ...779...59G} {779, 59}

\bibitem[\protect\citeauthoryear{{Guidi} et~al.,}{{Guidi}
  et~al.}{2018}]{Guidi2018}
{Guidi} G.,  et~al., 2018, \mn@doi [\mnras] {10.1093/mnras/sty1642}, \href
  {https://ui.adsabs.harvard.edu/abs/2018MNRAS.479.1505G} {479, 1505}

\bibitem[\protect\citeauthoryear{{Gullbring}, {Hartmann}, {Brice{\~n}o}  \&
  {Calvet}}{{Gullbring} et~al.}{1998}]{Gullbring1998}
{Gullbring} E.,  {Hartmann} L.,  {Brice{\~n}o} C.,   {Calvet} N.,  1998,
  \mn@doi [\apj] {10.1086/305032}, \href
  {http://cdsads.u-strasbg.fr/abs/1998ApJ...492..323G} {492, 323}

\bibitem[\protect\citeauthoryear{{Gutermuth}, {Megeath}, {Myers}, {Allen},
  {Pipher}  \& {Fazio}}{{Gutermuth} et~al.}{2009}]{Gutermuth2009}
{Gutermuth} R.~A.,  {Megeath} S.~T.,  {Myers} P.~C.,  {Allen} L.~E.,  {Pipher}
  J.~L.,   {Fazio} G.~G.,  2009, \mn@doi [\apjs] {10.1088/0067-0049/184/1/18},
  \href {https://ui.adsabs.harvard.edu/abs/2009ApJS..184...18G} {184, 18}

\bibitem[\protect\citeauthoryear{{Haffert}, {Bohn}, {de Boer}, {Snellen},
  {Brinchmann}, {Girard}, {Keller}  \& {Bacon}}{{Haffert}
  et~al.}{2019}]{Haffert2019}
{Haffert} S.~Y.,  {Bohn} A.~J.,  {de Boer} J.,  {Snellen} I.~A.~G.,
  {Brinchmann} J.,  {Girard} J.~H.,  {Keller} C.~U.,   {Bacon} R.,  2019,
  \mn@doi [Nature Astronomy] {10.1038/s41550-019-0780-5}, \href
  {https://ui.adsabs.harvard.edu/abs/2019NatAs...3..749H} {3, 749}

\bibitem[\protect\citeauthoryear{{Haffert} et~al.,}{{Haffert}
  et~al.}{2020}]{Haffert2020}
{Haffert} S.~Y.,  et~al., 2020, \mn@doi [\aap] {10.1051/0004-6361/202038706},
  \href {https://ui.adsabs.harvard.edu/abs/2020A&A...640L..12H} {640, L12}

\bibitem[\protect\citeauthoryear{{Hartmann}, {Herczeg}  \& {Calvet}}{{Hartmann}
  et~al.}{2016}]{Hartmann2016}
{Hartmann} L.,  {Herczeg} G.,   {Calvet} N.,  2016, \mn@doi [\araa]
  {10.1146/annurev-astro-081915-023347}, \href
  {https://ui.adsabs.harvard.edu/abs/2016ARA&A..54..135H} {54, 135}

\bibitem[\protect\citeauthoryear{{Hashimoto} et~al.,}{{Hashimoto}
  et~al.}{2011}]{Hashimoto2011}
{Hashimoto} J.,  et~al., 2011, \mn@doi [\apjl] {10.1088/2041-8205/729/2/L17},
  \href {https://ui.adsabs.harvard.edu/abs/2011ApJ...729L..17H} {729, L17}

\bibitem[\protect\citeauthoryear{{Hashimoto}, {Aoyama}, {Konishi}, {Uyama},
  {Takasao}, {Ikoma}  \& {Tanigawa}}{{Hashimoto} et~al.}{2020}]{Hashimoto2020}
{Hashimoto} J.,  {Aoyama} Y.,  {Konishi} M.,  {Uyama} T.,  {Takasao} S.,
  {Ikoma} M.,   {Tanigawa} T.,  2020, arXiv e-prints, \href
  {https://ui.adsabs.harvard.edu/abs/2020arXiv200307922H} {p. arXiv:2003.07922}

\bibitem[\protect\citeauthoryear{{Hauschildt}}{{Hauschildt}}{1992}]{Hauschildt1992}
{Hauschildt} P.~H.,  1992, \mn@doi [\jqsrt] {10.1016/0022-4073(92)90105-D},
  \href {https://ui.adsabs.harvard.edu/abs/1992JQSRT..47..433H} {47, 433}

\bibitem[\protect\citeauthoryear{{Helling} \& {Woitke}}{{Helling} \&
  {Woitke}}{2006}]{Helling2006}
{Helling} C.,  {Woitke} P.,  2006, \mn@doi [\aap] {10.1051/0004-6361:20054598},
  \href {https://ui.adsabs.harvard.edu/abs/2006A&A...455..325H} {455, 325}

\bibitem[\protect\citeauthoryear{{Helling}, {Dehn}, {Woitke}  \&
  {Hauschildt}}{{Helling} et~al.}{2008}]{Helling2008}
{Helling} C.,  {Dehn} M.,  {Woitke} P.,   {Hauschildt} P.~H.,  2008, \mn@doi
  [\apjl] {10.1086/533462}, \href
  {https://ui.adsabs.harvard.edu/abs/2008ApJ...675L.105H} {675, L105}

\bibitem[\protect\citeauthoryear{{Herczeg} \& {Hillenbrand}}{{Herczeg} \&
  {Hillenbrand}}{2014}]{Herczeg2014}
{Herczeg} G.~J.,  {Hillenbrand} L.~A.,  2014, \mn@doi [\apj]
  {10.1088/0004-637X/786/2/97}, \href
  {https://ui.adsabs.harvard.edu/abs/2014ApJ...786...97H} {786, 97}

\bibitem[\protect\citeauthoryear{{Isella}, {Benisty}, {Teague}, {Bae},
  {Keppler}, {Facchini}  \& {P{\'e}rez}}{{Isella} et~al.}{2019}]{Isella2019}
{Isella} A.,  {Benisty} M.,  {Teague} R.,  {Bae} J.,  {Keppler} M.,  {Facchini}
  S.,   {P{\'e}rez} L.,  2019, \mn@doi [\apjl] {10.3847/2041-8213/ab2a12},
  \href {https://ui.adsabs.harvard.edu/abs/2019ApJ...879L..25I} {879, L25}

\bibitem[\protect\citeauthoryear{{Johns-Krull}}{{Johns-Krull}}{2007}]{Johns-Krull2007}
{Johns-Krull} C.~M.,  2007, \mn@doi [\apj] {10.1086/519017}, \href
  {http://cdsads.u-strasbg.fr/abs/2007ApJ...664..975J} {664, 975}

\bibitem[\protect\citeauthoryear{{Johns-Krull}, {Valenti}  \&
  {Koresko}}{{Johns-Krull} et~al.}{1999}]{Johns-Krull1999}
{Johns-Krull} C.~M.,  {Valenti} J.~A.,   {Koresko} C.,  1999, \mn@doi [\apj]
  {10.1086/307128}, \href {http://cdsads.u-strasbg.fr/abs/1999ApJ...516..900J}
  {516, 900}

\bibitem[\protect\citeauthoryear{{Keppler} et~al.,}{{Keppler}
  et~al.}{2018}]{Keppler2018}
{Keppler} M.,  et~al., 2018, \mn@doi [\aap] {10.1051/0004-6361/201832957},
  \href {http://adsabs.harvard.edu/abs/2018A%26A...617A..44K} {617, A44}

\bibitem[\protect\citeauthoryear{{Keppler} et~al.,}{{Keppler}
  et~al.}{2020}]{Keppler2020}
{Keppler} M.,  et~al., 2020, \mn@doi [\aap] {10.1051/0004-6361/202038032},
  \href {https://ui.adsabs.harvard.edu/abs/2020A&A...639A..62K} {639, A62}

\bibitem[\protect\citeauthoryear{{Kratter} \& {Lodato}}{{Kratter} \&
  {Lodato}}{2016}]{Kratter2016}
{Kratter} K.,  {Lodato} G.,  2016, \mn@doi [\araa]
  {10.1146/annurev-astro-081915-023307}, \href
  {https://ui.adsabs.harvard.edu/abs/2016ARA&A..54..271K} {54, 271}

\bibitem[\protect\citeauthoryear{{Kraus}, {Ireland}, {Cieza}, {Hinkley},
  {Dupuy}, {Bowler}  \& {Liu}}{{Kraus} et~al.}{2014}]{Kraus2014}
{Kraus} A.~L.,  {Ireland} M.~J.,  {Cieza} L.~A.,  {Hinkley} S.,  {Dupuy} T.~J.,
   {Bowler} B.~P.,   {Liu} M.~C.,  2014, \mn@doi [\apj]
  {10.1088/0004-637X/781/1/20}, \href
  {http://cdsads.u-strasbg.fr/abs/2014ApJ...781...20K} {781, 20}

\bibitem[\protect\citeauthoryear{{Lagrange} et~al.,}{{Lagrange}
  et~al.}{2009}]{Lagrange2009}
{Lagrange} A.-M.,  et~al., 2009, \mn@doi [\aap] {10.1051/0004-6361:200811325},
  \href {http://cdsads.u-strasbg.fr/abs/2009A%26A...493L..21L} {493, L21}

\bibitem[\protect\citeauthoryear{{Lagrange} et~al.,}{{Lagrange}
  et~al.}{2010}]{Lagrange2010}
{Lagrange} A.-M.,  et~al., 2010, \mn@doi [Science] {10.1126/science.1187187},
  \href {http://cdsads.u-strasbg.fr/abs/2010Sci...329...57L} {329, 57}

\bibitem[\protect\citeauthoryear{{Launhardt} et~al.,}{{Launhardt}
  et~al.}{2020}]{Launhardt2020}
{Launhardt} R.,  et~al., 2020, \mn@doi [\aap] {10.1051/0004-6361/201937000},
  \href {https://ui.adsabs.harvard.edu/abs/2020A&A...635A.162L} {635, A162}

\bibitem[\protect\citeauthoryear{{Lenzen} et~al.,}{{Lenzen}
  et~al.}{2003}]{Lenzen2003}
{Lenzen} R.,  et~al., 2003, in {Iye} M.,  {Moorwood} A.~F.~M.,  eds,  Society
  of Photo-Optical Instrumentation Engineers (SPIE) Conference Series Vol.
  4841, Society of Photo-Optical Instrumentation Engineers (SPIE) Conference
  Series. pp 944--952, \mn@doi{10.1117/12.460044}

\bibitem[\protect\citeauthoryear{{Looper}, {Bochanski}, {Burgasser}, {Mohanty},
  {Mamajek}, {Faherty}, {West}  \& {Pitts}}{{Looper} et~al.}{2010}]{Looper2010}
{Looper} D.~L.,  {Bochanski} J.~J.,  {Burgasser} A.~J.,  {Mohanty} S.,
  {Mamajek} E.~E.,  {Faherty} J.~K.,  {West} A.~A.,   {Pitts} M.~A.,  2010,
  \mn@doi [\aj] {10.1088/0004-6256/140/5/1486}, \href
  {https://ui.adsabs.harvard.edu/abs/2010AJ....140.1486L} {140, 1486}

\bibitem[\protect\citeauthoryear{{Lovelace}, {Covey}  \& {Lloyd}}{{Lovelace}
  et~al.}{2011}]{Lovelace2011}
{Lovelace} R.~V.~E.,  {Covey} K.~R.,   {Lloyd} J.~P.,  2011, \mn@doi [\aj]
  {10.1088/0004-6256/141/2/51}, \href
  {http://cdsads.u-strasbg.fr/abs/2011AJ....141...51L} {141, 51}

\bibitem[\protect\citeauthoryear{{Luhman} et~al.,}{{Luhman}
  et~al.}{2008}]{Luhman2008}
{Luhman} K.~L.,  et~al., 2008, \mn@doi [\apj] {10.1086/527347}, \href
  {https://ui.adsabs.harvard.edu/abs/2008ApJ...675.1375L} {675, 1375}

\bibitem[\protect\citeauthoryear{{Maire} et~al.,}{{Maire}
  et~al.}{2016}]{Maire2016}
{Maire} A.-L.,  et~al., 2016, in Ground-based and Airborne Instrumentation for
  Astronomy VI. p. 990834 (\mn@eprint {arXiv} {1609.06681}),
  \mn@doi{10.1117/12.2233013}

\bibitem[\protect\citeauthoryear{{Marleau} \& {Cumming}}{{Marleau} \&
  {Cumming}}{2014}]{Marleau2014}
{Marleau} G.~D.,  {Cumming} A.,  2014, \mn@doi [\mnras]
  {10.1093/mnras/stt1967}, \href
  {https://ui.adsabs.harvard.edu/abs/2014MNRAS.437.1378M} {437, 1378}

\bibitem[\protect\citeauthoryear{{Marleau}, {Klahr}, {Kuiper}  \&
  {Mordasini}}{{Marleau} et~al.}{2017}]{Marleau2017}
{Marleau} G.-D.,  {Klahr} H.,  {Kuiper} R.,   {Mordasini} C.,  2017, \mn@doi
  [\apj] {10.3847/1538-4357/836/2/221}, \href
  {https://ui.adsabs.harvard.edu/abs/2017ApJ...836..221M} {836, 221}

\bibitem[\protect\citeauthoryear{{Marleau}, {Coleman}, {Leleu}  \&
  {Mordasini}}{{Marleau} et~al.}{2019a}]{Marleau2019a}
{Marleau} G.-D.,  {Coleman} G. A.~L.,  {Leleu} A.,   {Mordasini} C.,  2019a,
  \mn@doi [\aap] {10.1051/0004-6361/201833597}, \href
  {https://ui.adsabs.harvard.edu/abs/2019A&A...624A..20M} {624, A20}

\bibitem[\protect\citeauthoryear{{Marleau}, {Mordasini}  \& {Kuiper}}{{Marleau}
  et~al.}{2019b}]{Marleau2019}
{Marleau} G.-D.,  {Mordasini} C.,   {Kuiper} R.,  2019b, \mn@doi [\apj]
  {10.3847/1538-4357/ab245b}, \href
  {https://ui.adsabs.harvard.edu/abs/2019ApJ...881..144M} {881, 144}

\bibitem[\protect\citeauthoryear{{Marois}, {Lafreni{\`e}re}, {Doyon},
  {Macintosh}  \& {Nadeau}}{{Marois} et~al.}{2006}]{Marois2006}
{Marois} C.,  {Lafreni{\`e}re} D.,  {Doyon} R.,  {Macintosh} B.,   {Nadeau} D.,
   2006, \mn@doi [\apj] {10.1086/500401}, \href
  {http://adsabs.harvard.edu/abs/2006ApJ...641..556M} {641, 556}

\bibitem[\protect\citeauthoryear{{Marois}, {Macintosh}, {Barman}, {Zuckerman},
  {Song}, {Patience}, {Lafreni{\`e}re}  \& {Doyon}}{{Marois}
  et~al.}{2008}]{Marois2008}
{Marois} C.,  {Macintosh} B.,  {Barman} T.,  {Zuckerman} B.,  {Song} I.,
  {Patience} J.,  {Lafreni{\`e}re} D.,   {Doyon} R.,  2008, \mn@doi [Science]
  {10.1126/science.1166585}, \href
  {http://cdsads.u-strasbg.fr/abs/2008Sci...322.1348M} {322, 1348}

\bibitem[\protect\citeauthoryear{{Marois}, {Macintosh}  \&
  {V{\'e}ran}}{{Marois} et~al.}{2010}]{Marois2010b}
{Marois} C.,  {Macintosh} B.,   {V{\'e}ran} J.-P.,  2010, in Adaptive Optics
  Systems II. p. 77361J, \mn@doi{10.1117/12.857225}

\bibitem[\protect\citeauthoryear{{Mawet} et~al.,}{{Mawet}
  et~al.}{2014}]{Mawet2014}
{Mawet} D.,  et~al., 2014, \mn@doi [\apj] {10.1088/0004-637X/792/2/97}, \href
  {http://cdsads.u-strasbg.fr/abs/2014ApJ...792...97M} {792, 97}

\bibitem[\protect\citeauthoryear{{Mawet} et~al.,}{{Mawet}
  et~al.}{2016}]{Mawet2016}
{Mawet} D.,  et~al., 2016, in Adaptive Optics Systems V. p. 99090D,
  \mn@doi{10.1117/12.2233658}

\bibitem[\protect\citeauthoryear{{McCabe}, {Duch{\^e}ne}, {Pinte},
  {Stapelfeldt}, {Ghez}  \& {M{\'e}nard}}{{McCabe} et~al.}{2011}]{McCabe2011}
{McCabe} C.,  {Duch{\^e}ne} G.,  {Pinte} C.,  {Stapelfeldt} K.~R.,  {Ghez}
  A.~M.,   {M{\'e}nard} F.,  2011, \mn@doi [\apj] {10.1088/0004-637X/727/2/90},
  \href {https://ui.adsabs.harvard.edu/abs/2011ApJ...727...90M} {727, 90}

\bibitem[\protect\citeauthoryear{{Meyer} \& {Wilking}}{{Meyer} \&
  {Wilking}}{2009}]{Meyer2009}
{Meyer} M.~R.,  {Wilking} B.~A.,  2009, \mn@doi [\pasp] {10.1086/598804}, \href
  {https://ui.adsabs.harvard.edu/abs/2009PASP..121..350M} {121, 350}

\bibitem[\protect\citeauthoryear{{Milli}, {Mouillet}, {Lagrange}, {Boccaletti},
  {Mawet}, {Chauvin}  \& {Bonnefoy}}{{Milli} et~al.}{2012}]{Milli2012}
{Milli} J.,  {Mouillet} D.,  {Lagrange} A.-M.,  {Boccaletti} A.,  {Mawet} D.,
  {Chauvin} G.,   {Bonnefoy} M.,  2012, \mn@doi [\aap]
  {10.1051/0004-6361/201219687}, \href
  {http://cdsads.u-strasbg.fr/abs/2012A%26A...545A.111M} {545, A111}

\bibitem[\protect\citeauthoryear{{Milli} et~al.,}{{Milli}
  et~al.}{2017}]{Milli2017}
{Milli} J.,  et~al., 2017, \mn@doi [\aap] {10.1051/0004-6361/201629908}, \href
  {http://cdsads.u-strasbg.fr/abs/2017A%26A...597L...2M} {597, L2}

\bibitem[\protect\citeauthoryear{{Mizuno}}{{Mizuno}}{1980}]{Mizuno1980}
{Mizuno} H.,  1980, \mn@doi [Progress of Theoretical Physics]
  {10.1143/PTP.64.544}, \href
  {http://cdsads.u-strasbg.fr/abs/1980PThPh..64..544M} {64, 544}

\bibitem[\protect\citeauthoryear{{Mordasini}}{{Mordasini}}{2018}]{Mordasini2018}
{Mordasini} C.,  2018, preprint, \href
  {http://cdsads.u-strasbg.fr/abs/2018arXiv180401532M} {} (\mn@eprint {arXiv}
  {1804.01532})

\bibitem[\protect\citeauthoryear{{Mordasini}, {Alibert}, {Klahr}  \&
  {Henning}}{{Mordasini} et~al.}{2012}]{Mordasini2012a}
{Mordasini} C.,  {Alibert} Y.,  {Klahr} H.,   {Henning} T.,  2012, \mn@doi
  [\aap] {10.1051/0004-6361/201118457}, \href
  {http://cdsads.u-strasbg.fr/abs/2012A%26A...547A.111M} {547, A111}

\bibitem[\protect\citeauthoryear{{Mordasini}, {Marleau}  \&
  {Molli{\`e}re}}{{Mordasini} et~al.}{2017}]{Mordasini2017}
{Mordasini} C.,  {Marleau} G.-D.,   {Molli{\`e}re} P.,  2017, \mn@doi [\aap]
  {10.1051/0004-6361/201630077}, \href
  {http://cdsads.u-strasbg.fr/abs/2017A%26A...608A..72M} {608, A72}

\bibitem[\protect\citeauthoryear{{M{\"u}ller} et~al.,}{{M{\"u}ller}
  et~al.}{2018}]{Muller2018}
{M{\"u}ller} A.,  et~al., 2018, \mn@doi [\aap] {10.1051/0004-6361/201833584},
  \href {http://adsabs.harvard.edu/abs/2018A%26A...617L...2M} {617, L2}

\bibitem[\protect\citeauthoryear{{Muzerolle}, {Hartmann}  \&
  {Calvet}}{{Muzerolle} et~al.}{1998}]{Muzerolle1998}
{Muzerolle} J.,  {Hartmann} L.,   {Calvet} N.,  1998, \mn@doi [\aj]
  {10.1086/300636}, \href
  {https://ui.adsabs.harvard.edu/abs/1998AJ....116.2965M} {116, 2965}

\bibitem[\protect\citeauthoryear{{Natta}, {Testi}, {Muzerolle}, {Randich},
  {Comer{\'o}n}  \& {Persi}}{{Natta} et~al.}{2004}]{Natta2004}
{Natta} A.,  {Testi} L.,  {Muzerolle} J.,  {Randich} S.,  {Comer{\'o}n} F.,
  {Persi} P.,  2004, \mn@doi [\aap] {10.1051/0004-6361:20040356}, \href
  {https://ui.adsabs.harvard.edu/abs/2004A&A...424..603N} {424, 603}

\bibitem[\protect\citeauthoryear{{Naud} et~al.,}{{Naud}
  et~al.}{2014}]{Naud2014}
{Naud} M.-E.,  et~al., 2014, \mn@doi [\apj] {10.1088/0004-637X/787/1/5}, \href
  {https://ui.adsabs.harvard.edu/abs/2014ApJ...787....5N} {787, 5}

\bibitem[\protect\citeauthoryear{{Nelder \& Mead}}{{Nelder \&
  Mead}}{1965}]{NelderMead1965}
{Nelder \& Mead} 1965, Computer Journal, Vol 7, 308

\bibitem[\protect\citeauthoryear{{Neuh{\"a}user} et~al.,}{{Neuh{\"a}user}
  et~al.}{2000}]{Neuhauser2000}
{Neuh{\"a}user} R.,  et~al., 2000, \mn@doi [\aaps] {10.1051/aas:2000272}, \href
  {https://ui.adsabs.harvard.edu/abs/2000A&AS..146..323N} {146, 323}

\bibitem[\protect\citeauthoryear{{Nisini}, {Antoniucci}, {Giannini}  \&
  {Lorenzetti}}{{Nisini} et~al.}{2005}]{Nisini2005}
{Nisini} B.,  {Antoniucci} S.,  {Giannini} T.,   {Lorenzetti} D.,  2005,
  \mn@doi [\aap] {10.1051/0004-6361:20041409}, \href
  {https://ui.adsabs.harvard.edu/abs/2005A&A...429..543N} {429, 543}

\bibitem[\protect\citeauthoryear{{Nowak} et~al.,}{{Nowak}
  et~al.}{2020}]{Nowak2020}
{Nowak} M.,  et~al., 2020, \mn@doi [\aap] {10.1051/0004-6361/202039039}, \href
  {https://ui.adsabs.harvard.edu/abs/2020A&A...642L...2N} {642, L2}

\bibitem[\protect\citeauthoryear{{Nutter}, {Ward-Thompson}  \&
  {Andr{\'e}}}{{Nutter} et~al.}{2005}]{Nutter2005}
{Nutter} D.~J.,  {Ward-Thompson} D.,   {Andr{\'e}} P.,  2005, \mn@doi [\mnras]
  {10.1111/j.1365-2966.2005.08711.x}, \href
  {https://ui.adsabs.harvard.edu/abs/2005MNRAS.357..975N} {357, 975}

\bibitem[\protect\citeauthoryear{{Olofsson} et~al.,}{{Olofsson}
  et~al.}{2013}]{Olofsson2013}
{Olofsson} J.,  et~al., 2013, \mn@doi [\aap] {10.1051/0004-6361/201220675},
  \href {http://cdsads.u-strasbg.fr/abs/2013A%26A...552A...4O} {552, A4}

\bibitem[\protect\citeauthoryear{{Owen}}{{Owen}}{2016}]{Owen2016}
{Owen} J.~E.,  2016, \mn@doi [\pasa] {10.1017/pasa.2016.2}, \href
  {http://cdsads.u-strasbg.fr/abs/2016PASA...33....5O} {33, e005}

\bibitem[\protect\citeauthoryear{{Paardekooper} \& {Johansen}}{{Paardekooper}
  \& {Johansen}}{2018}]{Paardekooper2018}
{Paardekooper} S.-J.,  {Johansen} A.,  2018, \mn@doi [\ssr]
  {10.1007/s11214-018-0472-y}, \href
  {https://ui.adsabs.harvard.edu/abs/2018SSRv..214...38P} {214, 38}

\bibitem[\protect\citeauthoryear{{Pairet}, {Cantalloube}, {Gomez Gonzalez},
  {Absil}  \& {Jacques}}{{Pairet} et~al.}{2019}]{Pairet2019a}
{Pairet} B.,  {Cantalloube} F.,  {Gomez Gonzalez} C.~A.,  {Absil} O.,
  {Jacques} L.,  2019, \mn@doi [\mnras] {10.1093/mnras/stz1350}, \href
  {https://ui.adsabs.harvard.edu/abs/2019MNRAS.487.2262P} {487, 2262}

\bibitem[\protect\citeauthoryear{{Pairet}, {Cantalloube}  \&
  {Jacques}}{{Pairet} et~al.}{2020}]{Pairet2020}
{Pairet} B.,  {Cantalloube} F.,   {Jacques} L.,  2020, arXiv e-prints, \href
  {https://ui.adsabs.harvard.edu/abs/2020arXiv200805170P} {p. arXiv:2008.05170}

\bibitem[\protect\citeauthoryear{{Peterson} et~al.,}{{Peterson}
  et~al.}{2011}]{Peterson2011}
{Peterson} D.~E.,  et~al., 2011, \mn@doi [\apjs] {10.1088/0067-0049/194/2/43},
  \href {https://ui.adsabs.harvard.edu/abs/2011ApJS..194...43P} {194, 43}

\bibitem[\protect\citeauthoryear{{Pollack}, {Hubickyj}, {Bodenheimer},
  {Lissauer}, {Podolak}  \& {Greenzweig}}{{Pollack} et~al.}{1996}]{Pollack1996}
{Pollack} J.~B.,  {Hubickyj} O.,  {Bodenheimer} P.,  {Lissauer} J.~J.,
  {Podolak} M.,   {Greenzweig} Y.,  1996, \mn@doi [\icarus]
  {10.1006/icar.1996.0190}, \href
  {http://adsabs.harvard.edu/abs/1996Icar..124...62P} {124, 62}

\bibitem[\protect\citeauthoryear{{Price} et~al.,}{{Price}
  et~al.}{2018}]{Price2018}
{Price} D.~J.,  et~al., 2018, \mn@doi [\mnras] {10.1093/mnras/sty647}, \href
  {http://cdsads.u-strasbg.fr/abs/2018MNRAS.477.1270P} {477, 1270}

\bibitem[\protect\citeauthoryear{{Quanz}, {Amara}, {Meyer}, {Kenworthy},
  {Kasper}  \& {Girard}}{{Quanz} et~al.}{2013}]{Quanz2013}
{Quanz} S.~P.,  {Amara} A.,  {Meyer} M.~R.,  {Kenworthy} M.~A.,  {Kasper} M.,
  {Girard} J.~H.,  2013, \mn@doi [\apjl] {10.1088/2041-8205/766/1/L1}, \href
  {http://cdsads.u-strasbg.fr/abs/2013ApJ...766L...1Q} {766, L1}

\bibitem[\protect\citeauthoryear{{Rich} et~al.,}{{Rich}
  et~al.}{2019}]{Rich2019}
{Rich} E.~A.,  et~al., 2019, \mn@doi [\apj] {10.3847/1538-4357/ab0f3b}, \href
  {https://ui.adsabs.harvard.edu/abs/2019ApJ...875...38R} {875, 38}

\bibitem[\protect\citeauthoryear{{Romero}, {Schreiber}, {Cieza},
  {Rebassa-Mansergas}, {Mer{\'{\i}}n}, {Smith Castelli}, {Allen}  \&
  {Morrell}}{{Romero} et~al.}{2012}]{Romero2012}
{Romero} G.~A.,  {Schreiber} M.~R.,  {Cieza} L.~A.,  {Rebassa-Mansergas} A.,
  {Mer{\'{\i}}n} B.,  {Smith Castelli} A.~V.,  {Allen} L.~E.,   {Morrell} N.,
  2012, \mn@doi [\apj] {10.1088/0004-637X/749/1/79}, \href
  {http://cdsads.u-strasbg.fr/abs/2012ApJ...749...79R} {749, 79}

\bibitem[\protect\citeauthoryear{{Rousset} et~al.,}{{Rousset}
  et~al.}{2003}]{Rousset2003}
{Rousset} G.,  et~al., 2003, in {Wizinowich} P.~L.,  {Bonaccini} D.,  eds,
  Society of Photo-Optical Instrumentation Engineers (SPIE) Conference Series
  Vol. 4839, Society of Photo-Optical Instrumentation Engineers (SPIE)
  Conference Series. pp 140--149, \mn@doi{10.1117/12.459332}

\bibitem[\protect\citeauthoryear{{Sallum} et~al.,}{{Sallum}
  et~al.}{2015}]{Sallum2015}
{Sallum} S.,  et~al., 2015, \mn@doi [\nat] {10.1038/nature15761}, \href
  {http://adsabs.harvard.edu/abs/2015Natur.527..342S} {527, 342}

\bibitem[\protect\citeauthoryear{{Schmid}, {Joos}  \& {Tschan}}{{Schmid}
  et~al.}{2006}]{Schmid2006}
{Schmid} H.~M.,  {Joos} F.,   {Tschan} D.,  2006, \mn@doi [\aap]
  {10.1051/0004-6361:20053273}, \href
  {https://ui.adsabs.harvard.edu/abs/2006A&A...452..657S} {452, 657}

\bibitem[\protect\citeauthoryear{{Sicilia-Aguilar}, {Henning}, {Juh{\'a}sz},
  {Bouwman}, {Garmire}  \& {Garmire}}{{Sicilia-Aguilar}
  et~al.}{2008}]{Sicilia-Aguilar2008}
{Sicilia-Aguilar} A.,  {Henning} T.,  {Juh{\'a}sz} A.,  {Bouwman} J.,
  {Garmire} G.,   {Garmire} A.,  2008, \mn@doi [\apj] {10.1086/591932}, \href
  {https://ui.adsabs.harvard.edu/abs/2008ApJ...687.1145S} {687, 1145}

\bibitem[\protect\citeauthoryear{{Sicilia-Aguilar}, {Henning}, {Kainulainen}
  \& {Roccatagliata}}{{Sicilia-Aguilar} et~al.}{2011}]{Sicilia-Aguilar2011}
{Sicilia-Aguilar} A.,  {Henning} T.,  {Kainulainen} J.,   {Roccatagliata} V.,
  2011, \mn@doi [\apj] {10.1088/0004-637X/736/2/137}, \href
  {https://ui.adsabs.harvard.edu/abs/2011ApJ...736..137S} {736, 137}

\bibitem[\protect\citeauthoryear{{Sicilia-Aguilar}, {Henning}, {Linz},
  {Andr{\'e}}, {Stutz}, {Eiroa}  \& {White}}{{Sicilia-Aguilar}
  et~al.}{2013}]{Sicilia-Aguilar2013}
{Sicilia-Aguilar} A.,  {Henning} T.,  {Linz} H.,  {Andr{\'e}} P.,  {Stutz} A.,
  {Eiroa} C.,   {White} G.~J.,  2013, \mn@doi [\aap]
  {10.1051/0004-6361/201220170}, \href
  {https://ui.adsabs.harvard.edu/abs/2013A&A...551A..34S} {551, A34}

\bibitem[\protect\citeauthoryear{{Sokal}, {Deen}, {Mace}, {Lee}, {Oh}, {Kim},
  {Kidder}  \& {Jaffe}}{{Sokal} et~al.}{2018}]{Sokal2018}
{Sokal} K.~R.,  {Deen} C.~P.,  {Mace} G.~N.,  {Lee} J.-J.,  {Oh} H.,  {Kim} H.,
   {Kidder} B.~T.,   {Jaffe} D.~T.,  2018, \mn@doi [\apj]
  {10.3847/1538-4357/aaa1e4}, \href
  {https://ui.adsabs.harvard.edu/abs/2018ApJ...853..120S} {853, 120}

\bibitem[\protect\citeauthoryear{{Soummer}, {Pueyo}  \& {Larkin}}{{Soummer}
  et~al.}{2012}]{Soummer2012}
{Soummer} R.,  {Pueyo} L.,   {Larkin} J.,  2012, \mn@doi [\apjl]
  {10.1088/2041-8205/755/2/L28}, \href
  {https://ui.adsabs.harvard.edu/abs/2012ApJ...755L..28S} {755, L28}

\bibitem[\protect\citeauthoryear{{Sparks} \& {Ford}}{{Sparks} \&
  {Ford}}{2002}]{SparksFord2002}
{Sparks} W.,  {Ford} H.,  2002, \mn@doi [\apj] {10.1086/342401}, \href
  {http://cdsads.u-strasbg.fr/abs/2002ApJ...578..543S} {578, 543}

\bibitem[\protect\citeauthoryear{{Spiegel} \& {Burrows}}{{Spiegel} \&
  {Burrows}}{2012}]{Spiegel2012}
{Spiegel} D.~S.,  {Burrows} A.,  2012, \mn@doi [\apj]
  {10.1088/0004-637X/745/2/174}, \href
  {http://cdsads.u-strasbg.fr/abs/2012ApJ...745..174S} {745, 174}

\bibitem[\protect\citeauthoryear{{Spina} et~al.,}{{Spina}
  et~al.}{2017}]{Spina2017}
{Spina} L.,  et~al., 2017, \mn@doi [\aap] {10.1051/0004-6361/201630078}, \href
  {https://ui.adsabs.harvard.edu/abs/2017A&A...601A..70S} {601, A70}

\bibitem[\protect\citeauthoryear{{Stapelfeldt}, {Krist}, {M{\'e}nard},
  {Bouvier}, {Padgett}  \& {Burrows}}{{Stapelfeldt}
  et~al.}{1998}]{Stapelfeldt1998}
{Stapelfeldt} K.~R.,  {Krist} J.~E.,  {M{\'e}nard} F.,  {Bouvier} J.,
  {Padgett} D.~L.,   {Burrows} C.~J.,  1998, \mn@doi [\apjl] {10.1086/311479},
  \href {https://ui.adsabs.harvard.edu/abs/1998ApJ...502L..65S} {502, L65}

\bibitem[\protect\citeauthoryear{{Stolker} et~al.,}{{Stolker}
  et~al.}{2020a}]{Stolker2020a}
{Stolker} T.,  et~al., 2020a, \mn@doi [\aap] {10.1051/0004-6361/201937159},
  \href {https://ui.adsabs.harvard.edu/abs/2020A&A...635A.182S} {635, A182}

\bibitem[\protect\citeauthoryear{{Stolker}, {Marleau}, {Cugno}, {Molli{\`e}re},
  {Quanz}, {Todorov}  \& {K{\"u}hn}}{{Stolker} et~al.}{2020b}]{Stolker2020b}
{Stolker} T.,  {Marleau} G.~D.,  {Cugno} G.,  {Molli{\`e}re} P.,  {Quanz}
  S.~P.,  {Todorov} K.~O.,   {K{\"u}hn} J.,  2020b, \mn@doi [\aap]
  {10.1051/0004-6361/202038878}, \href
  {https://ui.adsabs.harvard.edu/abs/2020A&A...644A..13S} {644, A13}

\bibitem[\protect\citeauthoryear{{Szul{\'a}gyi}}{{Szul{\'a}gyi}}{2017}]{Szulagyi2017a}
{Szul{\'a}gyi} J.,  2017, \mn@doi [\apj] {10.3847/1538-4357/aa7515}, \href
  {http://adsabs.harvard.edu/abs/2017ApJ...842..103S} {842, 103}

\bibitem[\protect\citeauthoryear{{Taylor} \& {Storey}}{{Taylor} \&
  {Storey}}{1984}]{Taylor1984}
{Taylor} K.~N.~R.,  {Storey} J.~W.~V.,  1984, \mn@doi [\mnras]
  {10.1093/mnras/209.1.5P}, \href
  {https://ui.adsabs.harvard.edu/abs/1984MNRAS.209P...5T} {209, 5P}

\bibitem[\protect\citeauthoryear{{Tognelli}, {Prada Moroni}  \&
  {Degl'Innocenti}}{{Tognelli} et~al.}{2011}]{Tognelli2011}
{Tognelli} E.,  {Prada Moroni} P.~G.,   {Degl'Innocenti} S.,  2011, \mn@doi
  [\aap] {10.1051/0004-6361/200913913}, \href
  {https://ui.adsabs.harvard.edu/abs/2011A&A...533A.109T} {533, A109}

\bibitem[\protect\citeauthoryear{{Ubeira-Gabellini}, {Christiaens}, {Lodato},
  {Ancker}, {Fedele}, {Manara}  \& {Price}}{{Ubeira-Gabellini}
  et~al.}{2020}]{Ubeira-Gabellini2020}
{Ubeira-Gabellini} M.~G.,  {Christiaens} V.,  {Lodato} G.,  {Ancker} M. v.~d.,
  {Fedele} D.,  {Manara} C.~F.,   {Price} D.~J.,  2020, \mn@doi [\apjl]
  {10.3847/2041-8213/ab7019}, \href
  {https://ui.adsabs.harvard.edu/abs/2020ApJ...890L...8U} {890, L8}

\bibitem[\protect\citeauthoryear{{Wang}, {Mundt}, {Henning}  \& {Apai}}{{Wang}
  et~al.}{2004}]{Wang2004}
{Wang} H.,  {Mundt} R.,  {Henning} T.,   {Apai} D.,  2004, \mn@doi [\apj]
  {10.1086/425493}, \href
  {https://ui.adsabs.harvard.edu/abs/2004ApJ...617.1191W} {617, 1191}

\bibitem[\protect\citeauthoryear{{Wang} et~al.,}{{Wang}
  et~al.}{2020}]{Wang2020}
{Wang} J.~J.,  et~al., 2020, \mn@doi [\aj] {10.3847/1538-3881/ab8aef}, \href
  {https://ui.adsabs.harvard.edu/abs/2020AJ....159..263W} {159, 263}

\bibitem[\protect\citeauthoryear{{Wang} et~al.,}{{Wang}
  et~al.}{2021}]{Wang2021}
{Wang} J.~J.,  et~al., 2021, arXiv e-prints, \href
  {https://ui.adsabs.harvard.edu/abs/2021arXiv210104187W} {p. arXiv:2101.04187}

\bibitem[\protect\citeauthoryear{{Weingartner} \& {Draine}}{{Weingartner} \&
  {Draine}}{2001}]{Weingartner2001}
{Weingartner} J.~C.,  {Draine} B.~T.,  2001, \mn@doi [\apj] {10.1086/318651},
  \href {https://ui.adsabs.harvard.edu/abs/2001ApJ...548..296W} {548, 296}

\bibitem[\protect\citeauthoryear{{Wertz}, {Absil}, {G{\'o}mez Gonz{\'a}lez},
  {Milli}, {Girard}, {Mawet}  \& {Pueyo}}{{Wertz} et~al.}{2017}]{Wertz2017}
{Wertz} O.,  {Absil} O.,  {G{\'o}mez Gonz{\'a}lez} C.~A.,  {Milli} J.,
  {Girard} J.~H.,  {Mawet} D.,   {Pueyo} L.,  2017, \mn@doi [\aap]
  {10.1051/0004-6361/201628730}, \href
  {http://cdsads.u-strasbg.fr/abs/2017A%26A...598A..83W} {598, A83}

\bibitem[\protect\citeauthoryear{{Williams} \& {Cieza}}{{Williams} \&
  {Cieza}}{2011}]{Williams2011}
{Williams} J.~P.,  {Cieza} L.~A.,  2011, \mn@doi [\araa]
  {10.1146/annurev-astro-081710-102548}, \href
  {http://cdsads.u-strasbg.fr/abs/2011ARA%26A..49...67W} {49, 67}

\bibitem[\protect\citeauthoryear{{Winn} \& {Fabrycky}}{{Winn} \&
  {Fabrycky}}{2015}]{Winn2015}
{Winn} J.~N.,  {Fabrycky} D.~C.,  2015, \mn@doi [\araa]
  {10.1146/annurev-astro-082214-122246}, \href
  {http://cdsads.u-strasbg.fr/abs/2015ARA%26A..53..409W} {53, 409}

\bibitem[\protect\citeauthoryear{{Woitke} \& {Helling}}{{Woitke} \&
  {Helling}}{2003}]{Woitke2003}
{Woitke} P.,  {Helling} C.,  2003, \mn@doi [\aap] {10.1051/0004-6361:20021734},
  \href {https://ui.adsabs.harvard.edu/abs/2003A&A...399..297W} {399, 297}

\bibitem[\protect\citeauthoryear{{Woitke} \& {Helling}}{{Woitke} \&
  {Helling}}{2004}]{Woitke2004}
{Woitke} P.,  {Helling} C.,  2004, \mn@doi [\aap] {10.1051/0004-6361:20031605},
  \href {https://ui.adsabs.harvard.edu/abs/2004A&A...414..335W} {414, 335}

\bibitem[\protect\citeauthoryear{{Wright} et~al.,}{{Wright}
  et~al.}{2010}]{Wright2010}
{Wright} E.~L.,  et~al., 2010, \mn@doi [\aj] {10.1088/0004-6256/140/6/1868},
  \href {http://cdsads.u-strasbg.fr/abs/2010AJ....140.1868W} {140, 1868}

\bibitem[\protect\citeauthoryear{{Wu} \& {Sheehan}}{{Wu} \&
  {Sheehan}}{2017}]{Wu2017a}
{Wu} Y.-L.,  {Sheehan} P.~D.,  2017, \mn@doi [\apjl]
  {10.3847/2041-8213/aa8771}, \href
  {http://adsabs.harvard.edu/abs/2017ApJ...846L..26W} {846, L26}

\bibitem[\protect\citeauthoryear{{Zhu}}{{Zhu}}{2015}]{Zhu2015b}
{Zhu} Z.,  2015, \mn@doi [\apj] {10.1088/0004-637X/799/1/16}, \href
  {http://adsabs.harvard.edu/abs/2015ApJ...799...16Z} {799, 16}

\bibitem[\protect\citeauthoryear{{Zhu}, {Dong}, {Stone}  \& {Rafikov}}{{Zhu}
  et~al.}{2015}]{Zhu2015a}
{Zhu} Z.,  {Dong} R.,  {Stone} J.~M.,   {Rafikov} R.~R.,  2015, \mn@doi [\apj]
  {10.1088/0004-637X/813/2/88}, \href
  {http://adsabs.harvard.edu/abs/2015ApJ...813...88Z} {813, 88}

\bibitem[\protect\citeauthoryear{{van der Bliek}, {Manfroid}  \&
  {Bouchet}}{{van der Bliek} et~al.}{1996}]{van-der-Bliek1996}
{van der Bliek} N.~S.,  {Manfroid} J.,   {Bouchet} P.,  1996, \aaps, \href
  {https://ui.adsabs.harvard.edu/abs/1996A&AS..119..547V} {119, 547}

\bibitem[\protect\citeauthoryear{{van der Marel}, {Cazzoletti}, {Pinilla}  \&
  {Garufi}}{{van der Marel} et~al.}{2016}]{vanderMarel2016c}
{van der Marel} N.,  {Cazzoletti} P.,  {Pinilla} P.,   {Garufi} A.,  2016,
  \mn@doi [\apj] {10.3847/0004-637X/832/2/178}, \href
  {http://cdsads.u-strasbg.fr/abs/2016ApJ...832..178V} {832, 178}

\bibitem[\protect\citeauthoryear{{van der Marel} et~al.,}{{van der Marel}
  et~al.}{2021}]{van-der-Marel2021}
{van der Marel} N.,  et~al., 2021, \mn@doi [\aj] {10.3847/1538-3881/abc3ba},
  \href {https://ui.adsabs.harvard.edu/abs/2021AJ....161...33V} {161, 33}

\makeatother
\end{thebibliography}






\appendix

\section{VLT/NACO reduction pipeline for non-coronagraphic data}\label{app:NACOpip}

\begin{enumerate}
    \item First the pipeline automatically finds the stellar position in each science cube. This is done by a) subtracting the median of the closest three cubes in time (excluding the cube in question), which subtracts an estimate of both DARK current and sky background; b) removing spatial frequencies corresponding to either bad pixels or large scale detector variation (i.e.~using a low-pass and a high-pass filter); c) looking for maxima in the median image of each cube. We record the approximate ($x$,$y$) position of the star on the detector and the quadrant in which it is located.
    \item We then subtract the sky from the original frames of each cube, using the principal component analysis (PCA) based sky subtraction algorithm implemented in {\sc vip}. 
    For each science cube, the PCA library was set to the medians of each individual cube where the star is located in a different quadrant from the one considered. Both science and sky cubes are cropped to the quadrant of the detector in which the star is located. We tested 1 to 30 subtracted principal components ($n_{\rm pc}$), and noted no further improvement in dark current and sky background residuals for $n_{\rm pc}$ larger than 5. We hence adopted the latter value.
    \item The pipeline then calculates a master flat field from raw flats acquired during twilight at three different telescope altitudes, and apply it to all science cubes. It also computes a static bad pixel map based on flat values smaller than 0.67 and larger than 1.5. 
    \item We correct for NaN values, and subsequently bad pixels, using an iterative sigma filter algorithm implemented in {\sc vip} and designed to correct for clusters of bad pixels. The first pass of bad pixel correction uses the static bad pixel map inferred from the flat field, while the second pass looks iteratively for 8-$\sigma$ outliers to correct for cosmic rays. Bad pixels are replaced by the median of good neighbouring pixels.
    \item We subsequently find the centroid of the star by fitting a 2D Moffat function and shift each frame so that the star would fall exactly on the central pixel of each image (images are cropped to odd dimensions). 
    \item All recentered images are then gathered into a single master cube. For each frame of the master cube, the associated derotation angle required to align North up and East left is found by interpolation of the parallactic angle at start and end of the original cube it pertains to.
    \item From which we identify and remove bad frames, either corresponding to the opening/decrease in performance of the adaptive optics loop, or jittering during the integration which elongated the PSF. Bad frame trimming is based on the cross-correlation of each frame to the median of all frames in the cube. For each frame, the threshold used to consider a frame bad is a computed Pearson correlation factor (with the median frame) lower than 0.8, as calculated in a 7 FWHM x 7 FWHM cropped window centered on the star. For both NACO datasets, this removed $\sim$5\% of all images.
    \item Given the large number of images (over 26,000 and 55,000 for the 2017 and 2018 datasets, respectively), we then median-combine consecutive frames together 10 by 10 and 16 by 16 for the 2017 and 2018 datasets, respectively.
    \item The FWHM of the stellar point-spread function is then estimated by fitting a 2D gaussian profile to the median image of the whole cube. The flux of the star is then calculated by integration over a 1-FWHM aperture. 
    \item Finally, we post-process the calibrated cubes using both median-ADI \citep{Marois2006} in 5\arcsec $\times$ 5\arcsec~frames and PCA-ADI \citep[e.g.][]{Amara2012} in 2\arcsec $\times$ 2\arcsec frames, as implemented in {\sc vip}. For PCA-ADI, a range of 1 to 100 $n_{\rm pc}$ was explored.
\end{enumerate}

\section{IRDIS reduction pipeline for non-coronagraphic data}\label{app:IRDISpip}

\begin{enumerate}
\item The pipeline first calculates master sky background images using {\sc esorex}'s \verb+sph_ird_sky_bg+ recipe, upon provision of the raw sky images obtained during the sequence.

\item A master flat field and static bad pixel map are then calculated with the \verb+sph_ird_instrument_flat+ recipe, using raw flats and corresponding raw darks. 

\item The master sky is subtracted and the flat field is divided from all good pixels of science cubes using the \verb+sph_ird_science_dbi+ recipe.
Since this does not systematically yield an average background level of zero, we complemented by an additional manual subtraction of the residual sky so that the median pixel intensities at $>5\arcsec$ from the star is null.

\item Next, the pipeline uses {\sc vip} routines to correct for bad pixels using an iterative sigma filter algorithm. A first pass corrects for the static bad pixels identified in the master flat field, and a second pass corrects for all residual 8$\sigma$ outliers. 

\item A 2D Moffat profile is subsequently fit to the stellar PSF in each frame in order to find the centroid, and all images are shifted for the star to fall exactly on the central pixel of odd-size frames. 

\item All centered cubes are then collated in a single master cube, and corresponding derotation angles to align north up and east left are calculated. The derotation angles are interpolated for each frame of each cube, based on the parallactic angles at start and end of each cube, and consider the true north value of -1.75$\pm$0.08\degr~measured in \citet{Maire2016}.

\item Bad frames are then identified and removed according to both pixel intensities (rejecting stellar fluxes 1$\sigma$ below the median flux) and Pearson correlation coefficient calculated with respect to the median frame.

\item The anamorphism measured in \citet{Maire2016} is then corrected by rescaling the image along the $y$ dimension using a fourth-order Lanczos interpolation.

\item The pipeline finally post-processes the calibrated cube using median-ADI on full frames, and both PCA-ADI and sPCA \citep{Absil2013} on 2\arcsec$\times$2\arcsec cropped frames. sPCA performs PCA-ADI in concentric annuli, with the PCA library for the annulus of each image built from the same annulus in other images of the cube where a putative planet would have rotated by more than a given threshold.
The PCA-ADI algorithms are run with 1 to 50 $n_{\rm pc}$ subtracted, and an angular threshold ranging from 0.5$\times$FWHM (for the innermost annulus) to 1$\times$FWHM (for the outermost annulus) linear motion was chosen for sPCA.
\end{enumerate}

\section{IFS reduction pipeline for non-coronagraphic data}\label{app:IFSpip}

\begin{enumerate}
\item The pipeline first calculates master darks for all images (science or other calibrations) with different integration times using the \verb+sph_ifs_master_dark+ recipe. This is necessary because the {\sc esorex} IFS recipes (version 3.13.2) do not appear to scale master darks when applied to images obtained with different integration times. Master darks are subtracted manually to the raw flats, pairing them based on integration time. 

\item Master detector flats are then calculated using the \verb+sph_ifs_master_detector_flat+ recipe. This is done in four steps, where the output of each step is used as input in the following step: first a preamplifier flat is calculated using broad-band lamp raw flats; second large-scale coloured flats are calculated for each of the 4 narrow-band flat lamps; third a large-scale white lamp flat is calculated; and fourth a small-scale coloured flat is calculated after removing the large scale structures (captured with a smoothing length of 5 pixels). In all subsequent recipes, we provided either large-scale coloured flats and/or the small-scale white flat as detector flat inputs for optimal performance.

\item Next, master sky background images are calculated from the sky cubes acquired during the observation, using the  recipe.

\item The spectra positions on the IFS detector are then determined using the \verb+sph_ifs_spectra_positions+ recipe. We changed the default value for the \verb+distortion+ option and set it to False, as we noticed letting it to True led to significant negative/positive parallel stripes in some of the spectral channels of the final cubes.

\item Both the master detector flats and spectra positions are subsequently provided as input to the \verb+sph_ifs_instrument_flat+ recipe in order to compute a total instrument flat.

\item The exact wavelength solution is then found by providing both the outputs of the above steps and raw wave calibration files using the \verb+sph_ifs_wave_calib+ recipe.

\item A master IFU flat is calculated by passing the outputs of previous steps to the \verb+sph_ifs_instrument_flat+ recipe.

\item All science cubes are then calibrated by providing the outputs of steps ii to vii to the \verb+sph_ifs_science_dr+ recipe. Since sky subtraction is sub-optimal, we manually correct for the residual sky level in order to have a median level of zero in 0\farcs2~apertures near the four corners of the IFS field. 

\item Next, the pipeline uses {\sc vip} routines for bad pixel correction, centering, bad frame removal and anamorphism correction in a same way as for IRDIS (steps iv to vii in Appendix~\ref{app:IRDISpip}). For IFS, the correction of the anamorphism involves rescaling in both $y$ and $x$ owing to the rotation of the field-of-view.

\item Finally, the pipeline post-processes the calibrated 4D IFS cube leveraging on spectral differential imaging \citep[SDI;][]{SparksFord2002} and angular differential imaging. More specifically: (a) PCA-SDI was applied on each individual spectral cube, and the SDI images are then derotated and median-combined (simply referred to as PCA-SDI throughout this paper); (b) PCA-ADI was applied at each wavelength on the 3D cubes composed of each individual spectral channel sampled in the temporal direction; (c) and PCA-\rep{ASDI}, combining both the radial and azimuthal diversity of SDI and ADI together, was applied to reach the highest contrast at the expense of possible self-subtraction of extended disc features. As opposed to the version of PCA-\rep{ASDI} used in \citet{Christiaens2019a} on SINFONI data 
the algorithm only builds a single PCA library containing both spectral and angular diversity. Our tests on these SPHERE/IFS data suggest that PCA-\rep{ASDI} in two steps reduce the SNR of the companion and appear to significantly lower the algorithmic throughput without gain in achieved contrast. 
A range of 1 to 10 $n_{\rm pc}$ was explored for both PCA-SDI and PCA-ASDI. 


\end{enumerate}

\section{Improvements to MCMC-NEGFC}\label{app:MCMC-NEGFC_improvements}

\begin{figure*}
	\centering
	\includegraphics[width=\textwidth]{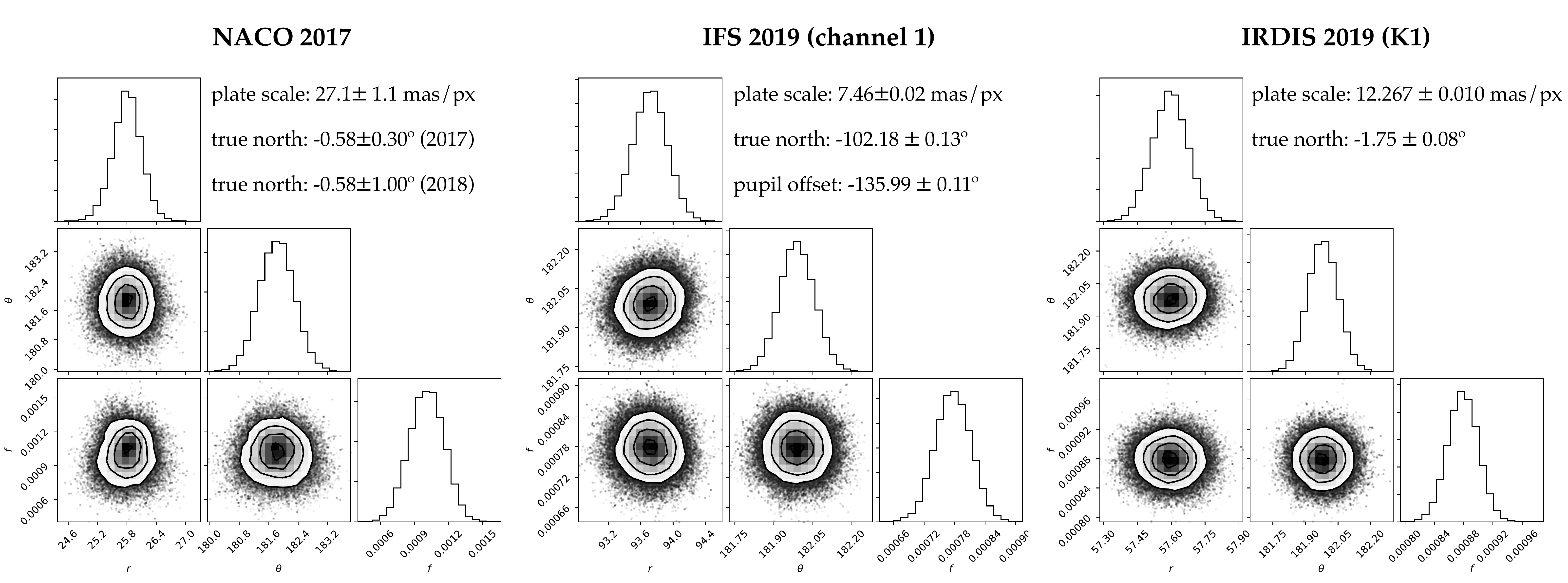}
    \caption{Corner plots for the radial separation, azimuth angle and contrast of the companion candidate retrieved by MCMC-NEGFC on the NACO 2017, IFS 2019 (first channel; lowest SNR) and IRDIS 2019 ($K1$ channel, highest SNR) datasets. The radial separations $r$ are in pixels and the azimuthal angles $\theta$ are measured counter-clockwise from the positive $x$ axis. We also show the systematic uncertainties considered for the calculation of final astrometric uncertainties for each instrument (NACO: \citealt{Milli2017}; SPHERE: \citealt{Maire2016}). 
    }
    \label{fig:MCMC-NEGFC_results}
\end{figure*}

In order for the MCMC algorithm to work (and converge) properly, 
we now use the following log-probability:
\begin{align}
\label{Eq:LogProba}
\log \mathcal{L} (D|M) =-\frac{1}{2} \sum_i \frac{(I_i - \mu)^2}{\sigma^2}
\end{align}
where $i$ are the indices of all the pixels in a 1.5-FWHM radius aperture centered on the location of our initial guess on the position of the companion; $I_i$ are the pixel intensities measured in that aperture in the post-processed PCA-ADI image; $\mu$ and $\sigma$ are the mean intensity and standard deviation measured in a 3-FWHM wide annulus at the same radial separation as the companion but excluding a region in azimuth near the location of the companion. The exclusion region is set to [PA$_b$-$\Delta$PA, PA$_b$+$\Delta$PA], where PA$_b$ is our initial estimate on the position angle of the companion, and $\Delta$PA is the range of parallactic angles covered by the ADI sequence (see Table~\ref{tab:CrA9_obs}).
We argue that Equation~\ref{Eq:LogProba} leads to a more robust estimate of the companion flux than the original expression (missing $\mu$ and $\sigma$). If the speckle residuals are not perfectly subtracted, as can be the case for a small $n_{\rm pc}$ subtracted, variable adaptive optics performance throughout the sequence, and/or companions very close to the central star, the mean residual level $\mu$ at the separation of the companion can be non-null \citep[e.g.][]{Christiaens2018}. It is thus necessary to subtract that component. Scaling by $\sigma^2$ measured in an annulus that avoids the area around the point source also allows to include the speckle noise as an additional source of variance in the posterior distributions. Furthermore, for very faint companions in raw detector units (e.g. in ADI datacubes that have been normalised by the integration time), the absence of scaling by $\sigma^2$ 
would lead to large $\log \mathcal{L} (D|M)$ likelihoods whether the companion is well subtracted or not subtracted at all. 
In turn this leads to an unreliable posterior distribution for the flux, hence associated best estimate and uncertainty. 
The MCMC-NEGFC algorithm stops once a given convergence criterion is met. 
In the original version of the algorithm, the convergence criterion relied on a Gelman-Rubin test; i.e.~based on a comparison of the variance calculated from two sections of the chains \citep[][]{Gelman1992}. 
However, since the samples in a Markov chain are not independent, the Gelman-Rubin test is inadequate and may break the chain too early, resulting in underestimated variances of the posterior distributions, hence uncertainties. Therefore, we now use a test based on the integrated auto-correlation time $\tau$ \citep[e.g.][]{Goodman2010,Foreman-Mackey2013}. 
The estimate of $\tau$ is performed at regular intervals as the chain progresses, 
and becomes reliable if the number of samples $N$ on which it is computed is sufficiently large compared to the estimated $\tau$ ($N >> \tau$).
In practice, we considered a threshold of $N/\max(\tau_f) > 50$ where $\tau_f$ is the autocorrelation time for parameter $f$ (i.e. $r$, PA or contrast), as recommended
in the documentation of {\sc emcee} \citep{Foreman-Mackey2013}. 
Since the sampling uncertainty on the true variance of a parameter is $\propto \sqrt{\tau_f/N}$, 
the criterion we chose also implies 
a relative accuracy of $\lesssim 14$\% on the variances (and thereby uncertainties) inferred for each parameter. Using this convergence criterion, 
we noticed that a significantly larger number of steps was required before the convergence criterion was met compared to the Gelman-Rubin-based criterion: between 1500 and 2500 iterations were required for all our datasets, while the Gelman-Rubin-based criterion typically suggested convergence within 200 steps. We used 128 walkers for all datasets and a burn-in factor of 0.5, leading to a total of $>$100,000 samples for the posterior distributions on $r$, PA and contrast of the point source in each dataset. 

Our third modification is motivated by the varying quality of the adaptive optics correction throughout all our ADI sequences, which resulted in significant scatter 
for the measured stellar fluxes (a proxy for the Strehl ratio).
Only the NACO 2017 dataset showed a low relative standard deviation of 3.2\% for stellar fluxes measured in all frames (with respect to the median flux). The NACO 2018, IFS 2019 and IRDIS 2019 (unsaturated) datasets showed relative standard deviations of 24.7\%, 13.9\% and 8.7\% (after bad frame removal), with individual flux measurements varying by up to a factor 2.
Fortunately, since the point source was recovered in all our unsaturated datasets, we could measure the stellar flux in each individual image, and use that information to inject the flux of the negative companion in each image proportionally. The injected flux in frame $i$ is:
\begin{align}
\label{Eq:InjectedFlux}
F_i = F \times \frac{F_{*\mathrm{,}i}}{F_{*\mathrm{,med}}}
\end{align}
where $F$ is the companion candidate flux sampled by the MCMC-NEGFC algorithm, $F_{*\mathrm{,}i}$ is the stellar flux measured in frame $i$ and $F_{*\mathrm{,med}}$ is the median stellar flux.
This modification allowed us to get up to an order of magnitude improvement in accuracy for the estimated contrast of the point source with respect to the star in the different datasets.
Since for coronagraphic datasets one does not have the luxury of knowing how the stellar flux (hence the companion flux) varies throughout the observing sequence, this modification was only implemented as an additional option in {\sc vip}'s {\sc mcmc\_negfc\_sampling} function.




Figure~\ref{fig:MCMC-NEGFC_results} shows three examples of the results obtained by MCMC-NEGFC among the 43 ADI sequences considered -- one for each filter or spectral channel of the NACO, IFS and IRDIS datasets.
We selected the best NACO $L'$ dataset (2017), the first channel of the IFS dataset (lowest SNR for the companion), and the $K1$ band of the IRDIS dataset (highest SNR for the companion).

\section{Images obtained with PCA-SDI, PCA-\rep{ASDI} and MAYO}\label{app:PCA-SDI_PCA-SADI}

\begin{figure*}
	\centering
	\includegraphics[width=\textwidth]{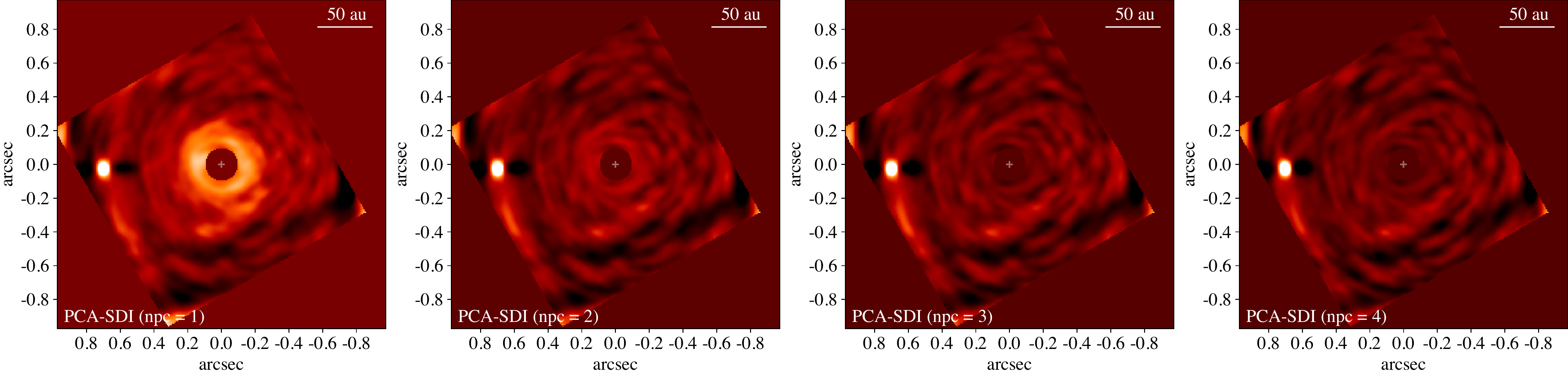}
	\includegraphics[width=\textwidth]{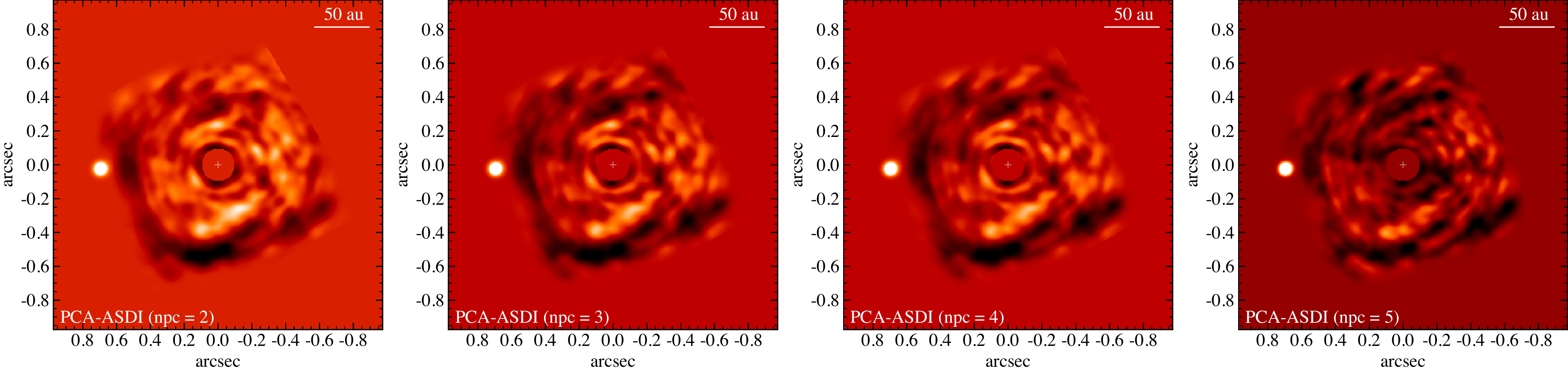}
    \caption{PCA-SDI (top row) and PCA-\rep{ASDI} (bottom row) images obtained on the SPHERE/IFS dataset. A tentative spiral pattern is detected, with a possible primary arm pointing towards the point source. 
    }
    \label{fig:PCA-SDI_PCA-SADI}
\end{figure*}

\begin{figure}
	\centering
	\includegraphics[width=0.9\columnwidth]{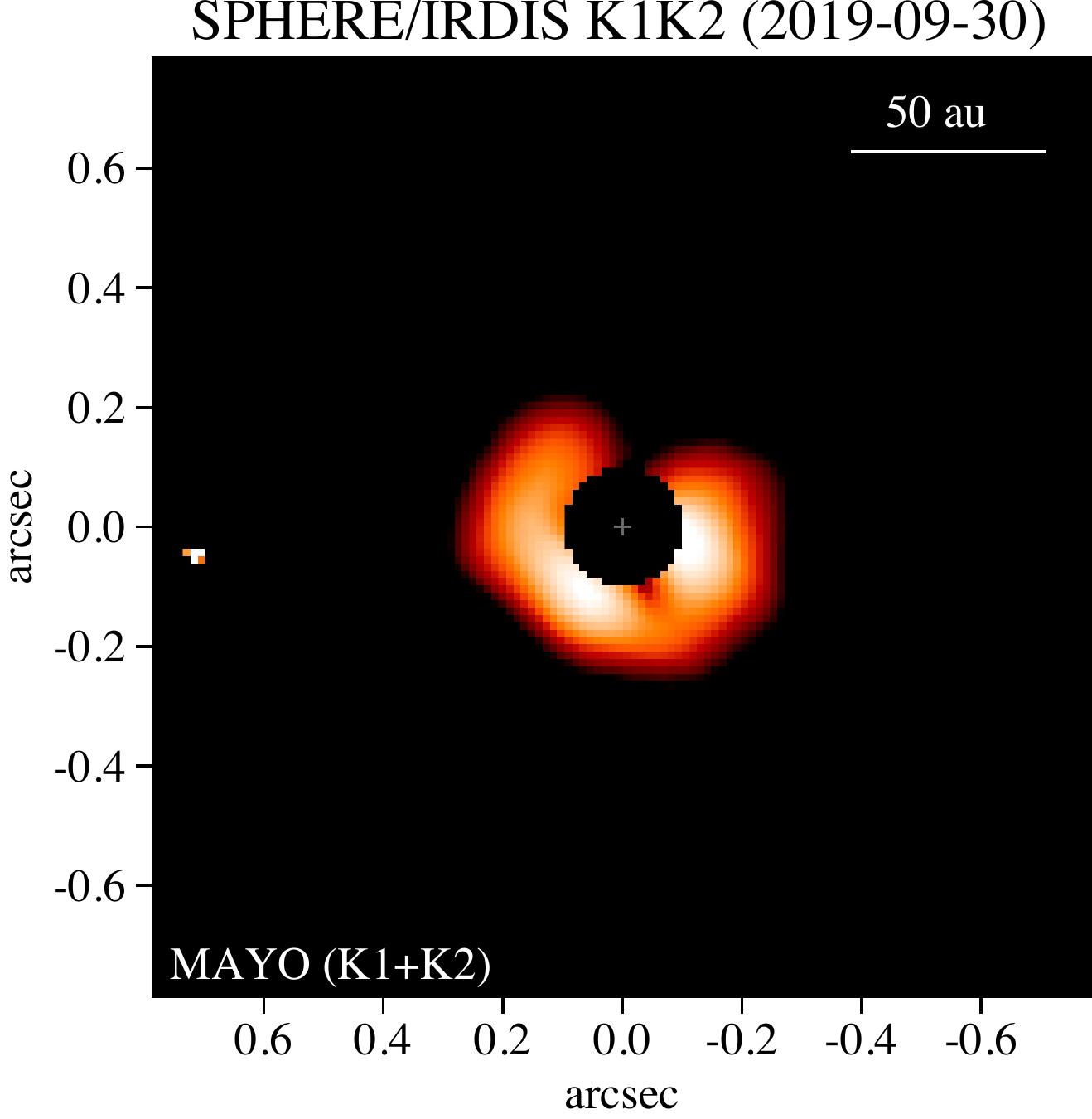}
    \caption{IRDIS $K1$+$K2$ image obtained using {\sc mayonnaise}.
    }
    \label{fig:MAYO}
\end{figure}

Figure~\ref{fig:PCA-SDI_PCA-SADI} shows the post-processed IFS images obtained with PCA-SDI and PCA-\rep{ASDI} for increasing number of principal components subtracted.
The purpose of the figure is to evaluate the reliability of the spiral pattern. As expected for an authentic disc feature, the morphology of the signal is preserved but gets progressively self-subtracted for increasing $n_{\rm pc}$. Compared to the PCA-SDI images, the PCA-\rep{ASDI} images removed all the azimuthally extended signals due to angular differential imaging. \rep{Since we were only interested in testing the reliability of the spiral pattern, we ran the PCA-ASDI algorithm using the} \verb+crop_ifs+ \rep{option of} {\sc vip}'s \verb+pca+ \rep{routine, which crops frames after rescaling in order to both save memory and decrease computation time. Cropping removes different amount of information radially in different spectral channels and accounts for the abrupt edges of the cropped field in the final median-combined frame, however it does not affect the signal at shorter separation.}

We also applied {\sc mayonnaise} on the IRDIS dataset \citep{Pairet2020}. Contrary to PCA, {\sc mayonnaise} projects sparse (planet-like) and extended (disc-like) signals on different bases, which allows both components to be restored and disentangled. {\sc mayonnaise} recovered sparse emission from the companion, and possibly extended emission from the protoplanetary disc (Figure~\ref{fig:MAYO}). 
However, given the absence of spirals in the IRDIS image and their low SNR in the IFS image ($\sim 3$ in the PCA-SDI $n_{\rm pc}=1$ image), 
higher sensitivity data are required to confirm the authenticity of the spirals.

\section{Corner plot of CrA-9}\label{app:specfit_corner_plots}

\begin{figure}
	\centering
	\includegraphics[width=\columnwidth]{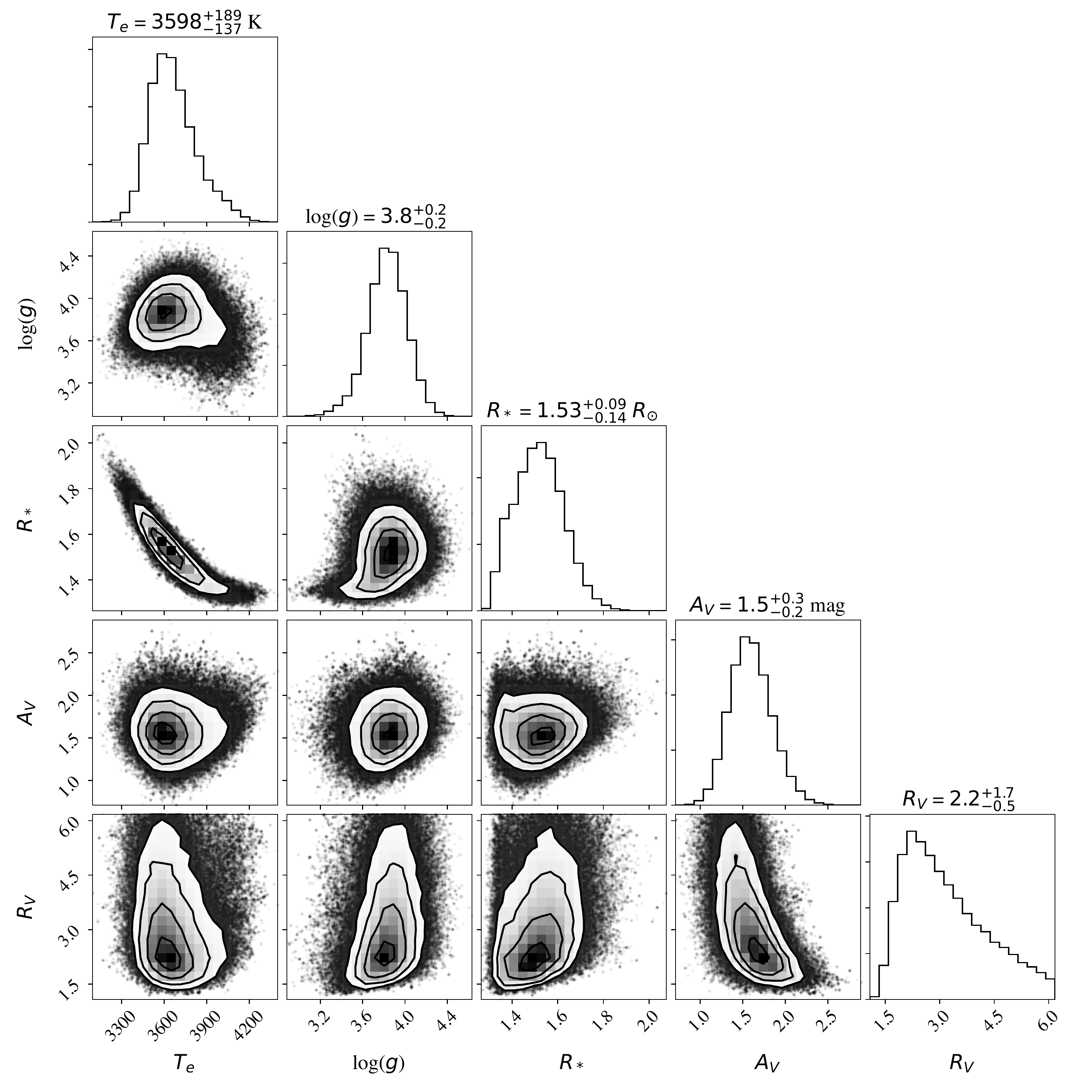}
    \caption{Corner plot showing the posterior distribution for the parameters derived by \texttt{specfit} for CrA-9. 
    We note minor degeneracies (i) between the effective temperature $T_e$, the radius $R_*$ and the optical extinction $A_V$, and (ii) between $A_V$ and the total-to-selective optical extinction ratio $R_V$.
    }
    \label{fig:SED_cornerplot}
\end{figure}

The corner plot showing the most likely physical parameters for the star inferred by {\sc specfit} is presented in Figure~\ref{fig:SED_cornerplot}.

\section{\texttt{specfit} tests on the companion spectrum}\label{app:specfit_tests}

\begin{figure*}
	\centering
	\includegraphics[width=0.67\textwidth]{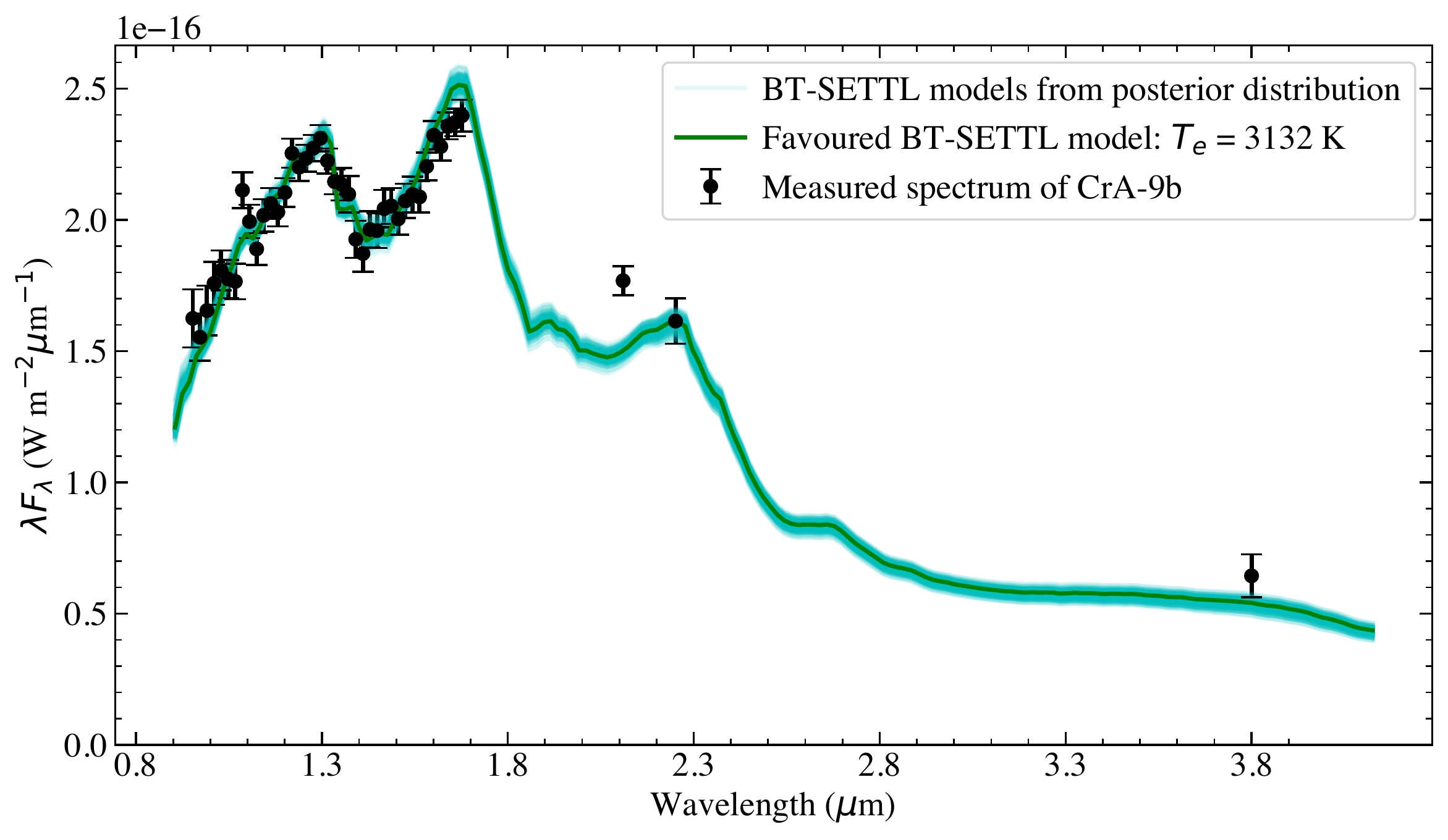}
    \caption{\rep{BT-SETTL models retrieved by \texttt{specfit} when using a gaussian prior of $\mu=3.7$ and $\sigma=0.2$ for $\log(g)$. This leads to only a slightly poorer fit than $\log(g) \sim 4.6$, obtained with uniform priors. Favoured parameters are $T_e = 3132$K, $\log(g) = 3.9$, $R_b = 0.57 R_J$ and $A_V = 2.0$ mag.}
    }
    \label{fig:specfit_test_logg}
\end{figure*}

\begin{figure*}
	\centering
	\includegraphics[width=0.67\textwidth]{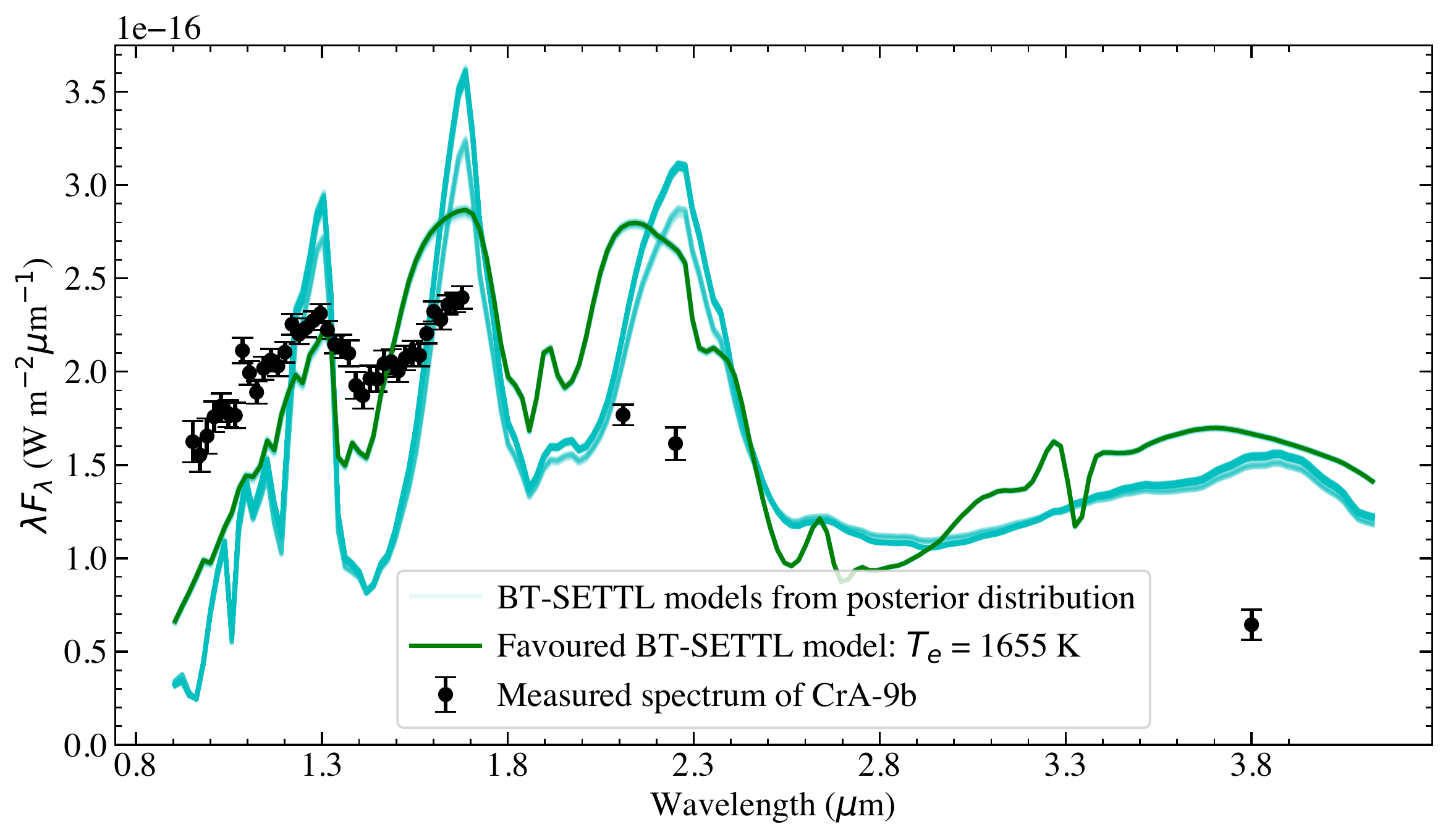}
	\\
	\includegraphics[width=0.67\textwidth]{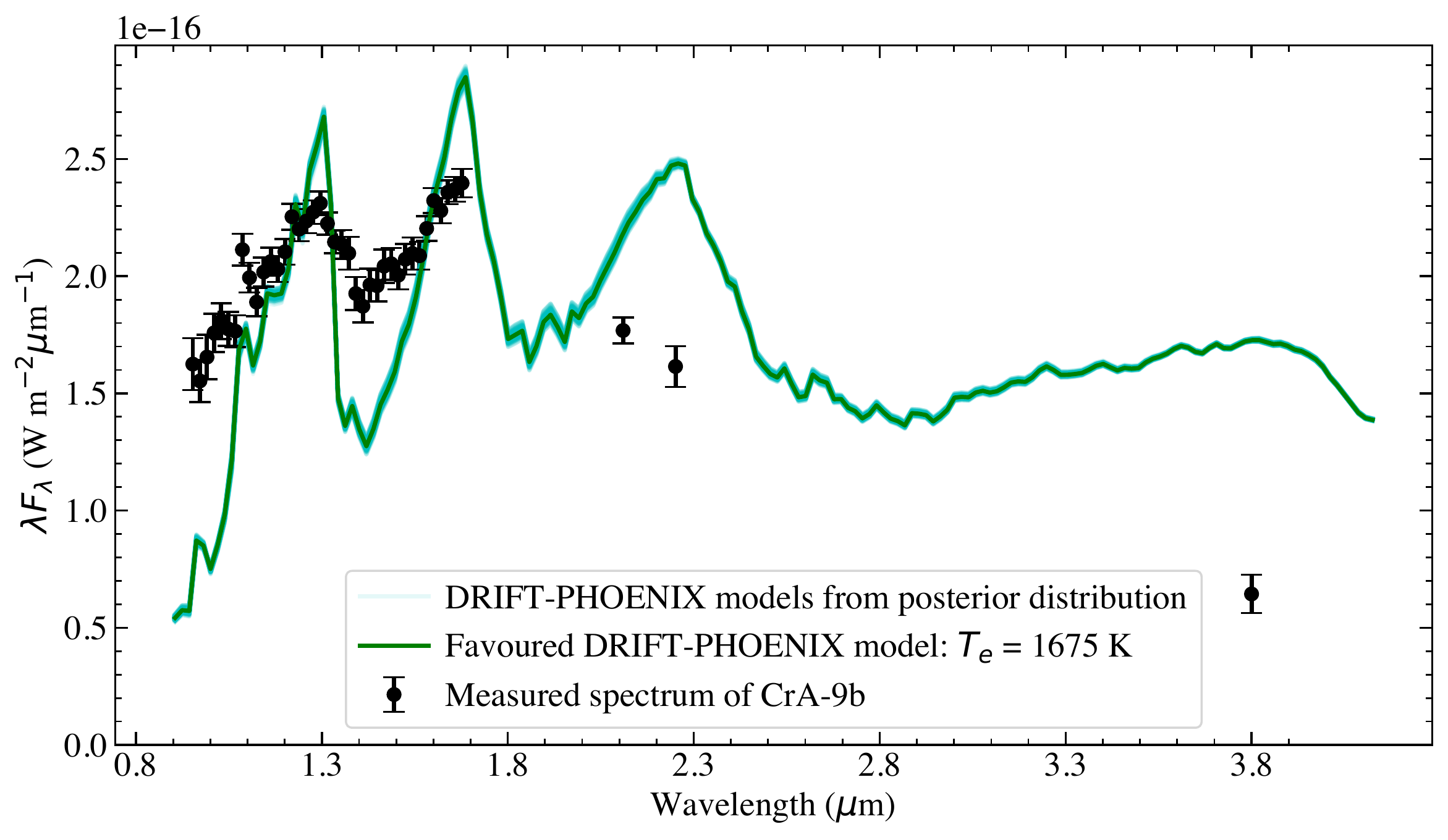}
    \caption{\rep{BT-SETTL (top) and DRIFT-PHOENIX (bottom) models retrieved by \texttt{specfit} when fixing the photometric radius to 1.8 $R_J$. Poor fits are obtained when fixing the photometric radius to the expected physical radius of a young Jovian planet with the absolute near-IR magnitudes of the companion.
    }}
    \label{fig:specfit_test_R}
\end{figure*}

\rep{Figure~\ref{fig:specfit_test_logg} shows the best-fit model obtained after running {\sc specfit} using a gaussian prior of $\mu=3.7$ and $\sigma=0.1$ for $\log(g)$ with the BT-SETTL grid of models. The favoured parameters are $T_e = 3132$K, $\log(g) = 3.9$, $R_b = 0.57 R_J$ and $A_V = 2.0$ mag. We notice only a slight decrease in the quality of the fit with respect to the results obtained after letting the surface gravity as a free parameter ($\Delta$AIC $\sim$21), with only the $H$ band measurements being slightly overpredicted.}

\rep{The BT-SETTL and DRIFT-PHOENIX models favoured by {\sc specfit} when the photometric radius is fixed to 1.8 $R_J$ are shown in Figure \ref{fig:specfit_test_R}. We did not test BT-DUSTY models with such constraint due to the incompleteness of this grid at lower temperatures than 3000 K. The plots show that this constraint leads to visually poor fits. This is also conveyed by the $\Delta$AIC values of $\sim$2590 and $\sim$2130 achieved by the BT-SETTL and DRIFT-PHOENIX models, respectively. For the BT-SETTL fit, we notice that {\sc specfit} converged on three (equally bad) clusters of parametric solutions with effective temperature lower than 2000~K, while a single kind of solution at an effective temperature of $\sim$1675 K was found with the DRIFT-PHOENIX grid. The favoured parameters are $T_e = 1655$K, $\log(g) = 5.5$, $R_b = 1.8 R_J$ and $A_V = 0.0$ mag for BT-SETTL, and $T_e = 1675$K, $\log(g) = 3.3$, $R_b = 1.8 R_J$ and $A_V = 0.0$ mag for DRIFT-PHOENIX.}

\bsp	
\label{lastpage}
\end{document}